\documentclass[aps, reprint,groupedaddress,longbibliography,superscriptaddress]{revtex4-2}
\usepackage{graphicx, tabularx, longtable, nicefrac, amsmath,enumerate, supertabular, eurosym}
\usepackage{hyperref,xcolor}
\hypersetup{breaklinks=true, colorlinks = true, citecolor = blue!30!black, linkcolor = blue!30!black}
\usepackage{booktabs}

\newcommand{\env}[1]{\texttt{#1}}
\usepackage[utf8]{inputenc}
\usepackage[T1]{fontenc}
\usepackage{tikz}
\usepackage{pgfplots}
\pgfplotsset{compat=1.8}
\usepgfplotslibrary{polar}
\definecolor{bblack}{HTML}{363033}
\usepackage{xcolor, booktabs}
\definecolor{grun}{rgb}{0.0, 0.5, 0.0}
\definecolor{amber}{rgb}{1, 0.49, 0.0}
\definecolor{alizarin}{rgb}{0.82, 0.1, 0.26}
\definecolor{dblau}{RGB}{21,50,104}
\definecolor{hblau}{RGB}{132,191,234}
\definecolor{chamois}{RGB}{243,226,216}
\definecolor{altweis}{RGB}{246,244,240}
\definecolor{dgrau}{RGB}{0,101,141}
\definecolor{hgrau}{RGB}{87,87,86}
\usetikzlibrary{patterns, shapes,arrows,positioning, shadows,fit,backgrounds}
\usepackage{setspace}
\usepackage{tabu}
\usepackage{multirow}
\usepackage{xcolor}
\usepackage{calc}
\usepackage{amsmath,amsthm,amssymb,euscript} 
\usepackage{enumerate,graphicx,setspace}
\usepackage{times}
\usepackage{longtable}
\usepackage{url}
\usepackage{enumitem}
\usepackage{comment}

\begin{document}

\title{Physics lab courses under digital transformation: A tri-national survey among university lab instructors about the role of new digital technologies and learning objectives}

\author{Simon Zacharias Lahme}
\email{simon.lahme@uni-goettingen.de}
\affiliation{Faculty of Physics, Physics Education Research, University of Göttingen, Friedrich-Hund-Platz 1, 37077 Göttingen, Germany}

\author{Pascal Klein}
\affiliation{Faculty of Physics, Physics Education Research, University of Göttingen, Friedrich-Hund-Platz 1, 37077 Göttingen, Germany}

\author{Antti Lehtinen}
\altaffiliation[Also at: ]{Department of Teacher Education}
\affiliation{Department of Physics, P.O. Box 35, 40014 University of Jyväskylä, Finland}

\author{Andreas Müller}
\affiliation{Faculty of Sciences, Department of Physics, University of Geneva, Boulevard du Pont d'Arve 40, 1211 Genève, Switzerland}

\author{Pekka Pirinen}
\affiliation{Department of Physics, P.O. Box 35, 40014 University of Jyväskylä, Finland}

\author{Lucija Rončević}
\affiliation{Department of Applied Physics, Faculty of Electrical Engineering and Computing, University of Zagreb, Unska 3, 10000 Zagreb, Croatia}

\author{Ana Sušac}
\affiliation{Department of Applied Physics, Faculty of Electrical Engineering and Computing, University of Zagreb, Unska 3, 10000 Zagreb, Croatia}

\date{\today}

\begin{abstract}
Physics lab courses permanently undergo transformations, in recent times especially to adapt to the emergence of new digital technologies and the Covid-19 pandemic in which digital technologies facilitated distance learning. Since these transformations often occur within individual institutions, it is useful to get an overview of these developments by capturing the status quo of digital technologies and the related acquisition of digital competencies in physics lab courses. Thus, we conducted a survey among physics lab instructors ($N=79$) at German, Finnish, and Croatian universities. The findings reveal that lab instructors already use a variety of digital technologies and that the pandemic particularly boosted the use of smartphones/tablets, simulations, and digital tools for communication/collaboration/organization. The participants generally showed a positive attitude toward using digital technologies in physics lab courses, especially due to their potential for experiments and students' competence acquisition, motivational effects, and contemporaneity. Acquiring digital competencies is rated as less important than established learning objectives, however, collecting and processing data with digital tools was rated as an important competency that students should acquire. The instructors perceived open forms of labwork and particular digital technologies for specific learning objectives (e.g., microcontrollers for experimental skills) as useful for reaching their learning objectives. Our survey contributes to the reflection of what impact the emergence of digital technologies in our society and the Covid-19 pandemic had on physics lab courses and reveals first indications for the future transformation of hands-on university physics education.
\end{abstract}

\maketitle

\section{Introduction}

For over 150 years, physics education in schools, colleges, and universities consists not only of lectures and exercises but also of practical elements in which students attend and/or conduct physics experiments. In the United States, "[t]hroughout the late 1800s and early 1900s physicists and teachers [...] expand the role of laboratories and projects that actively engage students [...], the use of hands-on student laboratory activities increased dramatically" \cite[p.53]{Otero.2017}. At the same time, European physicists like Friedrich Kohlrausch started to educate students in their laboratories which was the cornerstone for physics lab courses as known nowadays \cite{Sacher.2020}. This development was caused by the desire to foster "habits of scientific thought" with "powerful examples of the method by which science obtains its results" \cite[p.290]{TheNewMovementamongPhysicsTeachers.1907} for which today the term \textit{scientific practice} is used \cite{Otero.2017}.

Since then, the way of teaching and learning in lab courses has been subject to several transformations (cf. also Ref.~\cite{Hofstein.2003}). For example, the pursued learning objectives have been discussed and revised by different physics associations (e.g., in the United States \cite{AmericanAssociationofPhysicsTeachers.1998,AmericanAssociationofPhysicsTeachers.2014} or in Germany \cite{KonferenzderFachbereichePhysik.2010}) and empirically compiled through surveys among lab instructors (e.g., for six European countries \cite{Welzel.1998,Welzel.1998b}, at the University of Colorado \cite{Zwickl.2013}, or in Germany/Austria \cite{Nagel.2018,Vorholzer.2022}).
Another example of transformation is the organizational frame of the lab courses, especially in terms of student numbers and degree of openness. While in the late 1800s students were invited to the researchers' real laboratories, with the increasing number of physics students in the first half of the 1900s large-scale lab courses with demonstration experiments and highly structured, cookbook-like instructions were implemented \cite{Sacher.2020}. Today, a trend back to smaller groups, more open, authentic, problem-orientated, inquiry-based investigation can be observed that goes along with a shift from reinforcing concepts to fostering competencies like experimental skills. Two recent strands of development here are \textit{design labs} and \textit{undergraduate research projects} \cite{Holmes.2016} having in common a greater autonomy of students in carrying out experiments in laboratory, course, homework, or other educational settings. In \textit{design labs} (cf. Refs.~\cite{Karelina.2007,Karelina.2007b,Etkina.2015,Kontro.2018,Sacher.2020}), students design experiments themselves, instead of following the given cookbook-styled instructions, but with strong scaffolding by guidance about the key steps and requirements of a scientific experiment, and rubrics for self-assessment \cite{Karelina.2007b}. \textit{Undergraduate research projects}, in which usually groups of students work on their own small research projects (cf. Refs.~\cite{Barro.2023,Planinsic.2007}), are widely advocated in the literature on undergraduate education, in general \cite{RuizPrimo.2011,Russell.2007,NationalAcademiesofSciencesEngineeringandMedicine.2017,Abraham.2022,Ahmad.2022,Oliver.2023}, and specifically as a way to promote \textit{higher order thinking skills} (HOTs) such as autonomy, curiosity, creativity, problem-solving, and critical thinking \cite{Zohar.2015,Walsh.2019,Murtonen.2019,Mieg.2022}.

Exemplary causes for such transformations are changed requirements in the labor market either in research or private enterprise. With new digital technologies or employees' expectations of employers' hard and soft skills, new requirements are placed on the education of physics students. Physics lab courses are especially prone to adapt to such transformations more than lectures or exercises as their practical orientation rather resembles that of the work processes in the labor market. Another cause of transformation is research findings on the effectiveness of lab courses revealing that traditional, cookbook-styled, concept-based lab courses are hardly effective in teaching physics content \cite{Holmes.2017,Holmes.2018} and expert-like views on experimental physics \cite{Teichmann.2022}), and do not meet students' expectations \cite{Rehfeldt.2017}. Instead, 
addressee-specific lab courses (e.g., for physics \cite{Neumann.2004}, medicine \cite{Theyen.1999,Klug.2017}, or physics pre-service teacher students \cite{Andersen.2020}) and a focus on experimental skills are more beneficial, the latter especially for acquiring critical thinking skills and appropriate views on experimentation \cite{Walsh.2022}. For design labs and undergraduate research projects with a range of objectives such as fostering authentic research practices, scientific abilities, and higher-order thinking skills research has shown that they can indeed be conducive to learning in this sense \cite{RuizPrimo.2011,Etkina.2015,Holmes.2018,Karelina.2007,Karelina.2007b}.

In recent times, there have been significant transformations on the digital front, mainly due to two factors: The first one is the emergence of a huge variety of digital technologies within the past about 25 years which allow simplified, automated, and faster data collection and analysis with higher precision (e.g., with computers, microcontroller, programming software), can serve as didactic tools (e.g., Virtual Reality), or facilitate communication, collaboration, and organization (e.g., digital media for presentation). Therefore, digital technologies impact the experimental setups, the methods of how experiments are conducted and analyzed, and how the lab courses are organized. In addition, new skills have become relevant for students during their studies and beyond in preparation for the labor market and physics research in technologized laboratories (cf. astrophysics or large-scale laboratories like at CERN). They need to learn the competent use of new digital technologies; thus, the transformation goes along with new digital learning objectives, especially in programming and automating data collection and analysis.

The second recent factor for transformation was the Covid-19 pandemic from the beginning of 2020 for over two years in which social-distancing rules prohibited or noticeably impeded on-campus teaching and learning, so new lab formats needed to be implemented rather quickly \cite{Werth.2022,Pols.2020,Bauer.2021,Bradbury.2020b,Hut.2020,Jelicic.folgt,Klein.2021b}. New pursued approaches for lab courses in distance learning were for example the use of second hand-data \cite{Klein.2021b,Werth.2022}, experiments with smartphones \cite{Pols.2020,Lahme.2022} and household items \cite{Werth.2022}, simulations \cite{Werth.2022}, or videos of the experiments either for preparation and instruction \cite{Hut.2020} or for demonstrating how an instructor conducts the experiment \cite{Werth.2022}. Furthermore, digital tools like video conferencing were used for communication and collaboration \cite{Hut.2020,Werth.2022} or elements of distance, on-campus, and remote, synchronous, and asynchronous learning were combined \cite{Bauer.2021,Werth.2022}. Probably as a consequence, also the social form of learning shifted from group work to individual work, and the learning objectives from developing experimental skills toward reinforcing concepts \cite{Werth.2022}.

With the recent digital leap driven by technological development and the necessity imposed by the pandemic, the time has come to take a unifying look at the contemporary role of digital technologies in physics lab courses. As instructors are usually independently responsible for their lab courses, transformations are probably not uniformly done, so such an overview allows us to reflect on the previous impact of the emergence of digital technologies in our society and the Covid-19 pandemic on physics lab courses in the different institutions. In perspective, this can help to identify goals and paths for future systemic transformation.

Thus, we surveyed the university lab instructors in three European countries, Germany, Finland, and Croatia inspired by a former study \cite{Welzel.1998,Welzel.1998b,Haller.1999} among physics, chemistry, and biology lab instructors in schools and universities in six different European countries in the late 90s. The study investigated the instructors' perspective of goals of labwork in science education. Thus, besides a ranking of the importance of different learning objectives, the participating instructors had also rated different forms of labwork according to their suitability for achieving different learning objectives. At that time, common use of computers had only emerged recently, and using computers in physics labs (e.g., for data analysis or interactive screen experiments) was an innovative approach that was perceived as rather less conducive to learning, especially by the German subgroup of physics professors in the mentioned study published in Ref.~\cite{Haller.1999}. Thus, the only form of labwork presented in that survey which was related to digital technologies was \textit{experiments using modern technologies
(e.g. for data capture)} \cite[p.78]{Welzel.1998b}. As digital technologies and multimedia are increasingly integral parts of physics education and are particularly investigated in relationship with physics experiments \cite{Girwidz.2019}, it is now interesting to investigate to which extent the instructors' perception of digital technologies has changed. So, we modified the questionnaire of Refs.~\cite{Welzel.1998,Welzel.1998b} by adding digital competencies as learning objectives, specific modern digital technologies as possible forms of labwork, and further questions about the experiences and attitudes toward digital technologies in general to capture a snapshot of the current role of digital technologies and learning objectives in European physics lab courses.

\section{State of research and research questions}\label{research}

\subsection{Role of digital technologies in physics labs}\label{roletechn}

According to Ref.~\cite{Franke.2022b}, lab courses are digitalized either for organizational reasons (increasing efficiency and removing barriers e.g., regarding time and location) or due to didactic thoughts about improving the teaching quality, the expansion of course content, and the improvement of practical relevance. While organizational advantages of digital technologies like facilitated collection and analysis of large data sets or higher precision of measurements are self-evident, the didactic potential needs to be researched. For example, Ref.~\cite{Becker.2020c} has shown that video analysis with tablets can reduce extraneous cognitive load and therefore increase conceptual understanding, Ref.~\cite{Hochberg.2018} has shown for experiments with smartphones/tablets and Ref.~\cite{Pirker.2017} has shown for 3D Virtual Reality environments that the respective technologies increase students' motivation, interest, and engagement. A review \cite{Jong.2013} summarizes that the right combination of virtual and physical elements in a lab course can outperform purely virtual or purely physical ones.

Digitalization can occur in four areas \cite{Franke.2022b}: The first is the \textit{laboratory environment} itself which can be digitalized by using digital hardware instead of analog equipment, integrating hardware within a network for (remote) controlled and automated data collection, or using digital technologies and multimedia applications (like microcontrollers, smartphones, simulations, Virtual Reality, etc.) also to enable experimenting outside of laboratories. Further areas of digitalization are related to \textit{processes and procedures} (e.g., by documenting, illustrating, modelizing, or simulating processes with digital technologies) and \textit{data} (e.g., digital-aided collection, documentation, or processing of data as well as the use of databases). The fourth area of digitalization is the \textit{interpersonal communication} ranging from digital learning materials over digital tools for communication, collaboration, and organization to digital feedback and assessment.

The \textit{laboratory environment} can be digitalized using a huge variety of technologies with different purposes of use, advantages, and limitation that can be located in the \textit{reality-virtuality continuum} \cite{Milgram.1994} depending on how real the experiments (setup and data) and the physical interaction are \cite{Ma.2006}. Far on the \textit{reality} side of this continuum, tools for digital-aided (hands-on) measurement data collection are located. The students interact with the real world and a real setup and utilize digital technologies merely for collecting data. Possible tools are smartphones/tablets, microcontrollers, microcomputers, or other data logging systems. Smartphones/Tablets enable students to independently collect first-hand data with their own devices also outside of laboratories in everyday life settings as far as the internal sensors are sufficient or suitable external sensors available. Microcontrollers offer low-threshold and easy-affordable access, too.  Example work for using smartphones, tablets, and microcontrollers for physics experiments was done by Refs.~\cite{Kuhn.2022,Staacks.2018,Stampfer.2020,Klein.2014,Kaps.2022,Organtini.2021,Barro.2023,Monteiro.2022d,Lahme.2022} for microcomputer-based experiments by Refs.~\cite{Thornton.1990,Redish.1997}.

In the middle of the continuum, the \textit{mixed reality}, remote-controlled experiments, Augmented Reality environments, and interactive screen experiments are located. The former consists of real setups and real data is collected, however, students are usually not next to the setup with remote/automatized control, and data collection is often facilitated by computers. In Augmented Reality environments students directly interact with real setups and additional information (e.g., visualizations) is virtually added e.g., when looking through the camera of a tablet or special Augmented Reality glasses. For interactive screen experiments, setups are virtually represented with realistic photos in a computer application. Students can interact with setup elements and generate pre-set data within the application but unlike Augmented Reality environments (and similar to remote-controlled experiments), there is no contact or interaction with real equipment but photos and data sets are based on real experiments. Both Augmented Reality environments and interactive screen experiments allow the combination of experimental setups and multiple representations e.g., to depict invisible characteristics and processes in real-time enabling a continuous learning experience. Interactive screen experiments additionally allow an individual and safe conduction of experiments otherwise inaccessible for students or in massive (online) courses even though the interaction possibilities are limited and not necessarily authentic. Example work for Augmented Reality environments for physics experiments was done by Refs.~\cite{Thees.2022,Schlummer.2021}, for interactive screen experiments by Refs.~\cite{Theyen.1999,Haase.2018}, and for remote-controlled experiments by Refs.~\cite{AitTahar.2019,Thoms.2019}.

Far on the \textit{virtual} side of the continuum, the whole environment is artificially created, e.g., by a computer. Prototypical are Virtual Reality environments authentically representing artificial setup and data. Especially in 3D Virtual Reality environments also dangerous or expensive experiments can be conducted in a reality-mimicking, immersive setting. Example work for Virtual Reality environments for physics experiments was done by Refs.~\cite{Pirker.2017,Price.2019,Porter.2020,Brixner.2021}. Other virtual technologies are artificial computer simulations in which experiments and data are completely modeled, for the benefit of correctness, simplicity, or ease of use often without any claim to realistic visualizations. Such simulations can either be of mathematical nature, e.g., when the outcome of an experiment or process is mathematically modeled (cf. computational methods as often used also by theoretical physicists) \cite{Samsonau.2018,Spencer.2005} or with an emphasis on the visualization of laws, processes, or effects (e.g., PhET simulations \cite{NixOliver.}).

The preceding shows the variety of digital technologies and their potential for teaching and learning in physics courses. In accordance, a review \cite{Girwidz.2019} of teacher-oriented literature in ten European countries revealed that multimedia applications are associated in articles with experimental activities more often than any other teaching-learning activity (e.g., data and knowledge representation or visualizations) with large potential and diversity of ideas and concepts. Although using digital technologies for physics experiments has already been intensively discussed, as far as we know no overview of how often such digital technologies are implemented in physics lab courses (in Europe) exists. Our study contributes to this desideratum explicitly considering the whole reality-virtuality continuum from smartphones/tablets, remote-controlled experiments, Virtual Reality environments, and the usage and/or creation of computer simulations to microcontrollers.

\subsection{Role of digital competencies in physics labs}

The use of digital technologies in physics lab courses goes along with the need for students to acquire related digital competencies for competently and meaningfully using them. This is not only relevant for the successful conduction of physics experiments in the lab course but also for the future labor market. In Sec.~\ref{generalobjec}, we analyze several catalogs of learning objectives for physics lab courses published within the past 25 years regarding the extent to which they contain the acquisition of digital competencies. This reveals a lack of catalogs for specific digital learning objectives for physics students. However, in research about the education of pre-service (science) teachers especially in German-speaking countries, a variety of corresponding frameworks already exist. Their transferability for the education of physics majors is discussed in Sec.~\ref{digitalobjec}.

\subsubsection{Digital competencies as learning objectives in physics labs}\label{generalobjec}

We analyzed five catalogs of learning objectives for physics lab courses that were either developed normatively \cite{AmericanAssociationofPhysicsTeachers.1997,AmericanAssociationofPhysicsTeachers.2014,KonferenzderFachbereichePhysik.2010} or as part of a research process \cite{Welzel.1998,Welzel.1998b,Zwickl.2013,Nagel.2018} representing the United States \cite{AmericanAssociationofPhysicsTeachers.1998,AmericanAssociationofPhysicsTeachers.2014,Zwickl.2013}, Europe \cite{Welzel.1998}, and Germany/Austria, in particular, \cite{KonferenzderFachbereichePhysik.2010,Nagel.2018}; for Finland and Croatia, to our knowledge, no comparable catalog exists. TABLE~\ref{tab:digitalcompetencies} summarizes all included learning objectives/outcomes that can somehow be related to physics lab courses and explicitly mention any digital technology, i.e., learning objectives that are often related to digital technologies (e.g., statistical data analysis) but were not explicitly linked to them were not considered.

\begin{table*}[htb]
\caption{Overview of digital learning objectives in selected catalogs of learning objectives for physics lab courses in the United States, Europe, and Germany/Austria from the past 25 years. To our knowledge, there are no comparable catalogs for Finland or Croatia.}
\begin{ruledtabular}
\begin{tabular}{p{.35\textwidth}p{.61\textwidth}}
\textbf{Catalog}&\textbf{Included digital learning objectives (summarized and reformulated)}\\\hline
\textbf{United States (1997)}: List of learning objectives for physics lab courses published by the American Association of Physics Teachers (AAPT) \cite{AmericanAssociationofPhysicsTeachers.1998}, first published in \cite{AmericanAssociationofPhysicsTeachers.1997}&Using computers/microcomputers for data collection, analysis, \& graphical display of data\\\hline
\textbf{Europe (1998)}: List of learning objectives for science lab courses collected in a Delphi study among school \& university educators from Denmark, France, Germany, Great Britain, Greece, \& Italy \cite{Welzel.1998,Welzel.1998b}&No digital learning objectives were explicitly stated, only the use of modern technologies (e.g. for data capture or modeling) was mentioned as a possible form of labwork.\\\hline
\textbf{Germany (2010)}: List of learning objectives \& expected learning outcomes for physics studies in Germany published by the Konferenz der Fachbereiche Physik (KFP, conference of the physics departments of German universities) \cite{KonferenzderFachbereichePhysik.2010} &\vspace{-\topsep}\vspace{1pt}\begin{itemize}[label=\textbullet, leftmargin=*, topsep=0pt, itemsep=0pt, partopsep=0pt, parsep=0pt]
    \item Using modern physics measurement methods in experiments
    \item Students are familiar with computer-aided measurement data acquisition.
    \item Students are able to create an engaging presentation (PowerPoint or similar).
    \item Using computer-aided computational methods for solving complex problems (including writing own programs in at least one programming language)
    \item Correctly using electronics for data collection including familiarity with electronic components/circuits, features of control, regulation \& measurement technology, computer-aided data collection \& experiment control
\end{itemize}\vspace{-\topsep}~\vspace{-3pt}\\\hline
\textbf{United States (2013)}: List of learning objectives collected among faculty of the University of Colorado where a new lab course should be implemented \cite{Zwickl.2013} &\vspace{-\topsep}\vspace{1pt}\begin{itemize}[label=\textbullet, leftmargin=*, topsep=0pt, itemsep=0pt, partopsep=0pt, parsep=0pt]
    \item Using slide shows with a good PowerPoint style to support oral presentations
    \item Computer-aided data analysis: using analytical \& computational modeling tools; using computational packages like Mathematica for statistical tests; handling, plotting, \& fitting digital data; sampling \& analyzing data with time/frequency domain methods (e.g., FFT); using Mathematica notebooks
    \item Using LabVIEW for recording, visualizing, analyzing, \& interpreting data
    \item Using digital oscilloscopes
\end{itemize}\vspace{-.5\topsep}~\vspace{-7pt}\\\hline
\textbf{United States (2014)}: Revision of the list of learning objectives for physics lab courses from 1997 published by the American Association of Physics Teachers (AAPT) \cite{AmericanAssociationofPhysicsTeachers.2014}&\vspace{-\topsep}\vspace{1pt}\begin{itemize}[label=\textbullet, leftmargin=*, topsep=0pt, itemsep=0pt, partopsep=0pt, parsep=0pt]
    \item Appropriately using computers for computational modeling physical systems including measurement devices
    \item Developing technical \& practical laboratory skills: use of computers to interface to experimental apparatus for data collection; use of data gathering tools like videos to extract physical data
    \item Use of computers for data analysis
\end{itemize}\vspace{-\topsep}~\vspace{-3pt}\\\hline
\textbf{Germany/Austria (2016)}: List of learning objectives based on Ref.~\cite{Zwickl.2013} that was discursively modified by the group of German physics lab instructors \cite{Nagel.2018}&\vspace{-\topsep}\vspace{1pt}\begin{itemize}[label=\textbullet, leftmargin=*, topsep=0pt, itemsep=0pt, partopsep=0pt, parsep=0pt]
    \item Using computer programs for data analysis
    \item Using LabVIEW
    \item Using sensors \& actuators
    \item Explaining \& using measurement devices
\end{itemize}\vspace{-\topsep}~\vspace{-7pt}
\end{tabular}
\end{ruledtabular}
\label{tab:digitalcompetencies}
\end{table*}

It reveals that recent catalogs list more digital learning objectives than older ones and that those are mostly related to computer-aided control of setups as well as data collection and analysis. Further digital learning objectives are linked to computational modeling of physical processes, digital measuring devices, and presentations of findings with slide shows. Other digital technologies or areas of digitalization as described in Sec.~\ref{roletechn} are barely mentioned (e.g., video analysis and online databases only once, using digital tools for communication/collaboration/organization not at all). It is also striking that the catalogs rather list the use of specific technologies (e.g., using Mathematica notebooks, LabVIEW, or computers for data analysis) than precise learning objectives. To our knowledge, a coherent framework or list of digital competencies physics students should acquire during their studies, in particular in physics lab courses, is currently missing.

\subsubsection{Transferability of frameworks for pre-service teachers}\label{digitalobjec}

In the teacher education research community, especially in German-speaking countries, several frameworks have already been developed specifically focusing on digital competencies. One of the most cited frameworks for digital competencies of pre-service science teachers is the DiKoLAN-framework (\textit{Digitale Kompetenzen für das Lehramt in den Naturwissenschaften}, in English \textit{Digital Competencies for Teaching in Science Education}, \cite{Thoms.2021,Becker.2020b}). Based on the well-known and widely-spread TPACK model (\textit{Technological Pedagogical Content Knowledge}, \cite{Mishra.2006}) the framework lists explicit digital competencies that pre-service science teachers should achieve during their studies. These are classified into seven dimensions, the more general competencies \textit{documentation}, \textit{presentation}, \textit{communication/collaboration}, and \textit{information search and evaluation} as well as the more subject-specific competencies \textit{data acquisition}, \textit{data processing}, and \textit{simulation and modeling}. These competencies are accompanied by generic \textit{technical core competencies} and knowledge about the \textit{legal framework} of using digital technologies. Advantageous is that the framework is domain-specific for science as, e.g., it includes digital data collection and processing. While the specific competency descriptions are very specific for pre-service science teachers, therefore, not applicable to physics major students, the overall competency dimensions and their definitions (cf. Refs.~\cite{Becker.2020b,ArbeitsgruppeDigitaleBasiskompetenzenWorkgroupDigitalCoreCompetencies.2020}) can be applied to physics students in physics lab courses with almost no reformulating.

Another subject-related framework is the DiKoLeP-framework (\textit{Digitale Kompetenzen von Lehramtsstudierenden im Fach Physik}, translated \textit{Digital competencies of student teachers in the subject of physics}, \cite{GroeHeilmann.2021}) that is based on the DiKoLAN-framework but its dimensions (\textit{digital data collection, simulations, explanatory videos, subject-related basics}) focus rather on digital media than digital core competencies. A further subject-independent framework is the Austrian Digi.kompP-framework (\textit{Digitale Kompetenzen für PädagogInnen}, translated \textit{Digital competencies for educators}, \cite{Brandhofer.2019}) that still acknowledges subject specifics of digital teaching and learning but focuses not on science or physics in particular rather on work of in-service educators.

The last is the related DigCompEdu- and DigComp2.1-framework (\textit{Digital Competence Framework for Educators} \cite{Redecker.2017} and \textit{Digital Competence Framework for Citizens} \cite{Carretero.2017}). The latter lists digital competencies for every European citizen. The former also lists teachers' digital competencies and has been adapted for science teachers \cite{Ghomi.2020} but is too strongly related to the role of educators and thus, not applicable to students in physics lab courses. However, the generic DigComp2.1-framework can be applied since it is intended for every European citizen. Even though it is not domain-specific (e.g., data collection and processing are not included), many aspects can be mapped to the domain-specific DiKoLAN-framework (e.g., information and data literacy, communication and collaboration, digital content creating, safety, and problem-solving). There is also one digital competency missing in the DiKoLAN-framework: \textit{identifying digital competence gaps} referring to a life-long development of digital competencies and metacognition/awareness of one's own abilities and personal developments.

In our opinion, the dimensions of the DiKoLAN-framework supplemented by \textit{identifying digital competence gaps} from the DigComp2.1-framework provide a sufficient framework of digital competencies physics students could acquire during their studies, in particular in lab courses since they are (besides courses on computational and numerical methods of physics) best suitable for acquiring most of these objectives. It would be beneficial to obtain information on to what extent these learning objectives should be pursued in physics lab courses. Our survey contributes to this desideratum by asking lab instructors about their perceived importance of these digital learning objectives, also in comparison to other objectives, e.g., as identified in the Delphi study by Refs.~\cite{Welzel.1998,Welzel.1998b}.

\subsection{Goals and research questions}

Based on the state of research, we argue that there is a need for getting an overview of the status quo of the use of digital technologies and the importance of acquiring digital competencies in European physics lab courses. This enables the reflection to what extent the emergence and broad availability of new digital technologies in our society as well as the Covid-19 pandemic had an impact on the conception of and the technology usage in university physics lab courses.

In this survey, we focus on the instructors' perspective, i.e., all faculty staff independent from the academic status who is responsible for the conceptualization of university physics lab courses or engaged in teaching the students there. They should better than their students be familiar with the use of digital technologies in the lab course, its learning objectives, and past and projected transformations. Further, they are stakeholders who decide to what extent digital technologies and learning objectives are part of the lab course and therefore disseminators for implementing digital technologies in university physics education more so than instructors in lectures or exercises.
Hence, we target university physics lab instructors and not students or other instructors or lecturers. To consider possible country-specific differences and to take a \textit{European} perspective, we survey lab instructors in Germany the most populous country in the European Union and representative of Middle Europe as well as Finland and Croatia as middle to small-populated countries representing Northern and Middle Europe respectively. The research questions are:
\begin{itemize}[leftmargin=28pt, topsep=0pt, itemsep=0pt, partopsep=0pt, parsep=0pt]
    \item[RQ 1:] \textit{What prior experiences and attitudes toward the use of digital technologies in physics lab courses do German, Finnish, and Croatian lab instructors have?}
    \item[RQ 2:] \textit{How do German, Finnish, and Croatian lab instructors rate the importance of acquiring digital competencies, also in comparison to other learning objectives of physics lab courses?}
    \item[RQ 3:] \textit{What potential do German, Finnish, and Croatian lab instructors see in different (digital) lab formats to achieve the learning objectives of physics lab courses?}
\end{itemize}

\section{The questionnaire}\label{design}

We conducted an anonymous online survey with closed and open questions participants answered independently. By this, we get an overview of the status quo of digital technologies in physics lab courses with a low threshold for participation (e.g., in comparison to interviews), especially because participation is possible at any time, does not last too long, and the anonymity reduces the risk of socially desired responses. The survey (cf. supplemental material for questionnaire in all four languages) consists of seven sections:

\textbf{1.~Demographic questions:} We asked the lab instructors about their country and the lab course type they want to respond for. They could select between \textit{introductory lab courses} for physics major and/or pre-service physics teacher students at the beginning of their studies, \textit{advanced lab courses} for physics major and/or pre-service physics teacher students in higher semesters, \textit{minor lab courses}, e.g., for medicine, chemistry, or biology students, \textit{lab courses for engineering students} and \textit{other} lab courses which need to be specified. Instructors were able to select only one lab course type even though they might be responsible for more than one and should briefly describe the target group and conceptualization of their lab course. Additionally, they should state their current academic status as well as whether they are responsible for the conceptualization and/or organization of the lab course and personally engaged in instructing students in labwork.

\textbf{2.~Ranking of learning objectives for physics lab courses:} The participants ranked five main learning objectives for labwork from least to most important. Four were identified by Ref.~\cite{Welzel.1998b,Welzel.1998} in a Delphi study (\textit{link theory and practice}, \textit{get to know the methods of scientific thinking}, \textit{learn experimental skills}, and \textit{increase their motivation, personal development, and social competency}) and the fifth learning objective \textit{acquire digital competencies} was deductively added. A first ranking was requested based on the implementation in participants' specific lab courses as accurately as possible even if not all learning objectives are followed (\textit{implemented learning objectives}). A second ranking should have been according to their own opinion of how important these learning objectives are in general, independent from their lab courses (\textit{generally desired learning objectives}). The learning objectives to be rated were displayed in randomized order and briefly defined in a selectable info box based on subcategories listed by Refs.~\cite{Welzel.1998,Welzel.1998b} and Refs.~\cite{Becker.2020b,ArbeitsgruppeDigitaleBasiskompetenzenWorkgroupDigitalCoreCompetencies.2020}.

\textbf{3.~Specific subcategories of digital competencies:} The instructors rated several specific sub-competencies of digital competencies on a five-point-scale from \textit{not important} to \textit{very important} or had the opportunity to state \textit{I don’t understand the item}. Nine items are strongly based on the competency descriptions in the DiKoLaN-framework \cite{Becker.2020b,ArbeitsgruppeDigitaleBasiskompetenzenWorkgroupDigitalCoreCompetencies.2020} with only minor changes to transform those sub-competencies and their descriptions into items and to adapt the formulations for the new target group of physics major instead of pre-service science teachers originally. We supplemented this list of sub-competencies by the meta-reflective competency \textit{identifying digital competence gaps} that is part of the DigComp 2.1-framework \cite{Carretero.2017}. All items were presented in randomized order. If desired, the participants had the opportunity to add and rate up to four further sub-competencies.

\textbf{4.~Experience with the use of digital technologies in a physics lab course:} The participants could list up to seven modern digital technologies they have already used in their lab course. We stated clearly that by modern digital technologies, we do not mean standard lab equipment just with a digital scale or an analog to digital converter but, e.g., smartphones/tablets, Virtual or Augmented Reality environments, spreadsheets, and other software for (statistical) data analysis, remote-controlled labs, microcontrollers, software for programming, computer simulations, etc. Furthermore, participants should quantify how often they used the listed digital technologies before, during, and after the lockdown caused by the Covid-19 pandemic, the period when on-campus teaching and learning was prohibited or noticeably impeded. For quantification, the participants selected between \textit{not at all}, \textit{at least once}, and \textit{regularly}. The section ends with an open text field in which the participants could describe in their own words what positive or negative experiences they had while using the listed digital technologies in a physics lab course.

\textbf{5.~Relationship between special forms of labwork and the learning objectives:} As in Ref.~\cite{Welzel.1998}, the instructors judged eight different forms of labwork according to their usefulness in achieving the five learning objectives rated before. They address three different degrees of openness (\textit{a (strongly) guided labwork session}, \textit{an open-ended labwork session}, \textit{undergraduate research projects in groups}) and the use of five different digital technologies (\textit{experiments with smartphones/tablets}, \textit{remote-controlled experimenets}, \textit{Virtual Reality environment}, \textit{usage/and or creation of computer simulations}, and \textit{microcontroller}). For each form of labwork, a short description was presented in a selectable info box. The rating was done on a five-point scale from \textit{not useful} to \textit{very useful}.

\textbf{6.~Attitudes toward the use of modern digital technologies in a lab course:} The instructors rated 15~statements about attitudes and beliefs toward the use of modern digital technologies in physics lab courses on a five-point scale from \textit{strongly disagree} to \textit{strongly agree} or could select \textit{I don’t understand the item}. 13~items, based on two surveys among German school teachers \cite{Schmechting.2020,SchulzZander.2001} and modified for the context of university education, address aspects like the status of digital technologies in university teaching and learning or the belief in the effectiveness of digital technologies for students’ cognitive, affective, and social learning progress. Deductively, we added \textit{In general, modern digital technologies should be part of teaching and learning in a physics lab course.} and \textit{Modern digital technologies should only be used in a physics lab course if they cannot be avoided (e.g., in home labs during the Covid-19 pandemic).} Afterwards, participants should describe their attitude regarding the use of modern digital technologies in physics lab courses in their own words and what reasons they have for their overall attitude. Then they were asked about the existence of any plans regarding the (further) inclusion of digital technologies or the acquisition of digital competencies in their lab course and underlying rationales.

\textbf{7.~Final question:} In a final open text field, participants could comment on anything they would like to add or state.

All items are informed by literature, so already underwent an initial validation by other researchers. Since they were available in English and/or German, we first prepared the English and German survey versions simultaneously within the group of authors (English is the project language). A 2$^{nd}$ year student assistant who studies English checked the comparability of both versions, but only minor improvements were necessary. The German and English version of the survey was independently piloted in Germany with one lab instructor, two Ph.D. students in the field of physics education research, and three further students (2$^{nd}$ and 5$^{th}$ year) to check the understandability, usability, and linguistic correctness of the survey. The structure, layout, and wording of the survey were improved based on the provided feedback. After that, a Finnish and Croatian version of the survey was prepared based on the English version. Both translations were again checked within the corresponding bilingual subgroup of authors and additionally piloted with one independent lab instructor in Finland and two in-service teachers and one pre-service physics teacher in Croatia but no changes were necessary.

\section{Data collection and pre-processing}\label{data collection}

\subsection{Implementation and acquisition of participants}

The primary target group was instructors for introductory and advanced physics lab courses in Germany, Finland, and Croatia, but the survey was open to all lab course types (e.g., engineering physics labs) and we later check for significant differences in the responses for countries and lab course types. The questionnaire was implemented in English, German, Finnish, and Croatian in the open-source web tool \textit{LimeSurvey}. Participants got access to the survey via a link and selected their preferred language. Most responded in their local language but some (especially from Germany) also in English. The average time for participation was $M=28~$min, $SD=19~$min (two outliers excluded). Data were collected from the beginning of September to mid of December 2022. The survey was promoted among German-speaking lab instructors at the conference of physics lab instructors in Kiel in September 2022, twice (one reminder) via a mailing list of the working group for physics lab courses of the German physics society with more than 300 recipients (not all are lab instructors), and by sending a personal invitation to all lab instructors at the University of Göttingen. In Finland, an invitation was sent twice (one reminder) via e-mail to vice-heads in charge of teaching at the departments of physics at Finnish universities and to the Finnish network of university physics educators combined with the request to forward the survey to all physics lab instructors in the respective institutions. In Croatia, an invitation was sent twice (one reminder) to all responsible for physics lab courses (as far as they could be found during an internet search and by personal contacts) via e-mail combined with the request to forward the survey likewise.

\subsection{Characterization of participants}

Overall, 81 instructors, 50 from Germany, 16 from Finland, 14 from Croatia, and 1 from Austria participated in our survey at least partly, i.e., they have completed at least the first and second survey sections about demographic data and learning objectives. Two German participants were removed from the data set since they had not responded in an interpretable way. The Austrian participant was assigned to the German group since in practice, German and Austrian lab instructors build one community (Austrian instructors usually participate in the German conference of physics lab instructors where this survey was promoted). In total, 51 participants (65\%) are responsible for the concept/organization of their lab course, and 69 (87\%) are engaged in instructing students. The academic status varies but most are professors (15\%), hold another senior position (29\%), or are Ph.D. student teaching assistants (22\%). They responded mostly for the two lab course types of primer interest, introductory (33, i.e., 42\%) and advanced (21, i.e., 27\%) lab courses. Academic status and lab course type are similarly distributed over the three countries. TABLE \ref{Participants} provides an overview of the 79 participants of which 64 participated fully, i.e., answered all items on the survey. For an assessment of the sample size and a discussion of possible selection biases in sampling, see Sec.~\ref{limitations}.

\begin{table}[h!tb]
\caption{Overview of the $N=79$ instructors from Germany (DE), Finland (FI), and Croatia (HR) who participated in the survey.}
\begin{ruledtabular}
\begin{tabular}{lrrr p{.695\columnwidth}p{.095\columnwidth}p{.095\columnwidth}p{.095\columnwidth}}
\textbf{Participants}&\textbf{DE}&\textbf{FI}&\textbf{HR}\\\hline
Full participation&37&16&11\\
Partial participation&12&0&3\\\hline
\textbf{Responsible} for the concept/organization&31&9&11\\\hline
\textbf{Engaged} in instruction&40&15&14\\\hline
\textbf{Academic status}&&&\\
Professor&8&&4\\
Other senior position&16&6&1\\
Postdoc&4&3&\\
Ph.D. student teaching assistant&8&2&7\\
Student teaching assistant &4&5&\\
Other position&5&&2\\\hline
\textbf{Lab course type}&&&\\
Introductory lab&19&11&3\\
Advanced lab&15&4&2\\
Minor lab&3&&1\\
Engineering students lab&7&&3\\
Other lab&2&1&4\\
\end{tabular}
\end{ruledtabular}
\label{Participants}
\end{table}

\subsection{Data pre-processing}\label{preprocessing}

While for closed questions the survey tool already exports data in the main survey language, i.e., English, responses to open-text field questions were translated by the group of authors into English so that they can be analyzed together and compared. Negated items were inverted for further analysis.

For the ratings of the learning objectives and the potential of different lab formats for reaching the different learning objectives, we analyzed if there are any differences in the responses based on the participants' origin country or the lab course type they responded for. For this, we conducted a Shapiro-Wilk test ($\alpha=.05$) to check whether the responses to all related survey items are normally distributed. Since all items were non-normally distributed for at least one country/ lab course type, Kruskal-Wallis tests ($\alpha=.05$) were conducted to search for significant differences in the responses. In case of any differences, post hoc pairwise comparisons with Bonferroni correction were done. However, in comparison to the huge number of possible differences, only a few items (cf. Appx.~\ref{appendixa}) were differently rated. Thus, in the following, we do not distinguish the different countries or lab course types and report all related findings for the whole sample group.

For the 15 items about the attitude toward the use of digital technologies in physics lab courses, a principal component analysis with varimax rotation was conducted since the Kaiser-Meyer-Olkin measure of sampling adequacy is with $KMO=.85$ \textit{meritorious} \cite{Kaiser.1974} and Barlett's test of sphericity \cite{BARTLETT.1950} is significant ($p<.001$). Based on Guttman's criteria \cite{Guttman.1954} three factors with eigenvalues $\geq1$ were considered. Based on Kaiser’s criteria \cite{Kaiser.1960} and the scree-plot, we chose a single-factor model explaining $46\%$ of the total variance (the next two factors would only explain 9\% and 8\% of total variance). The Cronbach's Alpha for the single factor is $\alpha=.91$. The corrected item-scale-correlation describing the discriminatory power is over the threshold of $.3$ for all but one item ($.28$ for the item \textit{I find it difficult to adapt to technical innovations.}) but excluding it would not change Cronbach's Alpha. Thus, the 15 items provide a very good scale in terms of their psychometric properties. The mean of these items is treated as a new variable \textit{attitude}. Statistical analysis in the same way as before reveals no significant differences for the \textit{attitude} regarding the countries and lab course types, so in the following, the \textit{attitude} is considered for all participants together, too.

For all closed items used in the questionnaire and discussed in the following sections, descriptive statistical data (number of responses, minimum, maximum, mean, and standard errror) are given in Appx.~\ref{appendixstatis}.

For the open-text field responses including the lists of digital technologies instructors already have experience with, we do not search for any significant differences regarding the different countries or lab course types as those responses are very individual and to be considered rather anecdotal.

\section{Findings and discussion}\label{findings}

For better readability, the findings are presented step by step related to the three research questions and discussed immediately after. Findings and discussion paragraphs are marked with bold captions. The section ends with an overall discussion of study limitations.

\subsection{Experiences with digital technologies in physics labs (RQ1)}\label{experiences}

\subsubsection{Previous usage of digital technologies in physics labs}

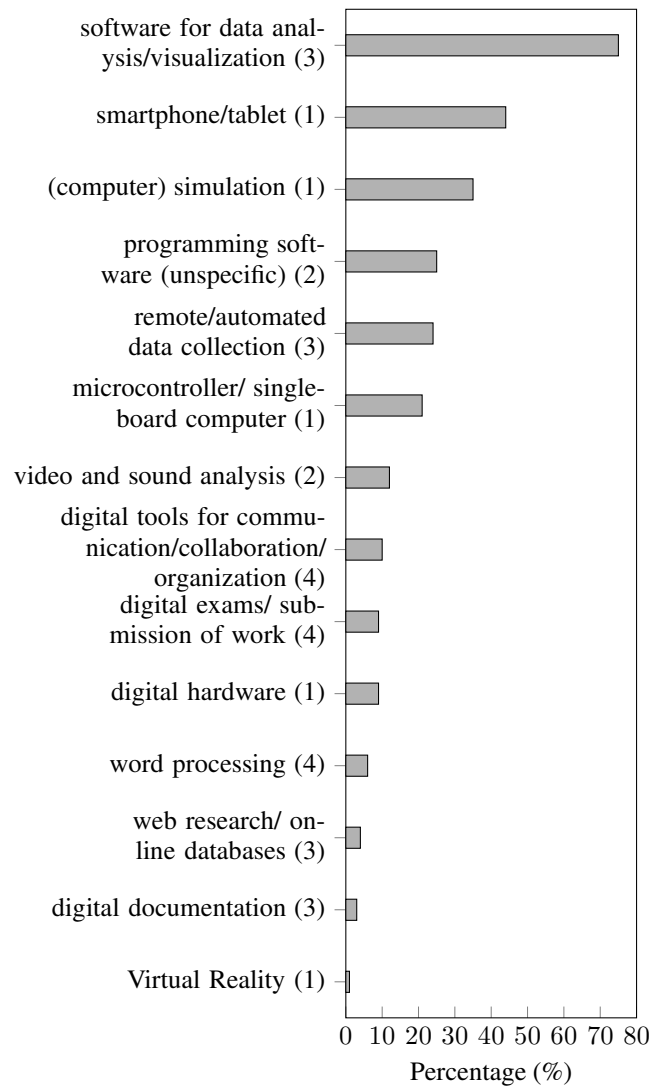
\begin{figure}[htb]
\flushleft
\begin{tikzpicture}
\begin{axis}[width=.63\columnwidth, height=15cm, xbar=3pt,
  ymax=2.9,  
  xmin=0, xmax=80,
  xtick={0,10,20,30,40,50,60,70,80},
  ymin =0.1,
  xlabel={Percentage (\%)},
  ytick = {.2,.4,.6,.8,1,1.2,1.4,1.6,1.8,2,2.2,2.4,2.6,2.8},
  yticklabel style={text width=.5\columnwidth,align=right, },
  yticklabels={Virtual Reality (1),digital documentation (3),web research/ online databases (3),word processing (4),digital hardware (1),digital exams/ submission of work (4),digital tools for communication/collaboration/
organization (4),video and sound analysis (2),microcontroller/ single-board computer (1),remote/automated data collection (3),programming software (unspecific) (2),(computer) simulation (1),smartphone/tablet (1),software for data analysis/visualization (3)},
   ytick pos=left,
xtick pos=left,
 legend columns=1, legend cell align = left,legend style = {draw = none},
 legend style = {at ={(0,1)}, anchor = south west},
 bar width = 8pt,
  ]

\addplot+[gray!60!,area legend,
    draw=black, error bars/.cd, x dir=both, x explicit, error mark options={black,mark size=2pt,line width=.7pt,rotate=90
     },  error bar style={line width=.7pt}
      ] 
		coordinates{
		(	75, 2.8	)
		(	44, 2.6	)
		(	35, 2.4	)
		(	25, 2.2	)
		(	24, 2	)
		(	21, 1.8)
		(   12, 1.6)
		(	10, 1.4)
		(   9, 1.2)
        (	9, 1	)
        (   6, .8)
        (	4, .6)
        (	3, .4)
        (   1, .2)
}; \label{meanexp}
\end{axis}
\end{tikzpicture}
\caption{Percentage of instructors ($N=68$) who used at least once in their lab course a technology coded in the respective category (no multiple coding with the same category per instructor). Numbers in brackets indicate the area of digitalization according to Ref.~\cite{Franke.2022b}, i.e., whether the technologies are related to the laboratory environment (1), processes and procedures (2), dealing with data (3), or interpersonal communication (4).}
\label{fig:experiences}
\end{figure}

Seven instructors stated that they have not used any digital technology in their lab course yet, the other 61 instructors listed in total 220 digital technologies (up to seven per person were possible). The technologies were categorized inductively by systematizing the responses and deductively by using the four areas of digitalization in lab courses compiled by Ref.~\cite{Franke.2022b}. FIG.~\ref{fig:experiences} presents the category system and the percentage of all instructors who have used in their lab course at least once a technology coded in the respective category.

\textbf{Findings (i):} Most frequently mentioned are \textit{software for data analysis/visualization} (75\% of all instructors), \textit{smartphones/tablets} (44\%), and \textit{(computer) simulations} (35\%), but also \textit{programming software} (25\%), \textit{remote/automated data collection} (24\%), and \textit{microcontrollers/ single-board computers} (21\%).

\textbf{Discussion of findings (i):} However, the data should be considered with caution as the responses are likely to be incomplete (e.g., one would expect digitally aided word processing in most lab courses for writing lab reports) and depend on the participants' perception of \textit{modern digital technologies}. Ref.~\cite[p.3]{Franke.2022b} already stated that “[t]he digitalization of practical laboratory courses is usually associated with remote experiments, working in virtual reality spaces, and digital data acquisition" even though "[d]igitalization already starts with the use of digital communication channels or teaching materials”.

\begin{figure}[htb]
\flushleft 
\begin{tikzpicture}
\begin{axis}[width=.73\columnwidth, height=16.5cm, xbar=3pt,
  ymax=2.1,  
  xmin=0, xmax=100,
  xtick={0,20,40,60,80,100},
  ymin =0.1,
  xlabel={Percentage (\%)},
  ytick = {0.2,0.4,.6,.8,1.,1.2,1.4,1.6,1.8,2.},
  yticklabel style={text width=.4\columnwidth,align=right, },
  yticklabels={video and sound analysis ($N=8$), programming software (unspecific) ($N=18$), 
  digital hardware ($N=8$), software for data analysis/visualization ($N=56$), microcontroller/ single-board computer ($N=16$), remote/automated data collection ($N=19$), smartphone/tablet ($N=35$), (computer) simulation ($N=26$), digital tools for communication/ collaboration/organization ($N=13$), digital exams/ submission of work ($N=7$)},
   ytick pos=left,
xtick pos=left,
 legend columns=1, legend cell align = left,legend style = {draw = none},
 legend style = {at ={(0,1)}, anchor = south west},
 bar width = 8pt,
  ]
\addlegendimage{empty legend}
\addlegendentry{\scalebox{1}[1]{\ref{before}} Before the pandemic}
\addlegendimage{empty legend}
\addlegendentry{\scalebox{1}[1]{\ref{during}} During the pandemic}
\addlegendimage{empty legend}
\addlegendentry{\scalebox{1}[1]{\ref{after}} After the pandemic}

\draw[dashed] (100,0.1) -- (100,200);

\draw[ultra thick] (0,120) -- (200,120);
\draw[ultra thick] (0,80) -- (200,80);

\addplot+[gray!20!,area legend,
    draw=black, error bars/.cd, x dir=both, x explicit, error mark options={black,mark size=2pt,line width=.7pt,rotate=90
     },  error bar style={line width=.7pt}
      ] 
		coordinates{
(	62.5, .2)
(	78.9, .4)
(   75, .6)
(	84.5, .8)

(	87.5, 1)
(89.5, 1.2)

(   42.9, 1.4)
(	53.8, 1.6)
(   76.9, 1.8)
(	85.7, 2)
}; \label{after}

\addplot+[gray!60!,area legend,
    draw=black, error bars/.cd, x dir=both, x explicit, error mark options={black,mark size=2pt,line width=.7pt,rotate=90
     },  error bar style={line width=.7pt}
      ] 
		coordinates{
(	62.5, .2)
(	77.8, .4)
(	75, .6)
(	78.6, .8)

(	56.3, 1)
(	57.9, 1.2)

(	71.4, 1.4)
(	76.9, 1.6)
(	100, 1.8)
(	85.7, 2)
}; \label{during}

\addplot+[gray,area legend,
    draw=black, error bars/.cd, x dir=both, x explicit, error mark options={black,mark size=2pt,line width=.7pt,rotate=90
     },  error bar style={line width=.7pt}
      ] 
		coordinates{
(	62.5, .2	)
(	66.7, .4	)
(	75, .6	)
(	73.2, .8)

(	81.3, 1	)
(	78.9, 1.2	)

(	25.7, 1.4	)
(	46.2, 1.6	)
(	61.5, 1.8)
(	28.6, 2	)
}; \label{before}

\end{axis}
\end{tikzpicture}
\caption{Percentage how often each digital technology was \textit{regularly} used in physics lab courses before, during, and after the Covid-19 pandemic. $N$ in brackets indicates how often a technology coded in the category was listed (multiple codings with the same category per instructor possible, only technologies with at least five mentions).}
\label{fig:technologies}
\end{figure}
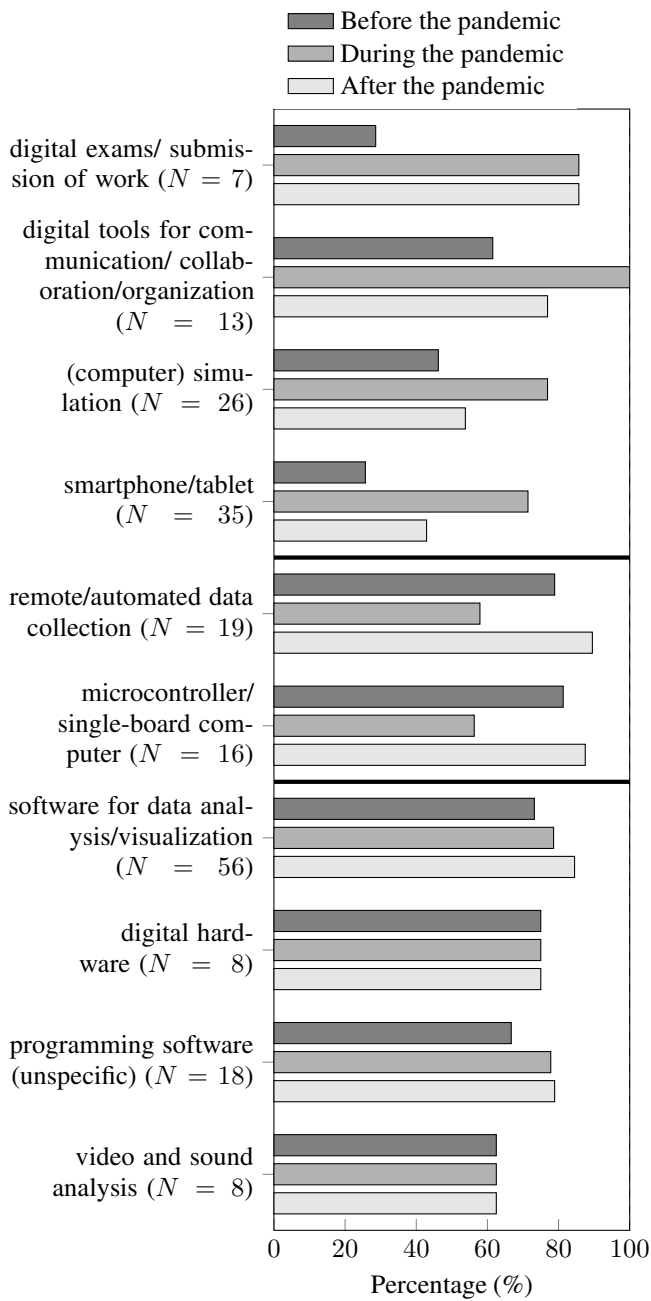

\textbf{Findings (ii):} Additionally, instructors rated how often (\textit{not at all - at least once - regularly}) they have used each technology before, during, and after the lockdown caused by the Covid-19 pandemic. In FIG.~\ref{fig:technologies} the percentage is displayed how often \textit{regularly} was chosen revealing three distinct developments (cf. Appx.~\ref{appendixb} for further visualization):

\begin{enumerate}[leftmargin=*, topsep=0pt, itemsep=0pt, partopsep=0pt, parsep=0pt]
    \item The use of \textit{digital exams/ submission of work}, \textit{digital tools for communication/organization/collaboration}, \textit{(computer) simulations}, and \textit{smartphones/tablets} has increased significantly during the pandemic. After the pandemic, the frequency of use of these technologies remained the same or decreased slightly but is still higher than before the pandemic.
    \item\textit{Remote/Automated data collection} and \textit{microcontrollers/ single-board computers} were used more rarely during the pandemic but almost equally often before and after.
    \item No major differences in the frequency of use can be found for \textit{software for data analysis/visualization}, \textit{digital hardware}, \textit{programming software (unspecific)}, and \textit{video and sound analysis}.
\end{enumerate}

\textbf{Discussion of findings (ii):} The different developments can be explained by the different usability of technologies during on-campus and distance learning. The technologies in the first group enabled instructors to conduct their lab courses even in distance learning forced by the pandemic, e.g., smartphones/tablets for data collection at home, digital exams, and communication tools for assessment and interaction. Thus, these technologies were used more often during the pandemic than before. The data suggest that instructors now appreciate the possibilities of these new-used technologies as the frequency of use has not returned to the initial value after the pandemic.
Contrary, using remote/automated data collection and microcontrollers (second group) requires equipment usually available in laboratories but not in students' private homes. Thus, their rarer use during the pandemic in comparison to before and after can be explained by the limited availability of these technologies for students in distance learning.
The technologies in the third group, especially software and video/sound analysis, are rather independent of the learning scenario (on-campus or distance learning); therefore, it is plausible that the frequency of use has not changed over the pandemic.

All in all, the data suggest that instructors have changed the use of digital technologies in their lab courses over time. During the pandemic, they especially used technologies that were accessible to their students also in distance learning settings, thus they introduced new technologies for data collection (e.g., smartphones/tablets) and the organization of the course (e.g., digital exams, tools for communication) while lab-specific equipment (e.g., microcontroller) was used less often. Contrary, technologies that are independent of the learning setting (e.g., tools for digital data analysis) have been used the same way also during the pandemic and afterward. Since the frequency of use of newly introduced technologies has not returned to the initial value after the pandemic, the data suggest that digital technologies are nowadays used more often in physics lab courses than before.

\subsubsection{Experiences with specific digital technologies}\label{expdigtechnqual}

\begin{table*}[htb]
\caption{Summary of instructors' past experiences with specific digital technologies in their lab courses based on 45 open text field responses (number of instructors with similar experiences in brackets, multiple counts per instructor possible), sorted by the categories in FIG.~\ref{fig:experiences}}
\begin{ruledtabular}
\begin{tabular}{p{.175\textwidth}p{.33\textwidth}p{.315\textwidth}p{.14\textwidth}}
\textbf{Technology}&\textbf{Positive experiences}&\textbf{Negative experiences}&\textbf{Other}\\\hline
\raggedright\hangindent=.3cm\textbf{Lab environment}&&&\\
\raggedright\hangindent=.3cm Digital hardware&\hangindent=.3cm&\hangindent=.3cm Students have problems with the technical setup (e.g., Linux \& Python) (1) &\hangindent=.3cm Some students have no PC (1)\\\cline{2-4}
\raggedright\hangindent=.3cm Smartphone/Tablet&\hangindent=.3cm Photos of the setup help students (2); experiments conductible at the students' private homes (2); motivating (1); everyday reference (1); availability of handy apps (1); better than traditional stopwatches (1); unspecific positive (1) &\hangindent=.3cm Insufficient opportunities for reaching the learning objectives (1); maintenance of tablets (1); difficulties with the use of the students' private smartphones (1); limited opportunities, e.g., for optics (1) &\hangindent=.3cm Increasing use due to technical progress (1)\\\cline{2-4}
\raggedright\hangindent=.3cm Microcontroller/ Single-board computer&\hangindent=.3cm Wide range of applications (3) \& facilitate data collection (4)&\hangindent=.3cm Malfunctions and operational difficulties often occur (2)&\\\cline{2-4}
\raggedright\hangindent=.3cm (Computer) Simulation&\hangindent=.3cm Support the understanding (3); allow experimentation without hardware (1); reduce students' fears (1); are suitable for distance learning (1); mostly work (1)&\hangindent=.3cm Bugs in the equipment (1) or a specific simulation software (1); a specific task using computer simulations is chaotic (1)&\hangindent=.3cm Essential for physics labs (1)\\\hline
\raggedright\hangindent=.3cm\textbf{Processes/Procedures}&\hangindent=.3cm&\hangindent=.3cm&\hangindent=.3cm\\
\raggedright\hangindent=.3cm Video/Sound analysis&\hangindent=.3cm Viana [software] works well (1)&&\\\cline{2-4}
\raggedright\hangindent=.3cm Programming software (unspecific)&\hangindent=.3cm Suitable for distance learning (1)&\hangindent=.3cm Students unfamiliar with programming software (2), overwhelmed (1) or do not want to learn it (1); bad provision \& support by the university (1); software has bugs (1)&\hangindent=.3cm\\\hline
\raggedright\hangindent=.3cm\textbf{Dealing with data}&\hangindent=.3cm&\hangindent=.3cm&\hangindent=.3cm\\
\raggedright\hangindent=.3cm Digital documentation&\hangindent=.3cm Videos of the experiments contribute to understanding \& are more engaging (1)&\hangindent=.3cm Videos of experiments without explanation unintelligible (1)&\\\cline{2-4}
\raggedright\hangindent=.3cm Remote/Automated data collection&\hangindent=.3cm Saves time \& enables better results (1); unspecific positive (3) &\hangindent=.3cm Cassy [interface for data acquisition] difficult to use \& a black box (1)&\hangindent=.3cm Nice to have but unnecessary (1)\\\cline{2-4}
\raggedright\hangindent=.3cm Software for data analysis/visualization&\hangindent=.3cm Wide usefulness (4); quicker data analysis (2); increasing understanding (2); unspecific positive (3)&\hangindent=.3cm Lack of students' basic skills in using (4) or lack of will to learn this software (1); students do not think about/understand the results (2); software (like Origin) difficult/clumsy (2); software not available for students' private computers (1) &\hangindent=.3cm Essential for physics labs (3); Relevant for students' future (1)\\\hline
\multicolumn{2}{l}{\textbf{Interpersonal communication}}&\hangindent=.3cm&\hangindent=.3cm\\
\raggedright\hangindent=.3cm Digital tools for communication/ collaboration/ organization&\hangindent=.3cm Video conferencing is helpful for quick meetings \& exams (2), so increases flexibility (1); learning platforms and Excel support organizing the labs (2)&\hangindent=.3cm Less helpful than face-to-face collaboration (1)&\\\cline{2-4}
\raggedright\hangindent=.3cm Digital exams/ Submission of work&\hangindent=.3cm Digital exams improved students' knowledge about task \& their preparation for lab day (1); digital submission of reports \& feedback faster (2) \& eco-friendly (1)&\hangindent=.3cm Digital submission of reports consumes more time (1)&\hangindent=.3cm Unspecific negative about digital exams (1)\\\cline{2-4}
\raggedright\hangindent=.3cm Word processing&\hangindent=.3cm LaTeX useful \& worth learning (2)&&
\end{tabular}
\end{ruledtabular}
\label{tab:exptechn}
\end{table*}

\textbf{Findings:} The instructors openly described positive and negative experiences with their listed technologies. TABLE~\ref{tab:exptechn} summarizes how the 45~responses were sorted based on the categories in FIG.~\ref{fig:experiences} and differentiated into positive, negative, and other experiences. The positive experiences are particularly related to the opportunities for easier and faster data collection and analysis, the benefit for students' understanding and motivation, and the facilitated interpersonal communication. The negative experiences are mainly related to occurring technical problems and new challenges both for instructors and students to familiarize themselves with new technologies and software. This is in accordance with the main challenges for the areas \textit{laboratory environment} and \textit{dealing with data} as described by Ref.~\cite{Franke.2022b} that students need to learn how to handle the new environment and, e.g., the software for data analysis with varying prior knowledge while instructors need to identify which related skills students need to learn, how they can learn them, and how, e.g., the process of data analysis can be explicated.

Additionally, instructors shared positive and negative experiences unrelated to specific digital technologies. Positively perceived was that digital technologies support students to work more independently (2 instructors), that they provide new/more varying opportunities for experimenting (2), are conducive to understanding (2), support the acquisition of digital skills (2), motivate students (1), and facilitate experimental processes (3) and organizing labs (1). Negative experiences are related to the different students' prior knowledge/skills (2), that digital technologies are black boxes impeding students' understanding (1) and can distract students from physics content (1), and that there is a certain (time-related) entry hurdle to implementing new technologies (2).

The instructors also stated that smartphones/tablets, microcontrollers, and simulations were particularly used during the pandemic (4), e.g., because they enabled students to experiment outside the laboratory (2). However, three described the burden of quickly choosing, learning, and implementing these technologies at the start of the pandemic.

\textbf{Discussion:} The findings here have rather anecdotal character pointing out the variety of positive and negative experiences which surely are subjective and depend on the local conditions in the instructors' institutions, their experimental tasks, target groups, etc. However, the list of negative experiences reveals two central dimensions that can be identified over all technologies. The first one is of rather technical nature referring to non-functioning, buggy software and equipment which might either be related to the manufacturers' production quality or the instructors' skills in competently using the technologies. Contrary, the second dimension is about a lack of (basic) skills in using the technologies among students. Thus, the reported experiences show how important the promotion and acquisition of digital competencies are, both for instructors and students. Therefore, digital competencies should also be learning objectives for physics lab courses and a first step for this learning process might be the identification of one's own digital competency gaps (cf. Sec.~\ref{competencies}).

\begin{table*}[h!bt]
\caption{Categories with descriptions and an anchor example as well as the coding-quantity ($N=63$ responses, double-coding possible) based on the analysis of the qualitative responses on the instructors' attitudes to digital technologies in physics lab courses.}
\begin{ruledtabular}
\begin{tabular}{p{.17\textwidth}p{.39\textwidth}p{.34\textwidth}c p{.07\textwidth}}
\textbf{Category}&\textbf{Description}&\textbf{Example (translated)}&\textbf{Quantity}\\\hline
\raggedright\textbf{Positive Attitude}&&&\\
\raggedright\hangindent=.3cm Potential for experiments \& competence acquisition&\hangindent=.3cm Digital technologies provide new opportunities or support the conduction of physics experiments and the experimenting-related acquisition of competencies for the students.&\hangindent=.3cm \textit{Digital technologies allow us to efficiently collect data and analyze it, which enables students to better understand the setup and physics being investigated.}&18\\\cline{2-4}
\raggedright\hangindent=.3cm Potential for teaching \& learning in general&\hangindent=.3cm Digital technologies can support/improve the teaching and learning (of physics), but no explicit relation to physics experiments was given.&\hangindent=.3cm \textit{Open for ideas, as the use of digital technologies can make teaching more efficient and diverse.}&5\\\cline{2-4}
\raggedright\hangindent=.3cm Motivational effects&\hangindent=.3cm Digital technologies motivate the students and increase their interest and engagement.&\hangindent=.3cm \textit{Use of digital media [...] is fun, motivating, and helps to avoid monotonous work.}&8\\\cline{2-4}
\raggedright\hangindent=.3cm "It's contemporary/ the future/ relevant for the labor market."&\hangindent=.3cm Digital technologies are used everywhere, also in real physics research and the labor market. Thus, it is contemporary and relevant to incorporate digital technologies in physics labs.&\hangindent=.3cm \textit{Digital technologies play an increasingly important role in the actual research and therefore should also be present in lab courses [...].}&18\\\cline{2-4}
\raggedright\hangindent=.3cm Unspecific positive&\hangindent=.3cm A positive attitude toward the use of (specific) digital technologies in physics labs is described, but no reasons/explanations are given.&\hangindent=.3cm \textit{I am very positive about new technologies.}&7\\\hline
\raggedright\textbf{Mixed Attitude}&&&\\
\raggedright\hangindent=.3cm "Not as an end itself"/"Only if it makes sense")&\hangindent=.3cm The potential of digital technologies is not denied but they should be used if beneficial in the specific use case and not as an end in themselves.&\hangindent=.3cm \textit{Modern digital media should not be used as an end in themselves. They are just tools, like others.}&19\\\hline
\raggedright\textbf{Negative Attitude}&&&\\
\raggedright\hangindent=.3cm Limitations/ Difficulties&\hangindent=.3cm Limitations that need to be acknowledged or difficulties/disadvantages that might occur when using digital technologies in a physics lab are stated.&\hangindent=.3cm \textit{[...]~current research does not use smartphones, tablets, or virtual reality. It, therefore, offers a false picture of reality in a physics laboratory. [...]}&12\\\cline{2-4}
\raggedright\hangindent=.3cm Unspecific negative&\hangindent=.3cm A negative attitude toward the use of (specific) digital technologies in physics labs is described, but no reasons/explanations are given.&\hangindent=.3cm \textit{Students are supposed to learn physics, not how to play with phones.}&7\\
\end{tabular}
\end{ruledtabular}
\label{tab:attitudesqual}
\end{table*}

\subsection{Attitude toward the use of digital technologies (RQ1)}\label{attitude}

\textbf{Findings (i):} The instructors' attitude toward the use of digital technologies in physics lab courses was recorded with 15 closed items. As described in Sec.~\ref{preprocessing}, these items form one scale, so the mean of each participant’s responses to all items is considered as the new variable \textit{attitude}. The mean participants' \textit{attitude} is $M=4.08$ ($SD=0.60$).

\textbf{Discussion of findings (i):} On average, instructors responded with \textit{agree} to each item implying an overall positive attitude toward digital technologies in physics lab courses.

\textbf{Findings (ii):} Related open-text responses by 63 instructors provide deeper insight into the reasons behind their attitudes and are supplemental to the quantitative responses also revealing reservations and concerns. The responses were assigned to eight categories presented in TABLE~\ref{tab:attitudesqual}. A positive attitude toward digital technologies was argued by 19 instructors with their potential for physics experiments and students' competence acquisition, by five instructors with the general potential for teaching and learning, by eight instructors with related motivational effects, and by 18 instructors with the contemporaneity of digital technologies and their relevance for students' future and labor market. 19 instructors revealed a mixed attitude toward digital technologies in lab courses stating that they should not be used as an end in themselves but only if they have a real benefit in the specific use case. A negative attitude was related to limitations, difficulties, or problems that (might) occur with the use of digital technologies in the lab course (12 instructors). Some instructors just stated their attitude (7 positive, 7 negative) without any explanations.

Potentials of digital technologies mentioned by the instructors were, for example, the opportunities and higher flexibility for data processing and visualization, the potential for deeper understanding, the facility to conduct otherwise unfeasible experiments, and the increasing students' motivation and interest. Otherwise, limitations of digital technologies mentioned by the instructors were for example the heterogeneous students' digital competencies, the restricted authenticity of technologies like smartphones in comparison to real physics research, the importance of hands-on activities which might be substituted by digital technologies, or the risk of less understanding (e.g., when a tool just becomes a black box).

\textbf{Discussion of findings (ii):} Most instructors did not refer to specific technologies but rather stated a general attitude toward digital technologies in physics labs, so interpreting the results largely depends on which technologies the instructors had in mind when responding to the question. However, one technology, Virtual Reality, was quite explicitly referred to by seven instructors, probably because it was one of the technologies that were mentioned in the survey multiple times. Besides one instructor who stated that \textit{Virtual Reality can make experiments possible that would otherwise not be feasible in a lab course}, they were quite consistently negative about this technology (e.g, \textit{Virtual experiments are not experiments.} or \textit{No virtual reality!}) suggesting that they do not see Virtual Reality as an adequate substitution for real experiments or superior to (2D) simulations.

\subsection{Role of digital competencies as learning objectives (RQ2)}\label{competencies}

\subsubsection{Digital competencies compared to other learning objectives}

The instructors ranked five general learning objectives of lab work by their importance once according to the current implementation in their lab course (\textit{implemented}) and once according to their personal preference (\textit{desired learning objectives}). For better visualization, the data were transformed so that $5$ is the highest and $1$ is the lowest ranking position.

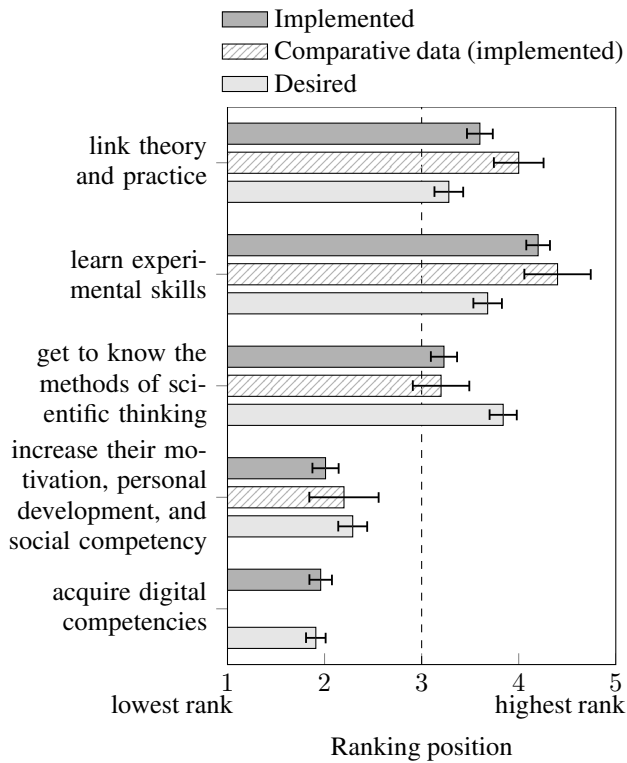
\begin{figure}[htb]
\flushleft
\begin{tikzpicture}
\begin{axis}[width=.78\columnwidth, height=9cm, xbar=3pt,
  ymax=1.1,  
  xmin=1, xmax=5,
  xtick={1,2,3,4,5},
  ymin =0.1,
  xlabel={Ranking position},
  extra x ticks={1,5},
extra x tick labels={lowest rank, highest rank},
extra x tick style={grid=none, tick style={draw=none}, tick label style={xshift=-21pt, yshift=-8pt}},
  ytick = {0.2,.4,.6,.8,1},
  yticklabel style={text width=.35\columnwidth,align=right, },
  yticklabels={acquire digital competencies, increase their motivation{,} personal development{,} and social competency, get to know the methods of scientific thinking, learn experimental skills, link theory and practice},
   ytick pos=left,
xtick pos=left,
 legend columns=1, legend cell align = left,legend style = {draw = none},
 legend style = {at ={(-0.05,1)}, anchor = south west},
 bar width = 8pt,
  ]
\addlegendimage{empty legend} \addlegendentry{\scalebox{1}[1]{\ref{implemented}} Implemented}
\addlegendimage{empty legend}
\addlegendentry{\scalebox{1}[1]{\ref{reference.obj}} Comparative data (implemented)}
\addlegendimage{empty legend}
\addlegendentry{\scalebox{1}[1]{\ref{desired}} Desired}

\draw[dashed] (200,0.1) -- (200,100);

\addplot+[gray!20!,area legend,
    draw=black, error bars/.cd, x dir=both, x explicit, error mark options={black,mark size=2pt,line width=.7pt,rotate=90
     },  error bar style={line width=.7pt}
      ] 
		coordinates{
(	3.28, 1) +-(.148, 1)
(	3.68, .8)  +-(.147, .8)
(	3.84, .6)  +-(.140, .6)
(	2.29, .4)  +-(.149, .8)
(	1.91, .2)  +-(.101, .2)
}; \label{desired}

\addplot+[gray!60!,pattern=north east lines, pattern color=gray!60!, area legend,
    draw=black, error bars/.cd, x dir=both, x explicit, error mark options={black,mark size=2pt,line width=.7pt,rotate=90
     },  error bar style={line width=.7pt}
      ] 
		coordinates{
(	4, 1) +-(.256, 1)
(   4.4, .8)  +-(.342, .8)
(	3.2, .6)  +-(.291, .6)
(	2.2, .4)  +-(.357, .4)
}; \label{reference.obj}

\addplot+[gray!60!,area legend,
    draw=black, error bars/.cd, x dir=both, x explicit, error mark options={black,mark size=2pt,line width=.7pt,rotate=90
     },  error bar style={line width=.7pt}
      ] 
		coordinates{
(	3.60, 1	) +-(.133, 1)
(	4.20, .8) +-(.122,.8)
(	3.23, .6) +-(.135,.6)
(	2.01, .4) +-(.135,.4)
(	1.96, .2	)+-(.116,.2)
}; \label{implemented}

\end{axis}
\end{tikzpicture}
\caption{Mean and standard error of the instructors' ($N=75$) ranking of implemented and desired learning objectives. The dashed line indicates the middle-ranking position. If applicable, comparative data ($N=10$) from Ref.~\cite{Haller.1999} is added.}
\label{fig:Comp}
\end{figure}

\textbf{Findings (i):} In FIG.~\ref{fig:Comp}, the mean ranking position of each learning objective is displayed for both implemented and desired learning objectives. The three most important implemented learning objectives are \textit{learn experimental skills} ($M=4.20, SE=0.12$), \textit{link theory to practice} ($M=3.60, SE=0.13$), and \textit{get to know the methods of scientific thinking} ($M=3.23, SE=0.23$). The fourth is \textit{increase their motivation, personal development, and social competency} ($M=2.01, SE=.014$) and the fifth \textit{acquire digital competencies} ($M=1.91, SE=0.10$). 

\textbf{Discussion of findings (i):} The data (also in Sec.~\ref{potential}) can be compared with the findings in Ref.~\cite{Haller.1999} since our survey is largely based on the well-known survey by Refs.~\cite{Welzel.1998,Welzel.1998b,Haller.1999} while only Ref.~\cite{Haller.1999} presents data for the relevant sub-sample of university physics educators, although unfortunately only for German physics professors. The comparison needs to be done with caution because the fifth objective to be rated there was \textit{for the teacher to evaluate the knowledge of the students} (which is an objective but not a learning objective) instead of \textit{acquire digital competencies} in our survey. However, to our knowledge, this is still the most suitable source of reference data available. The comparison of our findings with the ones in Ref.~\cite{Haller.1999} reveals that accordingly the three most important objectives there were also \textit{learn experimental skills} ($M=4.4, SE=0.3$), \textit{link theory to practice} ($M=4.0, SE=0.3$), and \textit{get to know the methods of scientific thinking} ($M=3.2, SE=0.3$), so within the scope of the standard errors, the ratings agree with each other. 

\textbf{Findings (ii):} Between the ranking of the implemented and desired learning objectives in our survey, no substantial differences can be seen.

\textbf{Discussion of findings (ii):} Within the scope of standard errors, mean values only differ for the learning objective \textit{get to know the methods of scientific thinking}, so the instructors would like to emphasize the objective \textit{getting to know the methods of scientific thinking} more than it is done in their lab courses so far ($M=3.84, SE=0.14$ for desired, $M=3.23, SE=0.14$ for implemented).

\textbf{Findings (iii):}The perceived importance of digital competencies in comparison to other learning objectives is rather independent of the instructors' attitude toward using digital technologies in physics lab courses (cf. Sec.~\ref{attitude}). The correlation (Spearman-Rho coefficient) between the ranking position and the \textit{attitude} is not significant for the ranking of the implemented learning objectives ($r(58)=.15, p=n.s.$) and significant but small for the desired learning objectives ($r(58)=.26, p=.044$).

\subsubsection{Importance of specific digital competencies}

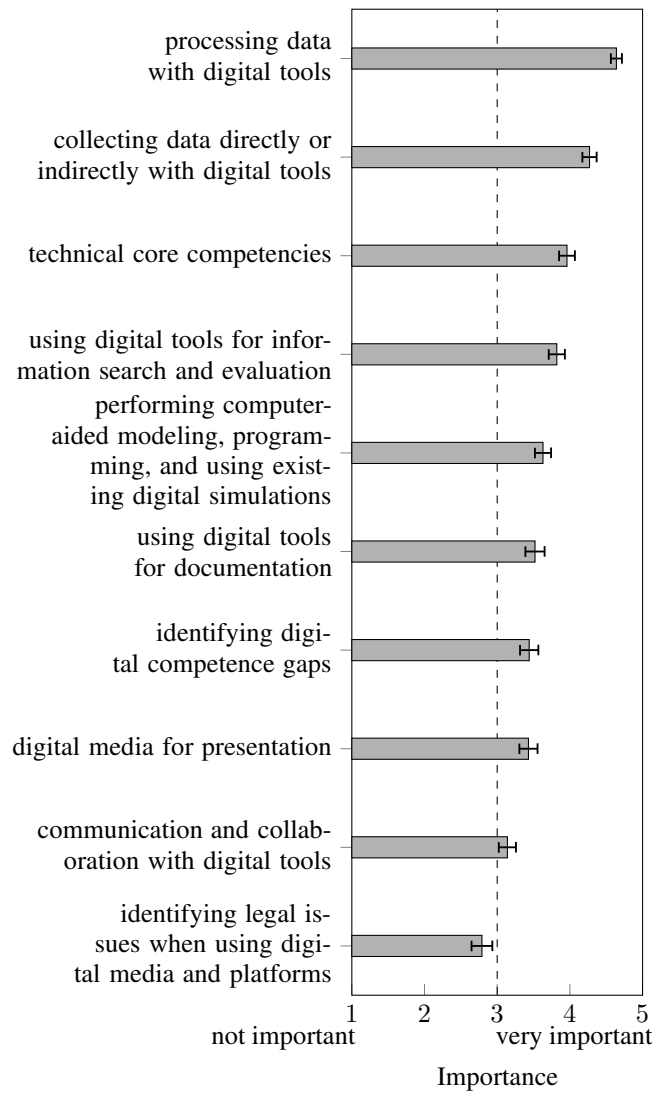
\begin{figure}[htb]
\flushleft
\begin{tikzpicture}
\begin{axis}[width=.63\columnwidth, height=14.7cm, xbar=3pt,
  ymax=2.1,  
  xmin=1, xmax=5,
  xtick={0,1,2,3,4,5},
  ymin =0.1,
  xlabel={Importance},
  extra x ticks={1,5},
extra x tick labels={not important, very important},
extra x tick style={grid=none, tick style={draw=none}, tick label style={xshift=-26pt, yshift=-8pt}},
  ytick = {.2,.4,.6,.8,1,1.2,1.4,1.6,1.8,2},
  yticklabel style={text width=.5\columnwidth,align=right, },
  yticklabels={identifying legal issues when using digital media and platforms, communication and collaboration with digital tools, digital media for presentation, identifying digital competence gaps, using digital tools for documentation, performing computer-aided modeling{,} programming{,} and using existing digital simulations, using digital tools for information search and evaluation, technical core competencies, collecting data directly or indirectly with digital tools, processing data with digital tools},
   ytick pos=left,
xtick pos=left,
 legend columns=1, legend cell align = left,legend style = {draw = none},
 legend style = {at ={(0,1)}, anchor = south west},
 bar width = 8pt,
  ]
\draw[dashed] (200,0.1) -- (200,300);

\addplot+[gray!60!,area legend,
    draw=black, error bars/.cd, x dir=both, x explicit, error mark options={black,mark size=2pt,line width=.7pt,rotate=90
     },  error bar style={line width=.7pt}
      ] 
		coordinates{
		(	4.64, 2	) +-(.076,2)
		(	4.27, 1.8	) +-(.098,1.8)
		(   3.96, 1.6)+-(.109,1.6)
		(	3.82, 1.4	) +-(.111,1.4)
		(   3.63, 1.2)+-(.111,1.2)
        (	3.52, 1	) +-(.132, 1)
        (   3.44, .8)+-(.126,.8)
        (	3.43, .6) +-(.125,.6)
        (	3.14, .4) +-(.118,.4)
        (   2.79, .2)+-(.143,.2)
}; \label{mean}
\end{axis}
\end{tikzpicture}
\caption{Mean and standard error of the perceived importance of specific digital competencies ($N=70-73$, some instructors did not rate all items). The dashed line indicates moderate importance.}
\label{fig:DigComp}
\end{figure}

\textbf{Findings (i):} The instructors also rated the importance of ten specific digital competencies mainly based on the DiKoLAN-framework. As displayed in FIG.~\ref{fig:DigComp}, all digital skills besides \textit{identifying legal issues when using digital media and platforms} were rated as at least \textit{moderately important} ($M\geq3$). The more subject-specific competencies from the DiKoLAN-framework, \textit{data acquisition} ($M=4.27, SE=0.10$), \textit {data processing} ($M=4.64, SE=0.08$), and \textit{simulation and modeling} ($M=3.63, SE=0.11$), tend to be rated as more important than rather general competencies (\textit{communication and collaboration} ($M=3.14, SE=0.12$), \textit{presentation} ($M=3.43, SE=0.13$), and {documentation} ($M=3.52, SE=0.13$)).

\textbf{Findings (ii):} However, the general competencies \textit{using digital tools for information search and evaluation} ($M=3.82, SE=0.11$) and \textit{technical core competencies} ($M=3.96, SE=0.11$) are rated as important, too.

\textbf{Discussion of findings (ii):} This might be explained with a subject-specific interpretation of these skills by the instructors, e.g., information search and evaluation associated with preparing for the lab day or technical core competencies with manipulating lab equipment.

\textbf{Findings (iii):} The rating of each digital competency is rather independent of the instructors' \textit{attitude} (cf. Sec.~\ref{attitude}) toward digital technologies in physics labs as the related Spearman-Rho correlation coefficient is significant only for \textit{performing computer-aided modeling, programming, and using existing digital simulations} ($r(58)=.27, p=.035$).

\textbf{Discussion of findings (iii):} So, instructors with a high attitude toward digital technologies in physics lab courses only more strongly desire programming skills.

\textbf{Findings (iv):} The instructors could add up to four digital competencies they consider as important to be acquired but were not listed in the predefined answers. Five out of 73 participants used this opportunity and listed one (one participant listed two) further digital learning objective(s) and rated their importance as \textit{important} or \textit{very important}. The added learning objectives were (shortened and translated): \textit{using digital tools to manage scientific writing, understanding how an analog-to-digital- and a digital-to-analog converter work, proper digitizing of data, learning good scientific practice in general, independent familiarization with digital tools}, and \textit{recognizing limits and approximations in digital tools}.

\textbf{Discussion of findings (iv):} We consider these responses as either not specific \textit{digital} learning objectives (e.g., learning good scientific practice in general) or assignable to the predefined learning objectives (e.g., digital tools to scientific writing can be mapped to \textit{using digital media for presentation}, objectives about meaningfully using digital tools for data collection can be mapped to \textit{collecting data directly or indirectly with digital tools}). Thus, the survey did not reveal any new digital learning objectives we have not considered before.

\subsection{Potential of different (digital) forms of labwork (RQ3)}\label{potential}

\textbf{Findings (i):} In addition to the rating of the learning objectives, the instructors evaluated eight different (digital) forms of labwork according to their usefulness to achieve the five main learning objectives. FIG.~\ref{fig:labformatsopenness} shows the usefulness of different degrees of openness of lab courses from the instructors' point of view. Overall, the instructors rated more open forms of lab work (\textit{an open-ended labwork session} and \textit{undergraduate research projects}) as more useful to achieve the learning objectives than \textit{a (strongly) guided labwork session}. While differences are rather small for the learning objectives \textit{linking theory to practice}, \textit{learn experimental skills}, and \textit{acquire digital competencies}, a high degree of openness is expected to outperform (strongly) guided lab sessions when \textit{methods of scientific thinking} should be taught or \textit{increasing the students' motivation, personal development and social competency} should be fostered.

\textbf{Discussion of findings (i):} Comparative data by Ref.~\cite{Haller.1999} show that within the scope of standard errors instructors nowadays rate the usefulness of \textit{a (strongly) guided labwork session} similar to German physics professors in the late 90s but rate \textit{an open-ended labwork session} as more useful nowadays, especially for \textit{link theory to practice} and \textit{get to know the methods of scientific thinking}. So, in comparison to 25 years ago, lab instructors nowadays perceive more open forms of labwork as more useful than guided forms of labwork. This shift is in accordance with previous research that has shown that a higher degree of openness is more conducive to learning than strong guidance \cite{Holmes.2018} and motivation and scientific thinking can be better promoted in open inquiry-based learning settings \cite{Etkina.2015}.

\begin{figure}[htb]
\flushleft 
\begin{tikzpicture}
\begin{axis}[width=.83\columnwidth, height=13cm, xbar=3pt,
  ymax=1.1,  
  xmin=1, xmax=5,
  xtick={1,1.5,2,2.5,3,3.5,4,4.5,5},
  ymin =0.1,
  xlabel={Perceived usefulness},
  extra x ticks={1,5},
extra x tick labels={not useful, very useful},
extra x tick style={grid=none, tick style={draw=none}, tick label style={xshift=-19pt, yshift=-8pt}},
  ytick = {0.2,0.4,.6,.8,1.},
  yticklabel style={text width=.3\columnwidth,align=right, },
  yticklabels={acquire digital competencies, increase their motivation{,} personal development{,} and social competency, get to know the methods of scientific thinking, learn experimental skills, link theory and practice},
   ytick pos=left,
xtick pos=left,
 legend columns=1, legend cell align = left,legend style = {draw = none},
 legend style = {at ={(-.5,1)}, anchor = south west},
 bar width = 8pt,
  ]
\addlegendimage{empty legend} \addlegendentry{\scalebox{1}[1]{\ref{guided}} A (strongly) guided labwork session}
\addlegendimage{empty legend}
\addlegendentry{\scalebox{1}[1]{\ref{reference guided}} Comparative data (a strongly guided labwork session)}
\addlegendimage{empty legend}
\addlegendentry{\scalebox{1}[1]{\ref{open}} An open-ended labwork session}
\addlegendimage{empty legend}
\addlegendentry{\scalebox{1}[1]{\ref{reference open}} Comparative data (an open-ended labwork session)}
\addlegendimage{empty legend}
\addlegendentry{\scalebox{1}[1]{\ref{undergraduate}} Undergraduate research projects in groups}

\draw[dashed] (200,0.1) -- (200,100);

\addplot+[gray!20!,area legend,
    draw=black, error bars/.cd, x dir=both, x explicit, error mark options={black,mark size=2pt,line width=.7pt,rotate=90
     },  error bar style={line width=.7pt}
      ] 
		coordinates{
(	3.94, 1) +-(.136, .1)
(   4.19, .8)  +-(.13, .8)
(	4.16, .6)  +-(.132, .6)
(	4.48, .4)  +-(.114, .4)
(3.47, .2) +-(.157, .2)
}; \label{undergraduate}

\addplot+[gray!60!,pattern=north west lines, pattern color=gray!60!,area legend,
    draw=black, error bars/.cd, x dir=both, x explicit, error mark options={black,mark size=2pt,line width=.7pt,rotate=90
     },  error bar style={line width=.7pt}
      ] 
		coordinates{
(	3.2, 1) +-(.57, .1)
(   3.8, .8)  +-(.51, .8)
(	3.5, .6)  +-(.52, .6)
(	3.3, .4)  +-(.97, .4)
}; \label{reference open}

\addplot+[gray!60!,area legend,
    draw=black, error bars/.cd, x dir=both, x explicit, error mark options={black,mark size=2pt,line width=.7pt,rotate=90
     },  error bar style={line width=.7pt}
      ] 
		coordinates{
(	3.95, 1) +-(.119, 1)
(	4.2, .8)  +-(.114, .8)
(	4.17, .6)  +-(.103, .6)
(	3.84, .4)  +-(.13, .4)
(	3.14, .2)  +-(.146, .2)
}; \label{open}

\addplot+[gray,pattern=north east lines,pattern color=gray,area legend,
    draw=black, error bars/.cd, x dir=both, x explicit, error mark options={black,mark size=2pt,line width=.7pt,rotate=90
     },  error bar style={line width=.7pt}
      ] 
		coordinates{
(	3.5, 1) +-(.37, .1)
(   4.1, .8)  +-(.28, .8)
(	3.4, .6)  +-(.48, .6)
(	2.47, .4)  +-(.81, .4)
}; \label{reference guided}

\addplot+[gray,area legend,
    draw=black, error bars/.cd, x dir=both, x explicit, error mark options={black,mark size=2pt,line width=.7pt,rotate=90
     },  error bar style={line width=.7pt}
      ] 
		coordinates{
(	3.57, 1	) +-(.156, 1)
(	3.75, 0.8) +-(.143,0.8)
(	3.06, .6) +-(.155,.6)
(	2.64, .4	) +-(.142,.4)
(	2.91, .2	)+-(.151,.2)
}; \label{guided}

\end{axis}
\end{tikzpicture}
\caption{Mean and standard error of the usefulness of different degrees of openness of lab courses to achieve different learning objectives from instructors' point of view ($N=63-64$, depending on the learning objective). The dashed line indicates moderate usefulness. If applicable, comparative data ($N=10$) from Ref.~\cite{Haller.1999} is added.}
\label{fig:labformatsopenness}
\end{figure}
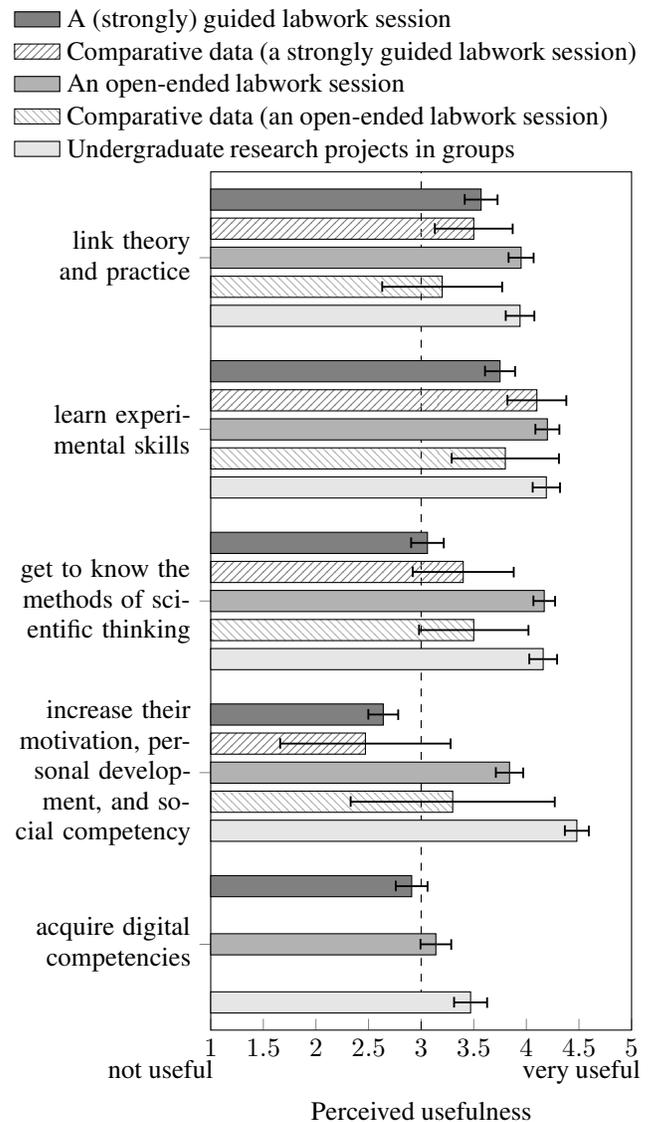

\begin{figure}[htb]
\flushleft
\begin{tikzpicture}
\begin{axis}[width=.83\columnwidth, height=13cm, xbar=3pt,
  ymax=1.1,  
  xmin=1, xmax=5,
  xtick={1,1.5,2,2.5,3,3.5,4,4.5,5},
  ymin =0.1,
  xlabel={Perceived usefulness},
  extra x ticks={1,5},
extra x tick labels={not useful, very useful},
extra x tick style={grid=none, tick style={draw=none}, tick label style={xshift=-19pt, yshift=-8pt}},
  ytick = {0.2,0.4,.6,.8,1.},
  yticklabel style={text width=.3\columnwidth,align=right, },
  yticklabels={acquire digital competencies, increase their motivation{,} personal development{,} and social competency, get to know the methods of scientific thinking, learn experimental skills, link theory and practice},
   ytick pos=left,
xtick pos=left,
 legend columns=1, legend cell align = left,legend style = {draw = none},
 legend style = {at ={(-.5,1)}, anchor = south west},
 bar width = 8pt,
  ]
\addlegendimage{empty legend}
\addlegendentry{\scalebox{1}[1]{\ref{smartphone}} Experiments with smartphones/tablets}
\addlegendimage{empty legend}
\addlegendentry{\scalebox{1}[1]{\ref{remote}} Remote-controlled experiments}
\addlegendimage{empty legend}
\addlegendentry{\scalebox{1}[1]{\ref{VR}} Virtual Reality environment}
\addlegendimage{empty legend}
\addlegendentry{\scalebox{1}[1]{\ref{simulation}} Usage and/or creation of computer simulations}
\addlegendimage{empty legend}
\addlegendentry{\scalebox{1}[1]{\ref{microcontroller}} Microcontrollers}

\draw[dashed] (200,0.1) -- (200,100);

\addplot+[gray!0!,area legend,
    draw=black, error bars/.cd, x dir=both, x explicit, error mark options={black,mark size=2pt,line width=.7pt,rotate=90
     },  error bar style={line width=.7pt}
      ] 
		coordinates{
(	3.06, 1	) +-(.139, 1)
(	3.58, 0.8) +-(.122,0.8)
(	2.87, .6) +-(.140,.6)
(	2.81, .4	) +-(.141,.4)
(	4.23, .2	)+-(.127,.2)
}; \label{microcontroller}

\addplot+[gray!20!,area legend,
    draw=black, error bars/.cd, x dir=both, x explicit, error mark options={black,mark size=2pt,line width=.7pt,rotate=90
     },  error bar style={line width=.7pt}
      ] 
		coordinates{
(	3.41, 1	) +-(.142, 1)
(	2.64, 0.8) +-(.167,0.8)
(	3.4, .6) +-(.135,.6)
(	2.73, .4	) +-(.14,.4)
(	4.34, .2	)+-(.1,.2)
}; \label{simulation}

\addplot+[gray!60!,area legend,
    draw=black, error bars/.cd, x dir=both, x explicit, error mark options={black,mark size=2pt,line width=.7pt,rotate=90
     },  error bar style={line width=.7pt}
      ] 
		coordinates{
(	2.6, 1	) +-(.14, 1)
(	2.31, 0.8) +-(.133,0.8)
(	2.48, .6) +-(.130,.6)
(	2.31, .4	) +-(.122,.4)
(	3.38, .2	)+-(.15,.2)
}; \label{VR}

\addplot+[gray,area legend,
    draw=black, error bars/.cd, x dir=both, x explicit, error mark options={black,mark size=2pt,line width=.7pt,rotate=90
     },  error bar style={line width=.7pt}
      ] 
		coordinates{
(	2.62, 1	) +-(.138, 1)
(	2.5, 0.8) +-(.144,0.8)
(	2.40, .6) +-(.127,.6)
(	2.09, .4	) +-(.113,.4)
(	3.39, .2	)+-(.15,.2)
}; \label{remote}

\addplot+[black!65!, area legend,
    draw=black, error bars/.cd, x dir=both, x explicit, error mark options={black,mark size=2pt,line width=.7pt,rotate=90
     },  error bar style={line width=.7pt}
      ] 
		coordinates{
(	3.02, 1	) +-(.133, 1)
(	3.11, 0.8) +-(.134,0.8)
(	2.75, .6) +-(.132,.6)
(	2.94, .4	) +-(.133,.4)
(	3.89, .2	)+-(.126,.2)
}; \label{smartphone}

\end{axis}
\end{tikzpicture}
\caption{Mean and standard error of the usefulness of different technologies in lab courses to achieve the different learning objectives from the instructors' point of view ($N=63-64$, depending on the learning objective). The dashed line indicates moderate usefulness.}
\label{fig:labformatstechn}
\end{figure}
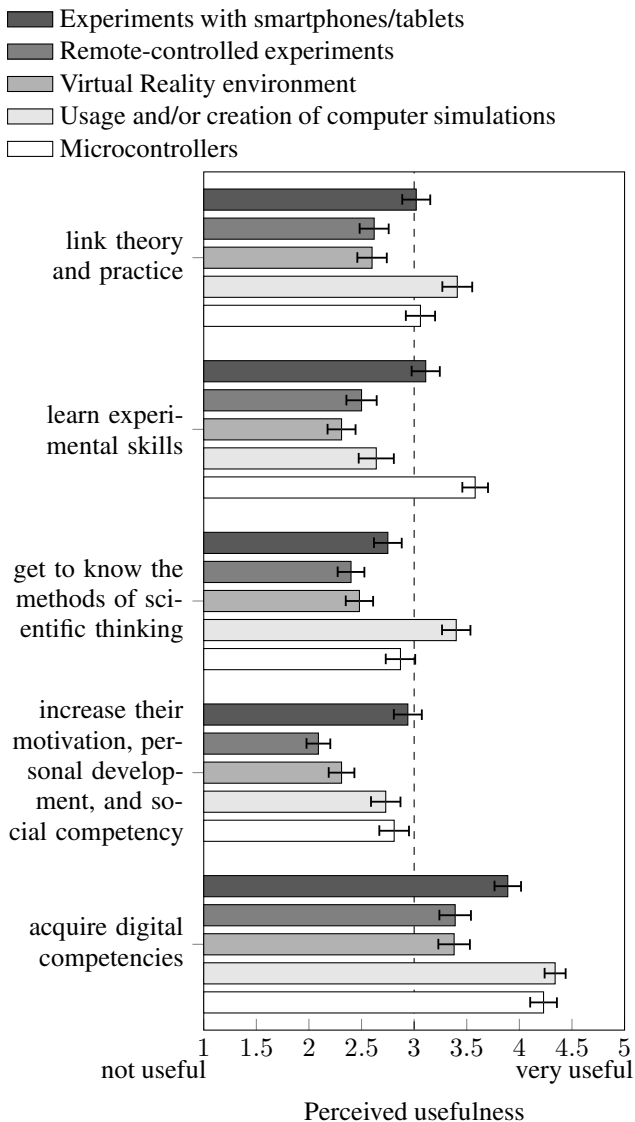

\textbf{Findings (ii):} FIG.~\ref{fig:labformatstechn} displays how the instructors perceived the usefulness of different digital technologies for achieving the learning objectives. In general, \textit{computer simulations}, \textit{microcontrollers}, and \textit{smartphones/tablets} are rated as more useful than \textit{remote-controlled experiments} and \textit{Virtual reality environments} which are rated as \textit{less useful} for achieving any of the five learning objectives besides \textit{acquiring digital competencies}.

However, even technologies rated as generally more useful are not perceived similarly for achieving all learning objectives. For \textit{linking theory and practice} and \textit{to get to know the methods of scientific thinking}, besides microcontrollers and smartphones/tablets especially the \textit{usage and/or creation of computer simulations} are helpful. Microcontrollers and smartphones/tablets can also better help to \textit{learn experimental skills} than other technologies like Virtual Reality. 
To \textit{increase the students' motivation, personal development, and social competency}, smartphones/tablets are perceived as more useful than remote-controlled experiments and Virtual Reality. Only for \textit{acquiring digital technologies}, all technologies are rated as \textit{useful} (but again with lower ratings for remote-controlled experiments and Virtual Reality).

\textbf{Discussion of findings (ii):} The instructors' differentiated perception of the potential of the technologies is in accordance with the literature review in Ref.~\cite{Chen.2012} where microcomputer-based laboratories and simulations were rated with a higher potential for supporting different learning activities and open inquiry than remote labs. The findings can also be mapped to a rather engineering-related literature review by Ref.~\cite{Ma.2006} revealing that instructors link remote-controlled experiments primarily to the learning objectives \textit{fostering conceptual understanding} (cf. \textit{link theory and practice}) and \textit{professional skills} (i.e., technical skills, cf. \textit{acquire digital competencies}) while simulated labs (cf. usage/and or creation of computer simulations) are more often linked to \textit{design skills} (i.e., scientific mind and ability to design and investigate, cf. \textit{get to know the methods of scientific thinking}), too.

\begin{figure}[htb]
\flushleft
\begin{tikzpicture}
\begin{axis}[width=.83\columnwidth, height=10cm, xbar=3pt,
  ymax=1.1,  
  xmin=1, xmax=5,
  xtick={1,1.5,2,2.5,3,3.5,4,4.5,5},
  ymin =0.1,
  xlabel={Perceived usefulness},
  extra x ticks={1,5},
extra x tick labels={not useful, very useful},
extra x tick style={grid=none, tick style={draw=none}, tick label style={xshift=-19pt, yshift=-8pt}},
  ytick = {0.2,0.4,.6,.8,1.},
  yticklabel style={text width=.3\columnwidth,align=right, },
  yticklabels={acquire digital competencies, increase their motivation{,} personal development{,} and social competency, get to know the methods of scientific thinking, learn experimental skills, link theory and practice},
   ytick pos=left,
xtick pos=left,
 legend columns=1, legend cell align = left,legend style = {draw = none},
 legend style = {at ={(-.5,1)}, anchor = south west},
 bar width = 8pt,
  ]
\addlegendimage{empty legend}
\addlegendentry{\scalebox{1}[1]{\ref{meanuse}} Mean over the different technologies}
\addlegendimage{empty legend}
\addlegendentry{\scalebox{1}[1]{\ref{reference}} Comparative data (modern technologies in general)}

\draw[dashed] (200,0.1) -- (200,100);

\addplot+[gray!60!,pattern=north east lines,area legend,
    draw=black, error bars/.cd, x dir=both, x explicit, error mark options={black,mark size=2pt,line width=.7pt,rotate=90
     },  error bar style={line width=.7pt}
      ] 
		coordinates{
(	2.5, 1	) +-(.56, 1)
(	2.3, 0.8) +-(.54,0.8)
(	2.6, .6) +-(.48,.6)
(	2.44, .4	) +-(.59,.4)
}; \label{reference}

\addplot+[gray!60!,area legend,
    draw=black, error bars/.cd, x dir=both, x explicit, error mark options={black,mark size=2pt,line width=.7pt,rotate=90
     },  error bar style={line width=.7pt}
      ] 
		coordinates{
(	2.94, 1	) +-(.11, 1)
(	2.83, 0.8) +-(.11,0.8)
(	2.78, .6) +-(.10,.6)
(	2.58, .4	) +-(.10,.4)
(	3.85, .2	) +-(.10,.2)
}; \label{meanuse}

\end{axis}
\end{tikzpicture}
\caption{Mean and standard error of the usefulness of digital technologies in lab courses to achieve the different learning objectives from the instructors' point of view ($N=63-64$, depending on the learning objective), determined for each instructor's mean of the rating of all five digital technologies as displayed in FIG.~\ref{fig:labformatstechn}. The dashed line indicates moderate usefulness. If applicable, comparative data ($N=10$) from Ref.~\cite{Haller.1999} is added.}
\label{fig:labformatstechncomp}
\end{figure}
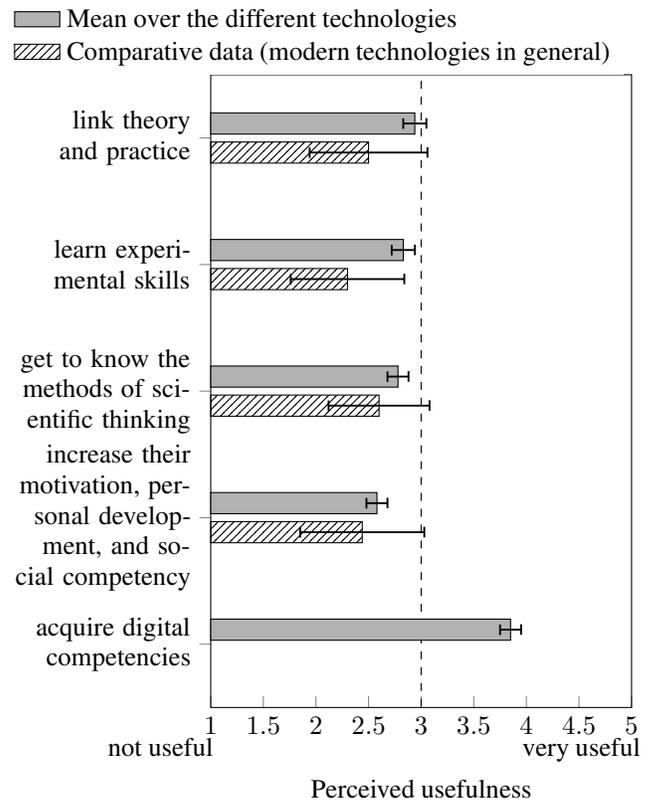

For comparison with data from the late 90s \cite{Haller.1999} where German physics professors rated the usefulness of \textit{experiments using modern technologies} (e.g., for data capture or modeling), we determined each instructor's rating of all five digital technologies per learning objective and derived the according mean and standard error for all instructors per learning objective. The comparison is displayed in FIG.~\ref{fig:labformatstechncomp} revealing that in tendency, digital technologies are nowadays perceived as more useful to reach the learning objectives, especially to \textit{link theory to practice} and \textit{to learn experimental skills}. However, due to high standard errors in the former study, these differences are not significant within the scope of standard errors.

\subsection{Instructors' future plans about digital technologies in labs}\label{futureplans}

\textbf{Findings:} 31 instructors described in their own words whether they have plans to (further) include any digital technologies or the acquisition of digital competencies in their lab course of whom 20 mentioned specific plans for the near future or the recent past, five described at least the desire for or some initial thoughts about such plans, and six negated the existence of any plans. Their non-existence was justified by the focus of their lab courses on traditional tasks, the preference of desktop computers for tablets for reasons of cost and theft protection, the lack of smartphones among some students, and liability issues when students use private smartphones. The instructors who neither mentioned nor negated any specific plans explained this with limited involvement in the lab course conception (e.g., as student teaching assistants) or just expressed positive views on digital technologies without formulating any goals. Specifically mentioned plans are primarily related to digitalized data collection, especially with microcontrollers/ single-board computers (mentioned by 6 instructors), external sensors (2), and digital data acquisition and analysis in general (3). By this, the instructors want to foster programming skills, deepen the understanding of the theory, enable more independent, engaging, and project-based experimental processes, or facilitate data collection and analysis. Further plans are related to digitalizing learning materials, e.g., including simulations (2), preparing explanatory videos (1), or using digital tools for communication and collaboration (2) to better prepare the students for the lab day. Other plans are related to hardware digitalization (personal computers and active boards), the use of electronic lab books, the preparation of experiments for distance learning for future pandemics (1), or more spatial flexibility for the students (1). 

\textbf{Discussion of findings:} To conclude, most instructors who responded to this question mentioned some ongoing or future plans for further implementing digital technologies in their lab courses. They primarily focus on digital technologies, especially single-board computers, rather than competencies to be fostered with the implementation of these technologies. However, 32 instructors skipped this question in the questionnaire (either because they have no related plans or because they did not want to write it down in the survey) and six further participants negated any related plans, so one can assume that there is a significant number of lab courses for which there are no specific plans for further implementing digital technologies.

\subsection{Discussion of study limitations}\label{limitations}

As our survey is about the instructors' perception of the role of digital technologies, the findings largely depend on the group of participants. We cannot fully rule out the possibility of socially desirable responses or a bias in participation that probably especially instructors with a positive attitude toward and/or a lot of experience with the use of digital technologies in their lab courses participated in our survey. Also the mode of data collection using an online survey tool might have affected the participation due to varying instructors' web affinity, albeit we assume that the resulting selection bias is of much lower relevance than the aforementioned similar, but more significant attitudes and experiences related to digital technologies in lab courses.

Furthermore, the responses depend on the instructors' perception of the used terms in our survey, especially what \textit{modern digital technologies} are or what the described learning objectives and forms of labwork are about. To mitigate this, we provided descriptions, definitions, and examples as far as possible and reasonable. However, the provided list of digital technologies might have also influenced the perception of digital technologies as one instructor mentioned in the final open text field that \textit{the survey focuses heavily on smartphones/tablets and completely ignores standalone computers as a digital medium}. Relating thereto, a further limitation of our study is that we have not narrowed down our survey to, e.g., one specific are of digitalization as identified by Ref.~\cite{Franke.2022b} but used the term \textit{modern digital technologies} in a broad meaning even though it is hard to compare so different forms of digitalization like virtual reality or smartphone experiments with software for word processing or data analysis. Future work should definitely focus specific areas of digitalization or even specific technologies. However, we accepted this limitation for our own study in favor of our study goal to capture a kind of screenshot of the use of digital technologies and the importance of acquiring digital competencies in the interested physics lab courses. This goal had the consequence that we needed to be open about what we understand under \textit{modern digital technologies} to get as many different responses from as many instructors as possible to investigate what they understand under \textit{modern digital technologies}. And Fig.~\ref{fig:experiences} and the shared experiences and attitudes in Tab.~\ref{tab:exptechn} and Tab.~\ref{tab:attitudesqual} show exactly this diverse perspective on digital technologies in physics lab courses. So, from our point of view, the broad concept of the term \textit{digital technologies} in our study is a limitation we could not avoid, if we wanted to provide this overview, especially as the prior work by Ref.~\cite{Franke.2022b} certainly gave us some indications but their findings have not been based on a large empirical base yet.

Another limitation we are aware of is the risk that despite several control cycles taken by us translations might have led to slightly different survey versions. Nevertheless, we decided to provide the questionnaire in the three national languages of our target countries to lower the threshold for participation by minimizing the risk of language-related obstacles.

Moreover, the sample size is crucial for the reliability of our findings. For data protection purposes and for reducing the risk of socially desirable responses, we could not ask the participants about their institutions. Thus, it is difficult to estimate the actual coverage rate of our survey within the main target group (instructors for introductory and advanced physics lab courses). To provide at least a cautious estimation, in TABLE~\ref{penetration} we compare the number of institutions (universities, universities of applied sciences, etc.) with at least one physics-related undergraduate degree program (e.g., physics major, physics teacher education, applied physics, physical engineering, etc.) in the three countries with the number of survey participants who are responsible for introductory or advanced lab courses. By this, we acknowledge that there is usually more than one instructor per lab course (e.g., with all the student assistants) but often only one is in charge of the lab course, i.e., actually responsible. However, we cannot exclude the possibility of instructors not being the head of a lab course but still feeling responsible for it (e.g. as they designed materials for the course) and there are also institutions with more than one introductory and advanced lab course. Overall, we suspect that we reached a good coverage of the target group especially in Finland and Croatia while in Germany the target group size impedes comparable proportions. Since we anyway found only slight differences between the three different countries (cf. TABLE \ref{tab:differences}), we would argue that the sample size is not a serious limitation of our survey.

\begin{table}[h!tb]
\caption{Sub-sample size of those survey participants who are responsible for the concept/organization of an introductory or advanced lab course in comparison to the number of institutions with physics-related undergraduate degree programs per country (Germany (DE), Finland (FI), and Croatia (HR)) to estimate the survey coverage rate.}
\begin{ruledtabular}
\begin{tabular}{p{.675\columnwidth}p{.095\columnwidth} p{.095\columnwidth}p{.095\columnwidth}}
&\textbf{DE}&\textbf{FI}&\textbf{HR}\\\hline
\raggedright\hangindent=.3cm Institutions with minimum one physics- related undergraduate degree program&86&9&4\\\hline
\raggedright\hangindent=.3cm \textbf{Survey participants responsible for}&&&\\
Introductory lab course&13&5&3\\
Advanced lab course&7&3&2\\
\end{tabular}
\end{ruledtabular}
\label{penetration}
\end{table}

\section{Conclusion and outlook}\label{conclusion}

\subsection{Review and outlook on digital technologies and competencies in physics lab courses}\label{review/outlook}

We surveyed physics lab instructors in Germany, Finland, and Croatia to investigate the status quo of digital technologies and related digital competencies in European physics lab courses. Here, we discuss our findings related to the impact of the emergence of new digital technologies and the Covid-19 pandemic on physics lab courses to examine implications that our findings might have regarding possible goals and paths for future transformations. This provides a review and outlook on digital technologies and competencies in physics lab courses.

Our findings show that digital technologies are indeed part of physics lab courses as 61 out of 68 participants stated that they have used digital technologies in their lab courses at least once and on average even more than three digital technologies are used. They concern all four areas of digitalization of lab courses \cite{Franke.2022b}, especially the data analysis with corresponding software, data collection with smartphones/tablets, microcontrollers, or other equipment for remote/automated data collection, and the use of computer simulations. The Covid-19 pandemic has boosted the use of digital technologies as the instructors reported a higher frequency of use of digital technologies after the pandemic in comparison to before (cf. Fig.~\ref{fig:technologies} \& Appx.~\ref{appendixb}). Especially digital technologies for digital exams/submission of work and communication/collaboration/organization as well as computer simulations and smartphones/tablets were newly introduced in many lab courses for their implementation under pandemic conditions and remained afterward.

The instructors reported an overall positive attitude toward digital technologies in physics lab courses that was primarily argued with the related potential for conducting experiments and the students' competence acquisition as well as their contemporaneity implying importance for the students' future and labor market. However, many instructors also stated clearly that digital technologies should not be used as an end in themselves but rather when their use is beneficial in the specific use case. Accordingly, they perceived different digital technologies as distinctly useful to follow the learning objectives (e.g., computer simulations for \textit{linking theory to practice} and \textit{getting to know the methods of scientific thinking}, microcontrollers, and smartphones/tablets for \textit{learning experimental skills} and for \textit{increasing the students' motivation, personal development, and social competency}).

The instructors also perceive a variety of digital technologies as useful to support the students' acquisition of digital competencies (even with the otherwise critically perceived Virtual Reality environments). However, in comparison to the other learning objectives (especially \textit{learning experimental skills} and \textit{linking theory to practice}) \textit{acquiring digital competencies} plays a subordinate role both in the instructors' current implementation in their lab courses and their desired (future) state. Even though the focused acquisition of digital competencies is perceived as rather unimportant in comparison to other major learning objectives, the instructors agree that students should learn (besides other digitally aided activities like information search and evaluation) how to collect and process data with digital tools. That is also in accordance with the digital technologies the instructors have already implemented and can be mapped to the highly rated importance of students' acquiring experimental skills. Instructors with a higher attitude toward the use of digital technologies in lab courses perceive the students' acquisition of digital competencies in general and \textit{performing of computer-aided modeling, programming, and use of digital simulations} in particular as more important than instructors with a lower attitude.

This review of the status quo shows that the digital transformation in our society accelerated by the Covid-19 pandemic has largely influenced physics lab courses, especially regarding the use of digital technologies that tend to be perceived as more useful for reaching the lab course learning objectives as in the former survey by Ref.~\cite{Haller.1999} in the late 90s (cf. Fig.~\ref{fig:labformatstechncomp}). However, according to the instructors, the choice of technologies should be made with care by considering the intended purpose of use to avoid use as an end in itself. The acquisition of digital competencies is perceived as less important than other learning objectives of physics lab courses. Correspondingly, the instructors' specific future plans are linked to the increasing use of new digital technologies, especially microcontrollers/ single-board computers, but not specifically to the students' acquisition of digital competencies.

For future transformations, the findings reveal that instructors would like to focus more on the students \textit{to get to know the methods of scientific thinking} and consider open forms of labwork and the usage and/or creation of computer simulations as useful to reach this learning objective. The same applies to the two main learning objectives, \textit{learning experimental skills} and \textit{linking theory to practice}, where microcontrollers and smartphones/tablets could support their achievement. For this, the versatility, portability, and availability of smartphones and other mobile devices as \textit{labs in a pocket} \cite{Stampfer.2020} makes them a very promising perspective also for more open forms of student experimentation like in design labs or undergraduate research projects being more conducive to learning \cite{RuizPrimo.2011,Etkina.2015,Holmes.2016}, which today can already draw on a broad extant body of literature (cf. Ref.~\cite{Monteiro.2022d} for a review, Ref.~\cite{Kuhn.2022} for a broad collection of inspiring examples, and Ref.~\cite{Barro.2023} for work specifically related to undergraduate research projects).

\subsection{Ideas for future investigation}

The presented survey can be seen as a basis for further investigations. Similar surveys in other countries both in Europe and other continents (e.g., the United States or Asian countries) would provide a more global, holistic view of the status quo of digital technologies and competencies in physics lab courses. With higher sample sizes, one could also more deeply investigate differences between different countries and lab course types since especially the learning objectives might be different for specific target groups (e.g., physics major vs medicine students). Furthermore, it would be worthwhile to mirror the instructors' perceptions as investigated in this survey with the students' perspective and research data from a validated instrument for measuring the acquisition of digital competencies. In the latter case, it would particularly be interesting to measure to what extent the students acquire digital competencies in physics lab courses and in comparative studies whether the different forms of labwork and digital technologies have the potential for reaching the different learning objectives as perceived by the instructors. Additionally, similar surveys could be conducted in a regular rhythm to investigate and map the developments and transformations of physics lab courses over time. We motivated our survey with a similar one done by Ref.~\cite{Welzel.1998,Welzel.1998b,Haller.1999} about the learning objectives of physics lab courses in the late 90s and how the use of digital technologies was perceived at that time. Our survey renewed this perspective with contemporary digital technologies but even while the presented survey was conducted, with artificial intelligence large language models like ChatGPT \cite{OpenAI.2022} new technologies have appeared that will have a significant impact on teaching and learning in general and probably also on physics lab courses (e.g., for lab reports, the students' preparation for a lab day, automated feedback and assessment, or general possibilities listed in \cite{Okonkwo.2021}). Thus, physics lab courses undergo a permanent transformation process led by different drivers like advancing digitalization and the emergence of ever-new digital technologies that should be observed and reflected in progress, to ensure modern and high-quality university physics education in the long term.

\section*{Acknowledgments}
We would like to thank all instructors who participated in our survey. Furthermore, we would like to thank our student assistants who helped us in implementing and testing the survey.
This work was done within the scope of the project \textit{Developing Digital Physics Laboratory Work for Distance Learning} (DigiPhysLab) funded by the Erasmus+ program of the European Union (2020-1-FI01-KA226-HE-092531).

\section*{Ethical statement}
The online survey was anonymous. In addition to an access time stamp, only personal data that participants were explicitly asked for during the survey was collected and stored. Data collection and storage were organized in accordance with the General Data Protection Regulation of the European Union and were coordinated with the data protection officer of the University of Göttingen. All participants were informed about the survey goals and the planned data collection and storage beforehand and consented to voluntarily participate.

\section*{Author contributions*}
\textbf{Pascal Klein}: Conceptualization (supporting), Funding Acquisition (supporting), Supervision (equal), Writing – review \& editing (supporting). \textbf{Simon Z. Lahme}: Conceptualization (lead), Data curation (lead), Formal Analysis, Investigation (lead), Methodology, Project administration (presented survey), Software (lead), Validation (lead), Visualization, Writing – original draft, Writing – review \& editing (lead). \textbf{Antti Lehtinen}: Conceptualization (supporting), Funding acquisition (lead), Project administration (DigiPhysLab), Supervision (equal), Writing – review \& editing (supporting). \textbf{Andreas Müller}: Supervision (equal), Writing – review \& editing (supporting). \textbf{Pekka Pirinen}: Conceptualization (supporting), Data curation (supporting), Investigation (supporting), Software (supporting), Validation (supporting), Writing – review \& editing (supporting). \textbf{Lucija Rončević}: Data curation (supporting), Investigation (supporting), Software (supporting), Validation (supporting). \textbf{Ana Sušac}: Conceptualization (supporting), Funding Acquisition (supporting), Supervision (equal), Writing – review \& editing (supporting).

* According to CREDIT (CRediT Contributor Roles Taxonomy), \url{https://credit.niso.org}


\appendix

\section{Overview of significant differences for countries and lab course types}\label{appendixa}

\squeezetable
\begin{longtable*}{p{.13\textwidth}p{.415\textwidth}p{.415\textwidth}}
\caption{Overview of significant differences within sections of the questionnaire regarding the participants' countries and lab course types. With a Kruskal-Wallis test ($\alpha=.05$) differences (related $p$-values in square brackets) were identified. Afterward, pairwise comparisons were done to search for significant differences (only Bonferroni corrected comparisons with $p_B<.05$ reported).}\\
\hline\hline
\textbf{Section}&\textbf{Differences between the countries}&\textbf{Differences between the lab course types}\\\hline
\raggedright Role of digital learning objectives in comparison to other learning objectives&No significant differences were found.&
\vspace{-\topsep}\vspace{1pt}\begin{itemize}[label=\textbullet, leftmargin=*, topsep=0pt, itemsep=0pt, partopsep=0pt, parsep=0pt]
    \item \textit{Learn experimental skills (implemented)} [$p=.047$] but all $p_B\geq.05$.
    \item \textit{Increase their motivation, personal development, and social competency (implemented)} [$p=.042$] was rated as more important for \textit{other} labs than introductory ($p_B=.035$).
\end{itemize}\vspace{-\topsep}~\vspace{-3pt}\\\hline
\raggedright Importance of specific digital competencies perceived by the instructors
&
\vspace{-\topsep}\vspace{1pt}\begin{itemize}[label=\textbullet, leftmargin=*, topsep=0pt, itemsep=0pt, partopsep=0pt, parsep=0pt]
    \item \textit{Using digital tools for information search and evaluation} [$p=.034$] but all $p_B\geq.05$.
    \item \textit{Identifying digital competence gaps} [$p=.010$] was rated as more important by Croatian instructors than by Finnish ($p_B=.008$).
\end{itemize}\vspace{-\topsep}~\vspace{-3pt}&\hangindent=.3cm No significant differences were found.\\\hline
\raggedright Potential of different (digital) lab formats for achieving different learning objectives&
\vspace{-\topsep}\vspace{1pt}\begin{itemize}[label=\textbullet, leftmargin=*, topsep=0pt, itemsep=0pt, partopsep=0pt, parsep=0pt]
    \item\textit{Learning experimental skills}: German instructors rated \textit{remote-controlled experiments} [$p<.001$] as less useful than Finnish ($p_B<.001$) and Croatian ($p_B=.005$) and \textit{microcontroller} [$p=.014$] as less useful than Croatian ($p_B=.046$). The \textit{usage and/or creation of computer simulations} was significant [$p=.032$] but all $p_B\geq.05$.
    \item\textit{Getting to know the methods of scientific thinking}. Croatian instructors rated \textit{microcontroller} [$p=.007$] as more useful than German ($p_B=.006$) and Finnish ($p_B=.036$).
    \item\textit{Acquire digital competencies}: German instructors rated \textit{remote-controlled labs} [$p=.016$] as less useful than Finnish ($p_B=.045$). The \textit{usage and/or creation of computer simulations} was significant [$p=.047$] but all $p_B\geq.05$.
\end{itemize}\vspace{-\topsep}~\vspace{-5pt}&
\vspace{-\topsep}\vspace{1pt}\begin{itemize}[label=\textbullet, leftmargin=*, topsep=0pt, itemsep=0pt, partopsep=0pt, parsep=0pt]
    \item\textit{Getting to know the methods of scientific thinking} was significant for \textit{smartphones/tablets} [$p=.044$] but all $p_B\geq.05$.
    \item\textit{Increasing the students' motivation, personal development, and social competency}: \textit{Remote-controlled experiments} [$p=.011$] were rated as less useful in engineering labs in comparison to introductory ($p_B=.011$) and advanced ($p_B=.021$) labs and that the \textit{usage and/or creation of computer simulations} [$p=.012$] is less useful in engineering labs in comparison to advanced labs ($p_B=.008$).
\end{itemize}\vspace{-\topsep}~\vspace{-7pt}\\
\\\hline\hline
\label{tab:differences}
\end{longtable*}


\section{Descriptive statistical parameters}\label{appendixstatis}

\squeezetable
\begin{longtable*}{p{.18\textwidth}p{.6\textwidth}p{.03\textwidth}p{.03\textwidth}p{.03\textwidth}p{.03\textwidth}p{.03\textwidth}}
\caption{Descriptive statistical parameters of the used closed items in the questionnaire. In brackets, the related figure or section where the data are discussed or displayed is given. For the full survey instrument in all four languages see supplemental material.}\\
\hline\hline\textbf{Section}&\textbf{Item}&\textbf{N}&\textbf{Min}&\textbf{Max}&\textbf{M}&\textbf{SE}\\\hline
\multicolumn{2}{l}{\textbf{Learning objectives for physics lab courses} (cf. Fig.~\ref{fig:Comp})}&&&&&\\
Implemented&link theory to practice&75&1&5&3.60&0.13\\
&learn experimental skills&75&1&5&4.20&0.12\\
&get to know the methods of scientific thinking&75&1&5&3.23&0.14\\
&increase their motivation, personal development, and social competency&75&1&5&2.01&0.14\\
&acquire digital competencies&75&1&5&1.96&0.12\\
Desired&link theory to practice&75&1&5&3.28&0.15\\
&learn experimental skills&75&1&5&3.68&0.15\\
&get to know the methods of scientific thinking&75&1&5&3.84&0.14\\
&increase their motivation, personal development, and social competency&75&1&5&2.29&0.15\\
&acquire digital competencies&75&1&4&1.91&0.10\\\hline
\multicolumn{2}{l}{\textbf{Specific subcategories of digital competencies} (cf. Fig.~\ref{fig:DigComp})}&&&&&\\
&\hangindent=.3cm using digital tools for documentation&73&1&5&3.52&0.13\\
&\hangindent=.3cm using digital media for presentation&70&1&5&3.43&0.13\\
&\hangindent=.3cm communication and collaboration with digital tools&72&1&5&3.14&0.12\\*
&\hangindent=.3cm using digital tools for information search and evaluation&73&2&5&3.82&0.11\\*
&\hangindent=.3cm collecting data directly or indirectly with digital tools&73&2&5&4.27&0.10\\*
&\hangindent=.3cm processing data with digital tools&73&2&5&4.64&0.08\\*
&\hangindent=.3cm performing computer-aided modelling, programming, and using existing digital simulations&73&2&5&3.63&0.11\\*
&\hangindent=.3cm technical core competencies&72&1&5&3.96&0.11\\*
&\hangindent=.3cm identifying legal issues when using digital media and platforms&73&1&5&2.79&0.14\\*
&\hangindent=.3cm identifying digital competence gaps&72&1&5&3.44&0.13\\\hline
\multicolumn{2}{l}{\textbf{Relationship between special forms of labwork and the learning objectives} (cf. Fig.~\ref{fig:labformatsopenness} \& Fig.~\ref{fig:labformatstechn})}&&&&\\
Linking theory and practice& A (strongly) guided labwork session&63&1&5&3.57&0.16\\
& An open ended labwork session&63&1&5&3.95&0.12\\
& Undergraduate research projects in groups&63&1&5&3.94&0.14\\
& Experiments with smartphones/tablets&63&1&5&3.02&0.13\\
& Remote-controlled experiments&63&1&5&2.62&0.14\\
& Virtual Reality environment&63&1&5&2.60&0.14\\
& Usage and/or creation of computer simulations&63&1&5&3.41&0.14\\
& Microcontroller&63&1&5&3.06&0.14\\\cline{2-7}
Learning experimental skills& A (strongly) guided labwork session&64&1&5&3.75&0.14\\
& An open ended labwork session&64&1&5&4.20&0.12\\
& Undergraduate research projects in groups&64&1&5&4.19&0.13\\
& Experiments with smartphones/tablets &64&1&5&3.11&0.13\\
& Remote-controlled experiments&64&1&5&2.50&0.14\\
& Virtual Reality environment&64&1&5&2.31&0.13\\
& Usage and/or creation of computer simulations&64&1&5&2.64&0.17\\
& Microcontroller&64&1&5&3.58&0.12\\\cline{2-7}
Getting to know the methods of scientific thinking& A (strongly) guided labwork session&63&1&5&3.06&0.16\\
& An open ended labwork session&63&2&5&4.17&0.10\\
& Undergraduate research projects in groups&63&1&5&4.16&0.13\\
& Experiments with smartphones/tablets &63&1&5&2.75&0.13\\
& Remote-controlled experiments&63&1&5&2.40&0.13\\
& Virtual Reality environment&63&1&4&2.48&0.13\\
& Usage and/or creation of computer simulations&63&1&5&3.4&0.14\\
& Microcontroller&63&1&5&2.87&0.14\\\cline{2-7}
Increasing the students' motivation, personal development, and social competency& A (strongly) guided labwork session&64&1&5&2.64&0.14\\
& An open ended labwork session&64&1&5&3.84&0.13\\
& Undergraduate research projects in groups&64&1&5&4.48&0.11\\
& Experiments with smartphones/tablets &64&1&5&2.94&0.13\\
& Remote-controlled experiments&64&1&4&2.09&0.11\\
& Virtual Reality environment&64&1&5&2.31&0.12\\
& Usage and/or creation of computer simulations&64&1&5&2.73&0.14\\
& Microcontroller&64&1&5&2.81&0.14\\\cline{2-7}
Acquiring digital competencies& A (strongly) guided labwork session&64&1&5&2.91&0.15\\
& An open ended labwork session&64&1&5&3.14&0.15\\
& Undergraduate research projects in groups&64&1&5&3.47&0.16\\
& Experiments with smartphones/tablets &64&1&5&3.89&0.13\\
& Remote-controlled experiments&64&1&5&3.39&0.15\\
& Virtual Reality environment&64&1&5&3.38&0.15\\
& Usage and/or creation of computer simulations&64&1&5&4.34&0.10\\
& Microcontroller&64&1&5&4.23&0.13\\\hline
\multicolumn{2}{l}{\textbf{Attitudes towards the use of modern digital technologies in a lab course} (cf. Sec.~\ref{attitude})}&&&&\\
&\hangindent=.3cm When used correctly, modern digital technologies make a physics lab course better.&62&2&5&4.32&0.10\\
&\hangindent=.3cm I do not want modern digital technologies to be included in my physics lab course. (inverted)&63&1&5&4.59&0.10\\
&\hangindent=.3cm Modern digital technologies are a trend that university teaching should not follow. (inverted)&61&1&5&3.89&0.17\\
&\hangindent=.3cm I find it difficult to adapt to technical innovations. (inverted)&63&2&5&4.46&0.10\\
&\hangindent=.3cm Modern digital technologies only cause disturbance in the physics lab course and distract from the learning content. (inverted)&63&1&5&4.16&0.12\\
&\hangindent=.3cm Dealing with modern digital technologies and media content is part of the educational mission of the university.&62&1&5&4.23&0.13\\
&\hangindent=.3cm Modern digital technologies should become a natural part of learning.&62&1&5&4.21&0.12\\
&\hangindent=.3cm Modern digital technologies expand the scope of action for designing physics lab courses.&62&1&5&4.35&0.10\\
&\hangindent=.3cm The use of modern digital technologies can have a positive impact on students’ learning success.&63&1&5&3.98&0.10\\
&\hangindent=.3cm The use of modern digital technologies can have a positive impact on students’ motivation and engagement.&63&1&5&4.11&0.10\\
&\hangindent=.3cm Modern digital technologies can enable students to work self-responsibly and independently.&62&1&5&3.73&0.12\\
&\hangindent=.3cm Modern digital technologies can enable students to develop further social skills and to work in groups.&63&1&5&3.35&0.11\\
&\hangindent=.3cm With modern digital technologies, one can foster students more adequately according to their skills.&62&1&5&3.50&0.13\\
&\hangindent=.3cm In general, modern digital technologies should be part of teaching and learning in a physics lab course.&62&1&5&4.15&0.12\\
&\hangindent=.3cm Modern digital technologies should only be used in a physics lab course if they cannot be avoided (e.g., in home labs during the Covid-19 pandemic). (inverted)&63&1&5&4.24&0.10\\
\hline\hline
\label{tab:questionnaire}
\centering
\end{longtable*}

\section{Visualization of the frequencies of use of different digital technologies over the course of the Covid-19 pandemic}\label{appendixb}

\begin{figure}
    \centering
\begin{tikzpicture}
\begin{axis}[
    enlargelimits=false,
    ymax=.4,  
  xmin=-.4, xmax=1,
  xtick={-.4,-.2,0,.2,.4,.6,.8,1},
  ymin =-.4,
  xlabel={During vs before the pandemic},
  ylabel={After vs during the pandemic},
  legend cell align={left}
]

\addlegendimage{empty legend} \addlegendentry{\scalebox{1}[1]{\ref{cluster1}} boosted}
\addlegendimage{empty legend}
\addlegendentry{\scalebox{1}[1]{\ref{cluster2}} impeded}
\addlegendimage{empty legend}
\addlegendentry{\scalebox{1}[1]{\ref{cluster3}} stable}

\draw[thick, draw=gray!80] (0,400) -- (200,400);
\draw[thick,draw=gray!80] (40,0) -- (40,1200);

\addplot+[only marks,mark options={black},
    mark=triangle*,
    mark size=2.9pt,]
	coordinates{
(	.57, 0	)
(	.38, -.23	)
(	.31, -.23	)
(	.46, -.29)};\label{cluster1}

\node[above right=10pt of {(78,290},outer sep=2pt] {Exams};
\node[above right=10pt of {(65,130},outer sep=2pt] {Com./Collab. tools};
\node[above right=10pt of {(33,55},outer sep=2pt] {Simulations};
\node[above right=10pt of {(81,24},outer sep=2pt] {Smartphones};

\addplot+[
    only marks,mark options={black, rotate=180},
    mark=triangle*,
    mark size=2.9pt]
	coordinates{
(	-.21, .32	)
(	-.25, .31	)};\label{cluster2}

\node[above right=10pt of {(-3,600},outer sep=2pt] {Microcontroller};
\node[above right=10pt of {(13.8,650},outer sep=2pt] {Remote data collection};

\addplot+[
    only marks,mark options={black, fill=black},mark=diamond*,
    mark size=2.9pt]
	coordinates{
(	.05, .06	)
(	0, 0	)
(	.11, .01)
(	0, 0	)
};\label{cluster3}

\node[above right=10pt of {(0,280},outer sep=2pt] {Video analysis};
\node[above right=10pt of {(2,350},outer sep=2pt] {Hardware};
\node[above right=10pt of {(20,410},outer sep=2pt] {Data analysis};
\node[above right=10pt of {(42,360},outer sep=2pt] {Programming};

\end{axis}
\end{tikzpicture}
\caption{Visualization of the three different developments (boosted, impeded, and stable) of the frequency of use of different digital technologies over the course of the Covid-19 pandemic. The coordinates of each digital technology were derived by subtracting the values from Fig.~\ref{fig:technologies} for before from during and during from after the pandemic.}
\label{fig:clusteranalysis}
\end{figure}

\bibliography{bibtex1}

\begin{thebibliography}{94}%
\makeatletter
\providecommand \@ifxundefined [1]{%
 \@ifx{#1\undefined}
}%
\providecommand \@ifnum [1]{%
 \ifnum #1\expandafter \@firstoftwo
 \else \expandafter \@secondoftwo
 \fi
}%
\providecommand \@ifx [1]{%
 \ifx #1\expandafter \@firstoftwo
 \else \expandafter \@secondoftwo
 \fi
}%
\providecommand \natexlab [1]{#1}%
\providecommand \enquote  [1]{``#1''}%
\providecommand \bibnamefont  [1]{#1}%
\providecommand \bibfnamefont [1]{#1}%
\providecommand \citenamefont [1]{#1}%
\providecommand \href@noop [0]{\@secondoftwo}%
\providecommand \href [0]{\begingroup \@sanitize@url \@href}%
\providecommand \@href[1]{\@@startlink{#1}\@@href}%
\providecommand \@@href[1]{\endgroup#1\@@endlink}%
\providecommand \@sanitize@url [0]{\catcode `\\12\catcode `\$12\catcode
  `\&12\catcode `\#12\catcode `\^12\catcode `\_12\catcode `\%12\relax}%
\providecommand \@@startlink[1]{}%
\providecommand \@@endlink[0]{}%
\providecommand \url  [0]{\begingroup\@sanitize@url \@url }%
\providecommand \@url [1]{\endgroup\@href {#1}{\urlprefix }}%
\providecommand \urlprefix  [0]{URL }%
\providecommand \Eprint [0]{\href }%
\providecommand \doibase [0]{https://doi.org/}%
\providecommand \selectlanguage [0]{\@gobble}%
\providecommand \bibinfo  [0]{\@secondoftwo}%
\providecommand \bibfield  [0]{\@secondoftwo}%
\providecommand \translation [1]{[#1]}%
\providecommand \BibitemOpen [0]{}%
\providecommand \bibitemStop [0]{}%
\providecommand \bibitemNoStop [0]{.\EOS\space}%
\providecommand \EOS [0]{\spacefactor3000\relax}%
\providecommand \BibitemShut  [1]{\csname bibitem#1\endcsname}%
\let\auto@bib@innerbib\@empty
\bibitem [{\citenamefont {Otero}\ and\ \citenamefont
  {Meltzer}(2017)}]{Otero.2017}%
  \BibitemOpen
  \bibfield  {author} {\bibinfo {author} {\bibfnamefont {V.~K.}\ \bibnamefont
  {Otero}}\ and\ \bibinfo {author} {\bibfnamefont {D.~E.}\ \bibnamefont
  {Meltzer}},\ }\bibfield  {title} {\bibinfo {title} {The past and future of
  physics education reform},\ }\href
  {https://physicstoday.scitation.org/doi/pdf/10.1063/PT.3.3555} {\bibfield
  {journal} {\bibinfo  {journal} {Physics Today}\ }\textbf {\bibinfo {volume}
  {50}},\ \bibinfo {pages} {50} (\bibinfo {year} {2017})}\BibitemShut {NoStop}%
\bibitem [{\citenamefont {Sacher}\ and\ \citenamefont
  {Bauer}(2020)}]{Sacher.2020}%
  \BibitemOpen
  \bibfield  {author} {\bibinfo {author} {\bibfnamefont {M.~D.}\ \bibnamefont
  {Sacher}}\ and\ \bibinfo {author} {\bibfnamefont {A.~B.}\ \bibnamefont
  {Bauer}},\ }\bibfield  {title} {\bibinfo {title} {Kompetenzf{\"o}rderung im
  {L}aborpraktikum},\ }in\ \href
  {https://www.wbv.de/openaccess/themenbereiche/hochschule-und-wissenschaft/shop/detail/name/_/0/1/6004804w/facet/6004804w///////nb/0/category/1754.html}
  {\emph {\bibinfo {booktitle} {Labore in der Hochschullehre}}},\ \bibinfo
  {editor} {edited by\ \bibinfo {editor} {\bibfnamefont {T.}~\bibnamefont
  {Haertel}}, \bibinfo {editor} {\bibfnamefont {S.}~\bibnamefont {Heix}},
  \bibinfo {editor} {\bibfnamefont {C.}~\bibnamefont {Terkowsky}}, \bibinfo
  {editor} {\bibfnamefont {S.}~\bibnamefont {Frye}}, \bibinfo {editor}
  {\bibfnamefont {T.~R.}\ \bibnamefont {Ortelt}}, \bibinfo {editor}
  {\bibfnamefont {K.}~\bibnamefont {Lensing}},\ and\ \bibinfo {editor}
  {\bibfnamefont {D.}~\bibnamefont {May}}}\ (\bibinfo  {publisher} {{wbv
  Media}},\ \bibinfo {address} {Bielefeld},\ \bibinfo {year} {2020})\ pp.\
  \bibinfo {pages} {51--66}\BibitemShut {NoStop}%
\bibitem [{\citenamefont {{The New Movement among Physics
  Teachers}}(1907)}]{TheNewMovementamongPhysicsTeachers.1907}%
  \BibitemOpen
  \bibfield  {author} {\bibinfo {author} {\bibnamefont {{The New Movement among
  Physics Teachers}}},\ }\bibfield  {title} {\bibinfo {title} {Circular {V}},\
  }\href {https://www.jstor.org/stable/1075139?seq=1} {\bibfield  {journal}
  {\bibinfo  {journal} {The School Review}\ }\textbf {\bibinfo {volume} {15}},\
  \bibinfo {pages} {290} (\bibinfo {year} {1907})}\BibitemShut {NoStop}%
\bibitem [{\citenamefont {Hofstein}\ and\ \citenamefont
  {Lunetta}(2003)}]{Hofstein.2003}%
  \BibitemOpen
  \bibfield  {author} {\bibinfo {author} {\bibfnamefont {A.}~\bibnamefont
  {Hofstein}}\ and\ \bibinfo {author} {\bibfnamefont {V.~N.}\ \bibnamefont
  {Lunetta}},\ }\bibfield  {title} {\bibinfo {title} {The laboratory in science
  education: Foundations for the twenty-first century},\ }\href
  {https://onlinelibrary.wiley.com/doi/10.1002/sce.10106} {\bibfield  {journal}
  {\bibinfo  {journal} {Science Education}\ }\textbf {\bibinfo {volume} {88}},\
  \bibinfo {pages} {28} (\bibinfo {year} {2003})}\BibitemShut {NoStop}%
\bibitem [{\citenamefont {{American Association of Physics
  Teachers}}(1998)}]{AmericanAssociationofPhysicsTeachers.1998}%
  \BibitemOpen
  \bibfield  {author} {\bibinfo {author} {\bibnamefont {{American Association
  of Physics Teachers}}},\ }\bibfield  {title} {\bibinfo {title} {Goals of the
  introductory physics laboratory},\ }\href {https://doi.org/10.1119/1.19042}
  {\bibfield  {journal} {\bibinfo  {journal} {American Journal of Physics}\
  }\textbf {\bibinfo {volume} {66}},\ \bibinfo {pages} {483} (\bibinfo {year}
  {1998})}\BibitemShut {NoStop}%
\bibitem [{\citenamefont {{American Association of Physics
  Teachers}}(2014)}]{AmericanAssociationofPhysicsTeachers.2014}%
  \BibitemOpen
  \bibfield  {author} {\bibinfo {author} {\bibnamefont {{American Association
  of Physics Teachers}}},\ }\href
  {https://www.aapt.org/Resources/upload/LabGuidlinesDocument_EBendorsed_nov10.pdf}
  {\bibinfo {title} {A{APT} recommendations for the undergraduate physics
  laboratory curriculum: Report prepared by a subcommittee of the {AAPT}
  {C}ommittee on {L}aboratories: Endorsed by the {AAPT} {E}xecutive {B}oard}}
  (\bibinfo {year} {2014})\BibitemShut {NoStop}%
\bibitem [{\citenamefont {{Konferenz der {F}achbereiche
  {P}hysik}}(2010)}]{KonferenzderFachbereichePhysik.2010}%
  \BibitemOpen
  \bibfield  {author} {\bibinfo {author} {\bibnamefont {{Konferenz der
  {F}achbereiche {P}hysik}}},\ }\href
  {https://www.kfp-physik.de/dokument/KFP_Handreichung_Konzeption-Studiengaenge-Physik-101108.pdf}
  {\bibinfo {title} {Zur konzeption von {B}achelor- und
  {M}aster-{S}tudieng{\"a}ngen in der {P}hysik: Handreichung der {K}onferenz
  der {F}achbereiche {P}hysik ({KFP})}} (\bibinfo {year} {2010})\BibitemShut
  {NoStop}%
\bibitem [{\citenamefont {Welzel}\ \emph
  {et~al.}(1998{\natexlab{a}})\citenamefont {Welzel}, \citenamefont {Haller},
  \citenamefont {Bandiera}, \citenamefont {Hammelev}, \citenamefont {Koumaras},
  \citenamefont {Niedderer}, \citenamefont {Paulsen}, \citenamefont
  {Robinault},\ and\ \citenamefont {von Aufschnaiter}}]{Welzel.1998}%
  \BibitemOpen
  \bibfield  {author} {\bibinfo {author} {\bibfnamefont {M.}~\bibnamefont
  {Welzel}}, \bibinfo {author} {\bibfnamefont {K.}~\bibnamefont {Haller}},
  \bibinfo {author} {\bibfnamefont {M.}~\bibnamefont {Bandiera}}, \bibinfo
  {author} {\bibfnamefont {D.}~\bibnamefont {Hammelev}}, \bibinfo {author}
  {\bibfnamefont {P.}~\bibnamefont {Koumaras}}, \bibinfo {author}
  {\bibfnamefont {H.}~\bibnamefont {Niedderer}}, \bibinfo {author}
  {\bibfnamefont {A.}~\bibnamefont {Paulsen}}, \bibinfo {author} {\bibfnamefont
  {K.}~\bibnamefont {Robinault}},\ and\ \bibinfo {author} {\bibfnamefont
  {S.}~\bibnamefont {von Aufschnaiter}},\ }\bibfield  {title} {\bibinfo {title}
  {Ziele, die {L}ehrende mit dem {E}xperimentieren in der
  naturwissenschaftlichen {A}usbildung verbinden: Ergebnisse einer
  europ{\"a}ischen {U}mfrage},\ }\href
  {https://www.researchgate.net/publication/241698001_Ziele_die_Lehrende_mit_experimentellem_Arbeiten_in_der_naturwissenschaftlichen_Ausbildung_verbinden-_Ergebnisse_einer_europaischen_Umfrage}
  {\bibfield  {journal} {\bibinfo  {journal} {Zeitschrift f{\"u}r Didaktik der
  Naturwissenschaften}\ }\textbf {\bibinfo {volume} {4}},\ \bibinfo {pages}
  {29} (\bibinfo {year} {1998}{\natexlab{a}})}\BibitemShut {NoStop}%
\bibitem [{\citenamefont {Welzel}\ \emph
  {et~al.}(1998{\natexlab{b}})\citenamefont {Welzel}, \citenamefont {Haller},
  \citenamefont {Bandiera}, \citenamefont {Hammelev}, \citenamefont {Koumaras},
  \citenamefont {Niedderer}, \citenamefont {Paulsen}, \citenamefont
  {Robinault},\ and\ \citenamefont {von Aufschnaiter}}]{Welzel.1998b}%
  \BibitemOpen
  \bibfield  {author} {\bibinfo {author} {\bibfnamefont {M.}~\bibnamefont
  {Welzel}}, \bibinfo {author} {\bibfnamefont {K.}~\bibnamefont {Haller}},
  \bibinfo {author} {\bibfnamefont {M.}~\bibnamefont {Bandiera}}, \bibinfo
  {author} {\bibfnamefont {D.}~\bibnamefont {Hammelev}}, \bibinfo {author}
  {\bibfnamefont {P.}~\bibnamefont {Koumaras}}, \bibinfo {author}
  {\bibfnamefont {H.}~\bibnamefont {Niedderer}}, \bibinfo {author}
  {\bibfnamefont {A.}~\bibnamefont {Paulsen}}, \bibinfo {author} {\bibfnamefont
  {K.}~\bibnamefont {Robinault}},\ and\ \bibinfo {author} {\bibfnamefont
  {S.}~\bibnamefont {von Aufschnaiter}},\ }\href
  {http://www.idn.uni-bremen.de/pubs/Niedderer/1998-LSE-WP6.pdf} {\bibinfo
  {title} {Teachers' objectives for labwork: Research tool and cross country
  results: Working paper 6}} (\bibinfo {year} {1998}{\natexlab{b}})\BibitemShut
  {NoStop}%
\bibitem [{\citenamefont {Zwickl}\ \emph {et~al.}(2013)\citenamefont {Zwickl},
  \citenamefont {Finkelstein},\ and\ \citenamefont
  {Lewandowski}}]{Zwickl.2013}%
  \BibitemOpen
  \bibfield  {author} {\bibinfo {author} {\bibfnamefont {B.~M.}\ \bibnamefont
  {Zwickl}}, \bibinfo {author} {\bibfnamefont {N.}~\bibnamefont
  {Finkelstein}},\ and\ \bibinfo {author} {\bibfnamefont {H.~J.}\ \bibnamefont
  {Lewandowski}},\ }\bibfield  {title} {\bibinfo {title} {The process of
  transforming an advanced lab course: Goals, curriculum, and assessments},\
  }\href {https://doi.org/10.1119/1.4768890} {\bibfield  {journal} {\bibinfo
  {journal} {American Journal of Physics}\ }\textbf {\bibinfo {volume} {81}},\
  \bibinfo {pages} {63} (\bibinfo {year} {2013})}\BibitemShut {NoStop}%
\bibitem [{\citenamefont {Nagel}\ \emph {et~al.}(2018)\citenamefont {Nagel},
  \citenamefont {Scholz},\ and\ \citenamefont {Weber}}]{Nagel.2018}%
  \BibitemOpen
  \bibfield  {author} {\bibinfo {author} {\bibfnamefont {C.}~\bibnamefont
  {Nagel}}, \bibinfo {author} {\bibfnamefont {R.}~\bibnamefont {Scholz}},\ and\
  \bibinfo {author} {\bibfnamefont {K.-A.}\ \bibnamefont {Weber}},\ }\bibfield
  {title} {\bibinfo {title} {Umfrage zu den {L}ehr/{L}ernzielen in
  physikalischen {P}raktika},\ }\href
  {http://www.phydid.de/index.php/phydid-b/article/view/829/974} {\bibfield
  {journal} {\bibinfo  {journal} {PhyDid B - Didaktik der Physik - Beitr{\"a}ge
  zur DPG-Fr{\"u}hjahrstagung - W{\"u}rzburg 2018}\ ,\ \bibinfo {pages} {79}}
  (\bibinfo {year} {2018})}\BibitemShut {NoStop}%
\bibitem [{\citenamefont {Vorholzer}\ \emph {et~al.}(2022)\citenamefont
  {Vorholzer}, \citenamefont {Ortmann},\ and\ \citenamefont
  {Graulich}}]{Vorholzer.2022}%
  \BibitemOpen
  \bibfield  {author} {\bibinfo {author} {\bibfnamefont {A.}~\bibnamefont
  {Vorholzer}}, \bibinfo {author} {\bibfnamefont {J.}~\bibnamefont {Ortmann}},\
  and\ \bibinfo {author} {\bibfnamefont {N.}~\bibnamefont {Graulich}},\
  }\bibfield  {title} {\bibinfo {title} {Naturwissenschaftliche {D}enk- und
  {A}rbeitsweisen in (physikalischen) {P}raktika},\ }\href
  {https://ojs.dpg-physik.de/index.php/phydid-b/article/view/1232/1512}
  {\bibfield  {journal} {\bibinfo  {journal} {PhyDid B - Beitr{\"a}ge zur
  Fr{\"u}hjahrstagung - virtuell 2022}\ ,\ \bibinfo {pages} {395}} (\bibinfo
  {year} {2022})}\BibitemShut {NoStop}%
\bibitem [{\citenamefont {Holmes}\ and\ \citenamefont
  {Wieman}(2016)}]{Holmes.2016}%
  \BibitemOpen
  \bibfield  {author} {\bibinfo {author} {\bibfnamefont {N.~G.}\ \bibnamefont
  {Holmes}}\ and\ \bibinfo {author} {\bibfnamefont {C.~E.}\ \bibnamefont
  {Wieman}},\ }\bibfield  {title} {\bibinfo {title} {Examining and contrasting
  the cognitive activities engaged in undergraduate research experiences and
  lab courses},\ }\bibfield  {journal} {\bibinfo  {journal} {Physical Review
  Physics Education Research}\ }\textbf {\bibinfo {volume} {12}},\ \href
  {https://doi.org/10.1103/PhysRevPhysEducRes.12.020103}
  {10.1103/PhysRevPhysEducRes.12.020103} (\bibinfo {year} {2016})\BibitemShut
  {NoStop}%
\bibitem [{\citenamefont {Karelina}\ \emph {et~al.}(2007)\citenamefont
  {Karelina}, \citenamefont {Etkina}, \citenamefont {Ruibal-Villasenor},
  \citenamefont {Rosengrant}, \citenamefont {{van Heuvelen}}, \citenamefont
  {Hmelo-Silver}, \citenamefont {Hsu}, \citenamefont {Henderson},\ and\
  \citenamefont {McCullough}}]{Karelina.2007}%
  \BibitemOpen
  \bibfield  {author} {\bibinfo {author} {\bibfnamefont {A.}~\bibnamefont
  {Karelina}}, \bibinfo {author} {\bibfnamefont {E.}~\bibnamefont {Etkina}},
  \bibinfo {author} {\bibfnamefont {M.}~\bibnamefont {Ruibal-Villasenor}},
  \bibinfo {author} {\bibfnamefont {D.}~\bibnamefont {Rosengrant}}, \bibinfo
  {author} {\bibfnamefont {A.}~\bibnamefont {{van Heuvelen}}}, \bibinfo
  {author} {\bibfnamefont {C.}~\bibnamefont {Hmelo-Silver}}, \bibinfo {author}
  {\bibfnamefont {L.}~\bibnamefont {Hsu}}, \bibinfo {author} {\bibfnamefont
  {C.}~\bibnamefont {Henderson}},\ and\ \bibinfo {author} {\bibfnamefont
  {L.}~\bibnamefont {McCullough}},\ }\bibfield  {title} {\bibinfo {title}
  {Design and non-design labs: Does transfer occur?},\ }\href
  {https://doi.org/10.1063/1.2820956} {\bibfield  {journal} {\bibinfo
  {journal} {AIP Conference Proceedings}\ }\textbf {\bibinfo {volume} {92}},\
  \bibinfo {pages} {92} (\bibinfo {year} {2007})}\BibitemShut {NoStop}%
\bibitem [{\citenamefont {Karelina}\ and\ \citenamefont
  {Etkina}(2007)}]{Karelina.2007b}%
  \BibitemOpen
  \bibfield  {author} {\bibinfo {author} {\bibfnamefont {A.}~\bibnamefont
  {Karelina}}\ and\ \bibinfo {author} {\bibfnamefont {E.}~\bibnamefont
  {Etkina}},\ }\bibfield  {title} {\bibinfo {title} {Acting like a physicist:
  Student approach study to experimental design},\ }\bibfield  {journal}
  {\bibinfo  {journal} {Physical Review Special Topics - Physics Education
  Research}\ }\textbf {\bibinfo {volume} {3}},\ \href
  {https://doi.org/10.1103/PhysRevSTPER.3.020106}
  {10.1103/PhysRevSTPER.3.020106} (\bibinfo {year} {2007})\BibitemShut
  {NoStop}%
\bibitem [{\citenamefont {Etkina}(2015)}]{Etkina.2015}%
  \BibitemOpen
  \bibfield  {author} {\bibinfo {author} {\bibfnamefont {E.}~\bibnamefont
  {Etkina}},\ }\bibfield  {title} {\bibinfo {title} {Millikan award lecture:
  Students of physics - listeners, observers or collaborative participants in
  physics scientific practices?},\ }\href
  {https://aapt.scitation.org/doi/pdf/10.1119/1.4923432} {\bibfield  {journal}
  {\bibinfo  {journal} {American Journal of Physics}\ }\textbf {\bibinfo
  {volume} {83}},\ \bibinfo {pages} {669} (\bibinfo {year} {2015})}\BibitemShut
  {NoStop}%
\bibitem [{\citenamefont {Kontro}\ \emph {et~al.}(2018)\citenamefont {Kontro},
  \citenamefont {Heino}, \citenamefont {Hendolin},\ and\ \citenamefont
  {Galambosi}}]{Kontro.2018}%
  \BibitemOpen
  \bibfield  {author} {\bibinfo {author} {\bibfnamefont {I.}~\bibnamefont
  {Kontro}}, \bibinfo {author} {\bibfnamefont {O.}~\bibnamefont {Heino}},
  \bibinfo {author} {\bibfnamefont {I.}~\bibnamefont {Hendolin}},\ and\
  \bibinfo {author} {\bibfnamefont {S.}~\bibnamefont {Galambosi}},\ }\bibfield
  {title} {\bibinfo {title} {Modernisation of the intermediate physics
  laboratory},\ }\href {https://doi.org/10.1088/1361-6404/aa9364} {\bibfield
  {journal} {\bibinfo  {journal} {European Journal of Physics}\ }\textbf
  {\bibinfo {volume} {39}},\ \bibinfo {pages} {025702} (\bibinfo {year}
  {2018})}\BibitemShut {NoStop}%
\bibitem [{\citenamefont {Barro}\ \emph {et~al.}(2023)\citenamefont {Barro},
  \citenamefont {Beguin}, \citenamefont {Brouzet}, \citenamefont {Charosky},
  \citenamefont {Darmendrail},\ and\ \citenamefont {M{\"u}ller}}]{Barro.2023}%
  \BibitemOpen
  \bibfield  {author} {\bibinfo {author} {\bibfnamefont {S.}~\bibnamefont
  {Barro}}, \bibinfo {author} {\bibfnamefont {C.}~\bibnamefont {Beguin}},
  \bibinfo {author} {\bibfnamefont {D.}~\bibnamefont {Brouzet}}, \bibinfo
  {author} {\bibfnamefont {L.}~\bibnamefont {Charosky}}, \bibinfo {author}
  {\bibfnamefont {L.}~\bibnamefont {Darmendrail}},\ and\ \bibinfo {author}
  {\bibfnamefont {A.}~\bibnamefont {M{\"u}ller}},\ }\href
  {https://arxiv.org/ftp/arxiv/papers/2305/2305.07483.pdf} {\bibinfo {title}
  {Smartphone experiments in undergraduate research}} (\bibinfo {year}
  {2023})\BibitemShut {NoStop}%
\bibitem [{\citenamefont {Planin{\v{s}}i{\v{c}}}(2007)}]{Planinsic.2007}%
  \BibitemOpen
  \bibfield  {author} {\bibinfo {author} {\bibfnamefont {G.}~\bibnamefont
  {Planin{\v{s}}i{\v{c}}}},\ }\bibfield  {title} {\bibinfo {title} {Project
  laboratory for first-year students},\ }\href
  {https://doi.org/10.1088/0143-0807/28/3/S07} {\bibfield  {journal} {\bibinfo
  {journal} {European Journal of Physics}\ }\textbf {\bibinfo {volume} {28}},\
  \bibinfo {pages} {S71} (\bibinfo {year} {2007})}\BibitemShut {NoStop}%
\bibitem [{\citenamefont {Ruiz-Primo}\ \emph {et~al.}(2011)\citenamefont
  {Ruiz-Primo}, \citenamefont {Briggs}, \citenamefont {Iverson}, \citenamefont
  {Talbot},\ and\ \citenamefont {Shepard}}]{RuizPrimo.2011}%
  \BibitemOpen
  \bibfield  {author} {\bibinfo {author} {\bibfnamefont {M.~A.}\ \bibnamefont
  {Ruiz-Primo}}, \bibinfo {author} {\bibfnamefont {D.}~\bibnamefont {Briggs}},
  \bibinfo {author} {\bibfnamefont {H.}~\bibnamefont {Iverson}}, \bibinfo
  {author} {\bibfnamefont {R.}~\bibnamefont {Talbot}},\ and\ \bibinfo {author}
  {\bibfnamefont {L.~A.}\ \bibnamefont {Shepard}},\ }\bibfield  {title}
  {\bibinfo {title} {Impact of undergraduate science course innovations on
  learning},\ }\href {https://doi.org/10.1126/science.1198976} {\bibfield
  {journal} {\bibinfo  {journal} {Science}\ }\textbf {\bibinfo {volume}
  {331}},\ \bibinfo {pages} {1269} (\bibinfo {year} {2011})}\BibitemShut
  {NoStop}%
\bibitem [{\citenamefont {Russell}\ \emph {et~al.}(2007)\citenamefont
  {Russell}, \citenamefont {Hancock},\ and\ \citenamefont
  {McCullough}}]{Russell.2007}%
  \BibitemOpen
  \bibfield  {author} {\bibinfo {author} {\bibfnamefont {S.~H.}\ \bibnamefont
  {Russell}}, \bibinfo {author} {\bibfnamefont {M.~P.}\ \bibnamefont
  {Hancock}},\ and\ \bibinfo {author} {\bibfnamefont {J.}~\bibnamefont
  {McCullough}},\ }\bibfield  {title} {\bibinfo {title} {Benefits of
  undergraduate research experiences},\ }\href
  {https://doi.org/10.1126/science.1140384} {\bibfield  {journal} {\bibinfo
  {journal} {Science}\ }\textbf {\bibinfo {volume} {316}},\ \bibinfo {pages}
  {548} (\bibinfo {year} {2007})}\BibitemShut {NoStop}%
\bibitem [{\citenamefont {{National Academies of Sciences, Engineering, and
  Medicine}}(2017)}]{NationalAcademiesofSciencesEngineeringandMedicine.2017}%
  \BibitemOpen
  \bibfield  {author} {\bibinfo {author} {\bibnamefont {{National Academies of
  Sciences, Engineering, and Medicine}}},\ }\href
  {https://doi.org/10.17226/24622} {\emph {\bibinfo {title} {Undergraduate
  Research Experiences for {STEM} Students: Successes, Challenges, and
  Opportunities}}}\ (\bibinfo  {publisher} {{National Academies Press}},\
  \bibinfo {address} {Washington, D.C.},\ \bibinfo {year} {2017})\BibitemShut
  {NoStop}%
\bibitem [{\citenamefont {Abraham}\ \emph {et~al.}(2022)\citenamefont
  {Abraham}, \citenamefont {Hudgings}, \citenamefont {Jackson},\ and\
  \citenamefont {Rogers}}]{Abraham.2022}%
  \BibitemOpen
  \bibfield  {author} {\bibinfo {author} {\bibfnamefont {N.~B.}\ \bibnamefont
  {Abraham}}, \bibinfo {author} {\bibfnamefont {J.}~\bibnamefont {Hudgings}},
  \bibinfo {author} {\bibfnamefont {M.}~\bibnamefont {Jackson}},\ and\ \bibinfo
  {author} {\bibfnamefont {W.~F.}\ \bibnamefont {Rogers}},\ }\bibfield  {title}
  {\bibinfo {title} {Undergraduate research in physics},\ }in\ \href@noop {}
  {\emph {\bibinfo {booktitle} {The Cambridge Handbook of Undergraduate
  Research}}},\ \bibinfo {editor} {edited by\ \bibinfo {editor} {\bibfnamefont
  {H.~A.}\ \bibnamefont {Mieg}}, \bibinfo {editor} {\bibfnamefont
  {E.}~\bibnamefont {Ambos}}, \bibinfo {editor} {\bibfnamefont
  {A.}~\bibnamefont {Brew}}, \bibinfo {editor} {\bibfnamefont {D.}~\bibnamefont
  {Galli}},\ and\ \bibinfo {editor} {\bibfnamefont {J.}~\bibnamefont
  {Lehmann}}}\ (\bibinfo  {publisher} {{University Press}},\ \bibinfo {address}
  {Cambridge},\ \bibinfo {year} {2022})\ pp.\ \bibinfo {pages}
  {191--198}\BibitemShut {NoStop}%
\bibitem [{\citenamefont {Ahmad}\ and\ \citenamefont
  {Al-Thani}(2022)}]{Ahmad.2022}%
  \BibitemOpen
  \bibfield  {author} {\bibinfo {author} {\bibfnamefont {Z.}~\bibnamefont
  {Ahmad}}\ and\ \bibinfo {author} {\bibfnamefont {N.~J.}\ \bibnamefont
  {Al-Thani}},\ }\bibfield  {title} {\bibinfo {title} {Undergraduate research
  experience models: A systematic review of the literature from 2011 to 2021},\
  }\href {https://doi.org/10.1016/j.ijer.2022.101996} {\bibfield  {journal}
  {\bibinfo  {journal} {International Journal of Educational Research}\
  }\textbf {\bibinfo {volume} {114}},\ \bibinfo {pages} {101996} (\bibinfo
  {year} {2022})}\BibitemShut {NoStop}%
\bibitem [{\citenamefont {Oliver}\ \emph {et~al.}(2023)\citenamefont {Oliver},
  \citenamefont {Werth},\ and\ \citenamefont {Lewandowski}}]{Oliver.2023}%
  \BibitemOpen
  \bibfield  {author} {\bibinfo {author} {\bibfnamefont {K.~A.}\ \bibnamefont
  {Oliver}}, \bibinfo {author} {\bibfnamefont {A.}~\bibnamefont {Werth}},\ and\
  \bibinfo {author} {\bibfnamefont {H.~J.}\ \bibnamefont {Lewandowski}},\
  }\bibfield  {title} {\bibinfo {title} {Student experiences with authentic
  research in a remote, introductory course-based undergraduate research
  experience in physics},\ }\href
  {https://doi.org/10.1103/PhysRevPhysEducRes.19.010124} {\bibfield  {journal}
  {\bibinfo  {journal} {Physical Review Physics Education Research}\ }\textbf
  {\bibinfo {volume} {19}},\ \bibinfo {pages} {010124} (\bibinfo {year}
  {2023})}\BibitemShut {NoStop}%
\bibitem [{\citenamefont {Zohar}\ and\ \citenamefont
  {Barzilai}(2015)}]{Zohar.2015}%
  \BibitemOpen
  \bibfield  {author} {\bibinfo {author} {\bibfnamefont {A.}~\bibnamefont
  {Zohar}}\ and\ \bibinfo {author} {\bibfnamefont {S.}~\bibnamefont
  {Barzilai}},\ }\bibfield  {title} {\bibinfo {title} {Metacognition and
  teaching higher order thinking ({HOT}) in science education},\ }in\
  \href@noop {} {\emph {\bibinfo {booktitle} {The Routledge International
  Handbook of Research on Teaching Thinking}}},\ \bibinfo {editor} {edited by\
  \bibinfo {editor} {\bibfnamefont {R.}~\bibnamefont {Wegerif}}, \bibinfo
  {editor} {\bibfnamefont {L.}~\bibnamefont {Li}},\ and\ \bibinfo {editor}
  {\bibfnamefont {J.}~\bibnamefont {{C. Kaufman}}}}\ (\bibinfo  {publisher}
  {Routledge},\ \bibinfo {address} {London},\ \bibinfo {year} {2015})\ pp.\
  \bibinfo {pages} {253--266}\BibitemShut {NoStop}%
\bibitem [{\citenamefont {Walsh}\ \emph {et~al.}(2019)\citenamefont {Walsh},
  \citenamefont {Quinn}, \citenamefont {Wieman},\ and\ \citenamefont
  {Holmes}}]{Walsh.2019}%
  \BibitemOpen
  \bibfield  {author} {\bibinfo {author} {\bibfnamefont {C.}~\bibnamefont
  {Walsh}}, \bibinfo {author} {\bibfnamefont {K.~N.}\ \bibnamefont {Quinn}},
  \bibinfo {author} {\bibfnamefont {C.}~\bibnamefont {Wieman}},\ and\ \bibinfo
  {author} {\bibfnamefont {N.~G.}\ \bibnamefont {Holmes}},\ }\bibfield  {title}
  {\bibinfo {title} {Quantifying critical thinking: Development and validation
  of the physics lab inventory of critical thinking},\ }\bibfield  {journal}
  {\bibinfo  {journal} {Physical Review Physics Education Research}\ }\textbf
  {\bibinfo {volume} {15}},\ \href
  {https://doi.org/10.1103/physrevphyseducres.15.010135}
  {10.1103/physrevphyseducres.15.010135} (\bibinfo {year} {2019})\BibitemShut
  {NoStop}%
\bibitem [{\citenamefont {Murtonen}\ and\ \citenamefont
  {Balloo}(2019)}]{Murtonen.2019}%
  \BibitemOpen
  \bibinfo {editor} {\bibfnamefont {M.}~\bibnamefont {Murtonen}}\ and\ \bibinfo
  {editor} {\bibfnamefont {K.}~\bibnamefont {Balloo}},\ eds.,\ \href
  {https://doi.org/10.1007/978-3-030-24215-2} {\emph {\bibinfo {title}
  {Redefining Scientific Thinking for Higher Education: Higher-Order Thinking,
  Evidence-Based Reasoning and Research Skills}}},\ \bibinfo {edition} {1st}\
  ed.\ (\bibinfo  {publisher} {{Palgrave Macmillan}},\ \bibinfo {address}
  {Cham},\ \bibinfo {year} {2019})\BibitemShut {NoStop}%
\bibitem [{\citenamefont {Mieg}\ \emph {et~al.}(2022)\citenamefont {Mieg},
  \citenamefont {Ambos}, \citenamefont {Brew}, \citenamefont {Galli},\ and\
  \citenamefont {Lehmann}}]{Mieg.2022}%
  \BibitemOpen
  \bibinfo {editor} {\bibfnamefont {H.~A.}\ \bibnamefont {Mieg}}, \bibinfo
  {editor} {\bibfnamefont {E.}~\bibnamefont {Ambos}}, \bibinfo {editor}
  {\bibfnamefont {A.}~\bibnamefont {Brew}}, \bibinfo {editor} {\bibfnamefont
  {D.}~\bibnamefont {Galli}},\ and\ \bibinfo {editor} {\bibfnamefont
  {J.}~\bibnamefont {Lehmann}},\ eds.,\ \href
  {https://doi.org/10.1017/9781108869508} {\emph {\bibinfo {title} {The
  Cambridge Handbook of Undergraduate Research}}}\ (\bibinfo  {publisher}
  {{University Press}},\ \bibinfo {address} {Cambridge},\ \bibinfo {year}
  {2022})\BibitemShut {NoStop}%
\bibitem [{\citenamefont {Holmes}\ \emph {et~al.}(2017)\citenamefont {Holmes},
  \citenamefont {Olsen}, \citenamefont {Thomas},\ and\ \citenamefont
  {Wieman}}]{Holmes.2017}%
  \BibitemOpen
  \bibfield  {author} {\bibinfo {author} {\bibfnamefont {N.~G.}\ \bibnamefont
  {Holmes}}, \bibinfo {author} {\bibfnamefont {J.}~\bibnamefont {Olsen}},
  \bibinfo {author} {\bibfnamefont {J.~L.}\ \bibnamefont {Thomas}},\ and\
  \bibinfo {author} {\bibfnamefont {C.~E.}\ \bibnamefont {Wieman}},\ }\bibfield
   {title} {\bibinfo {title} {Value added or misattributed? {A}
  multi-institution study on the educational benefit of labs for reinforing
  physics content},\ }\bibfield  {journal} {\bibinfo  {journal} {Physical
  Review Physics Education Research}\ }\textbf {\bibinfo {volume} {13}},\ \href
  {https://doi.org/10.1103/PhysRevPhysEducRes.13.010129}
  {10.1103/PhysRevPhysEducRes.13.010129} (\bibinfo {year} {2017})\BibitemShut
  {NoStop}%
\bibitem [{\citenamefont {Holmes}\ and\ \citenamefont
  {Wieman}(2018)}]{Holmes.2018}%
  \BibitemOpen
  \bibfield  {author} {\bibinfo {author} {\bibfnamefont {N.~G.}\ \bibnamefont
  {Holmes}}\ and\ \bibinfo {author} {\bibfnamefont {C.~E.}\ \bibnamefont
  {Wieman}},\ }\bibfield  {title} {\bibinfo {title} {Introductory physics labs:
  We can do better},\ }\href {https://doi.org/10.1063/PT.3.3816} {\bibfield
  {journal} {\bibinfo  {journal} {Physics Today}\ }\textbf {\bibinfo {volume}
  {71}},\ \bibinfo {pages} {38} (\bibinfo {year} {2018})}\BibitemShut {NoStop}%
\bibitem [{\citenamefont {Teichmann}\ \emph {et~al.}(2022)\citenamefont
  {Teichmann}, \citenamefont {Lewandowski},\ and\ \citenamefont
  {Alemani}}]{Teichmann.2022}%
  \BibitemOpen
  \bibfield  {author} {\bibinfo {author} {\bibfnamefont {E.}~\bibnamefont
  {Teichmann}}, \bibinfo {author} {\bibfnamefont {H.~J.}\ \bibnamefont
  {Lewandowski}},\ and\ \bibinfo {author} {\bibfnamefont {M.}~\bibnamefont
  {Alemani}},\ }\bibfield  {title} {\bibinfo {title} {Investigating students'
  views of experimental physics in {G}erman laboratory classes},\ }\href
  {https://doi.org/10.1103/PhysRevPhysEducRes.18.010135} {\bibfield  {journal}
  {\bibinfo  {journal} {Physical Review Physics Education Research}\ }\textbf
  {\bibinfo {volume} {18}},\ \bibinfo {pages} {010135} (\bibinfo {year}
  {2022})}\BibitemShut {NoStop}%
\bibitem [{\citenamefont {Rehfeldt}(2017)}]{Rehfeldt.2017}%
  \BibitemOpen
  \bibfield  {author} {\bibinfo {author} {\bibfnamefont {D.}~\bibnamefont
  {Rehfeldt}},\ }\href@noop {} {\emph {\bibinfo {title} {Erfassung der
  Lehrqualit{\"a}t naturwissenschaftlicher Experimentalpraktika}}},\ \bibinfo
  {series} {Studien zum Physik- und Chemielernen}, Vol.\ \bibinfo {volume}
  {246}\ (\bibinfo  {publisher} {Logos},\ \bibinfo {address} {Berlin},\
  \bibinfo {year} {2017})\BibitemShut {NoStop}%
\bibitem [{\citenamefont {Neumann}(2004)}]{Neumann.2004}%
  \BibitemOpen
  \bibfield  {author} {\bibinfo {author} {\bibfnamefont {K.}~\bibnamefont
  {Neumann}},\ }\href@noop {} {\emph {\bibinfo {title} {Didaktische
  {R}ekonstruktion eines physikalischen Praktikums f{\"u}r Physiker}}},\
  \bibinfo {series} {Studien zum Physiklernen}, Vol.~\bibinfo {volume} {38}\
  (\bibinfo  {publisher} {Logos},\ \bibinfo {address} {Berlin},\ \bibinfo
  {year} {2004})\BibitemShut {NoStop}%
\bibitem [{\citenamefont {They{\ss}en}(1999)}]{Theyen.1999}%
  \BibitemOpen
  \bibfield  {author} {\bibinfo {author} {\bibfnamefont {H.}~\bibnamefont
  {They{\ss}en}},\ }\href@noop {} {\emph {\bibinfo {title} {Ein Physikpraktikum
  f{\"u}r Studierende der Medizin: Darstellung der Entwicklung und Evaluation
  eines adressatenspezifischen Praktikums nach dem Modell der Didaktischen
  Rekonstruktion}}},\ \bibinfo {series} {Studien zum Physiklernen},
  Vol.~\bibinfo {volume} {9}\ (\bibinfo  {publisher} {Logos},\ \bibinfo
  {address} {Berlin},\ \bibinfo {year} {1999})\BibitemShut {NoStop}%
\bibitem [{\citenamefont {Klug}(2017)}]{Klug.2017}%
  \BibitemOpen
  \bibfield  {author} {\bibinfo {author} {\bibfnamefont {T.}~\bibnamefont
  {Klug}},\ }\href@noop {} {\emph {\bibinfo {title} {Wirkung
  kontextorientierter physikalischer Praktikumsversuche auf Lernprozesse von
  Studierenden der Medizin}}},\ \bibinfo {series} {Studien zum Physik- und
  Chemielernen}, Vol.\ \bibinfo {volume} {234}\ (\bibinfo  {publisher}
  {Logos},\ \bibinfo {address} {Berlin},\ \bibinfo {year} {2017})\BibitemShut
  {NoStop}%
\bibitem [{\citenamefont {Andersen}(2020)}]{Andersen.2020}%
  \BibitemOpen
  \bibfield  {author} {\bibinfo {author} {\bibfnamefont {J.}~\bibnamefont
  {Andersen}},\ }\emph {\bibinfo {title} {Entwicklung und Evaluierung eines
  spezifischen Anf{\"a}ngerpraktikums f{\"u}r Lehramtsstudierende im Fach
  Physik}},\ \href {https://macau.uni-kiel.de/receive/macau_mods_00001409}
  {\bibinfo {type} {Phd thesis}},\ \bibinfo  {school}
  {{Christian-Albrechts-Universit{\"a}t Kiel}}, \bibinfo {address} {Kiel}
  (\bibinfo {year} {2020})\BibitemShut {NoStop}%
\bibitem [{\citenamefont {Walsh}\ \emph {et~al.}(2022)\citenamefont {Walsh},
  \citenamefont {Lewandowski},\ and\ \citenamefont {Holmes}}]{Walsh.2022}%
  \BibitemOpen
  \bibfield  {author} {\bibinfo {author} {\bibfnamefont {C.}~\bibnamefont
  {Walsh}}, \bibinfo {author} {\bibfnamefont {H.~J.}\ \bibnamefont
  {Lewandowski}},\ and\ \bibinfo {author} {\bibfnamefont {N.~G.}\ \bibnamefont
  {Holmes}},\ }\bibfield  {title} {\bibinfo {title} {Skills-focused lab
  instruction improves critical thinking skills and experimentation views for
  all students},\ }\href {https://doi.org/10.1103/PhysRevPhysEducRes.18.010128}
  {\bibfield  {journal} {\bibinfo  {journal} {Physical Review Physics Education
  Research}\ }\textbf {\bibinfo {volume} {18}},\ \bibinfo {pages} {010128}
  (\bibinfo {year} {2022})}\BibitemShut {NoStop}%
\bibitem [{\citenamefont {Werth}\ \emph {et~al.}(2022)\citenamefont {Werth},
  \citenamefont {Hoehn}, \citenamefont {Oliver}, \citenamefont {Fox},\ and\
  \citenamefont {Lewandowski}}]{Werth.2022}%
  \BibitemOpen
  \bibfield  {author} {\bibinfo {author} {\bibfnamefont {A.}~\bibnamefont
  {Werth}}, \bibinfo {author} {\bibfnamefont {J.~R.}\ \bibnamefont {Hoehn}},
  \bibinfo {author} {\bibfnamefont {K.}~\bibnamefont {Oliver}}, \bibinfo
  {author} {\bibfnamefont {M.~F.~J.}\ \bibnamefont {Fox}},\ and\ \bibinfo
  {author} {\bibfnamefont {H.~J.}\ \bibnamefont {Lewandowski}},\ }\bibfield
  {title} {\bibinfo {title} {Instructor perspectives on the emergency
  transition to remote instruction of physics labs},\ }\href
  {https://doi.org/10.1103/PhysRevPhysEducRes.18.020129} {\bibfield  {journal}
  {\bibinfo  {journal} {Physical Review Physics Education Research}\ }\textbf
  {\bibinfo {volume} {18}},\ \bibinfo {pages} {020129} (\bibinfo {year}
  {2022})}\BibitemShut {NoStop}%
\bibitem [{\citenamefont {Pols}(2020)}]{Pols.2020}%
  \BibitemOpen
  \bibfield  {author} {\bibinfo {author} {\bibfnamefont {F.}~\bibnamefont
  {Pols}},\ }\bibfield  {title} {\bibinfo {title} {A physics lab course in
  times of {COVID}-19},\ }\href
  {https://files.eric.ed.gov/fulltext/EJ1261593.pdf} {\bibfield  {journal}
  {\bibinfo  {journal} {Electronic Journal for Research in Science {\&}
  Mathematics Education}\ }\textbf {\bibinfo {volume} {24}},\ \bibinfo {pages}
  {172} (\bibinfo {year} {2020})}\BibitemShut {NoStop}%
\bibitem [{\citenamefont {Bauer}\ \emph {et~al.}(2021)\citenamefont {Bauer},
  \citenamefont {Sacher}, \citenamefont {Habig},\ and\ \citenamefont
  {Fechner}}]{Bauer.2021}%
  \BibitemOpen
  \bibfield  {author} {\bibinfo {author} {\bibfnamefont {A.~B.}\ \bibnamefont
  {Bauer}}, \bibinfo {author} {\bibfnamefont {M.~D.}\ \bibnamefont {Sacher}},
  \bibinfo {author} {\bibfnamefont {S.}~\bibnamefont {Habig}},\ and\ \bibinfo
  {author} {\bibfnamefont {S.}~\bibnamefont {Fechner}},\ }\bibfield  {title}
  {\bibinfo {title} {Laborpraktika auf {D}istanz: Ans{\"a}tze in den
  {N}aturwissenschaften},\ }in\ \href
  {https://www.transcript-verlag.de/media/pdf/e3/4a/a9/oa9783839456903.pdf}
  {\emph {\bibinfo {booktitle} {Hochschule auf Abstand}}},\ \bibinfo {series
  and number} {Hochschulbildung: Lehre und Forschung},\ \bibinfo {editor}
  {edited by\ \bibinfo {editor} {\bibfnamefont {I.}~\bibnamefont {Neiske}},
  \bibinfo {editor} {\bibfnamefont {J.}~\bibnamefont {Osthushenrich}}, \bibinfo
  {editor} {\bibfnamefont {N.}~\bibnamefont {Schaper}}, \bibinfo {editor}
  {\bibfnamefont {U.}~\bibnamefont {Trier}},\ and\ \bibinfo {editor}
  {\bibfnamefont {N.}~\bibnamefont {V{\"o}ing}}}\ (\bibinfo  {publisher}
  {transcript},\ \bibinfo {address} {Bielefeld},\ \bibinfo {year} {2021})\ pp.\
  \bibinfo {pages} {155--168}\BibitemShut {NoStop}%
\bibitem [{\citenamefont {Bradbury}\ and\ \citenamefont
  {Pols}(2020)}]{Bradbury.2020b}%
  \BibitemOpen
  \bibfield  {author} {\bibinfo {author} {\bibfnamefont {F.~R.}\ \bibnamefont
  {Bradbury}}\ and\ \bibinfo {author} {\bibfnamefont {F.}~\bibnamefont
  {Pols}},\ }\bibfield  {title} {\bibinfo {title} {A pandemic-resilient
  open-inquiry physical science lab course which leverages the maker
  movement},\ }\href {https://ejrsme.icrsme.com/article/view/20416} {\bibfield
  {journal} {\bibinfo  {journal} {Electronic Journal for Research in Science
  {\&} Mathematics Education}\ }\textbf {\bibinfo {volume} {24}},\ \bibinfo
  {pages} {60} (\bibinfo {year} {2020})}\BibitemShut {NoStop}%
\bibitem [{\citenamefont {Hut}\ \emph {et~al.}(2020)\citenamefont {Hut},
  \citenamefont {Pols},\ and\ \citenamefont {Verschuur}}]{Hut.2020}%
  \BibitemOpen
  \bibfield  {author} {\bibinfo {author} {\bibfnamefont {R.~W.}\ \bibnamefont
  {Hut}}, \bibinfo {author} {\bibfnamefont {C.~F.~J.}\ \bibnamefont {Pols}},\
  and\ \bibinfo {author} {\bibfnamefont {D.~J.}\ \bibnamefont {Verschuur}},\
  }\bibfield  {title} {\bibinfo {title} {Teaching a hands-on course during
  corona lockdown: from problems to opportunities},\ }\bibfield  {journal}
  {\bibinfo  {journal} {Physics Education}\ }\textbf {\bibinfo {volume} {55}},\
  \href {https://doi.org/10.1088/1361-6552/abb06a} {10.1088/1361-6552/abb06a}
  (\bibinfo {year} {2020})\BibitemShut {NoStop}%
\bibitem [{\citenamefont {Jelicic}\ \emph {et~al.}(2022)\citenamefont
  {Jelicic}, \citenamefont {Geyer}, \citenamefont {Ivanjek}, \citenamefont
  {Klein}, \citenamefont {K{\"u}chemann}, \citenamefont {Dahlkemper},\ and\
  \citenamefont {Susac}}]{Jelicic.folgt}%
  \BibitemOpen
  \bibfield  {author} {\bibinfo {author} {\bibfnamefont {K.}~\bibnamefont
  {Jelicic}}, \bibinfo {author} {\bibfnamefont {M.-A.}\ \bibnamefont {Geyer}},
  \bibinfo {author} {\bibfnamefont {L.}~\bibnamefont {Ivanjek}}, \bibinfo
  {author} {\bibfnamefont {P.}~\bibnamefont {Klein}}, \bibinfo {author}
  {\bibfnamefont {S.}~\bibnamefont {K{\"u}chemann}}, \bibinfo {author}
  {\bibfnamefont {M.~N.}\ \bibnamefont {Dahlkemper}},\ and\ \bibinfo {author}
  {\bibfnamefont {A.}~\bibnamefont {Susac}},\ }\bibfield  {title} {\bibinfo
  {title} {Lab courses for prospective physics teachers: what could we learn
  from the first {COVID-19} lockdown?},\ }\href
  {https://doi.org/10.1088/1361-6404/ac6ea1} {\bibfield  {journal} {\bibinfo
  {journal} {European Journal of Physics}\ }\textbf {\bibinfo {volume} {43}},\
  \bibinfo {pages} {055701} (\bibinfo {year} {2022})}\BibitemShut {NoStop}%
\bibitem [{\citenamefont {Klein}\ \emph {et~al.}(2021)\citenamefont {Klein},
  \citenamefont {Ivanjek}, \citenamefont {Dahlkemper}, \citenamefont
  {Jeli{\v{c}}i{\'c}}, \citenamefont {Geyer}, \citenamefont {K{\"u}chemann},\
  and\ \citenamefont {Susac}}]{Klein.2021b}%
  \BibitemOpen
  \bibfield  {author} {\bibinfo {author} {\bibfnamefont {P.}~\bibnamefont
  {Klein}}, \bibinfo {author} {\bibfnamefont {L.}~\bibnamefont {Ivanjek}},
  \bibinfo {author} {\bibfnamefont {M.~N.}\ \bibnamefont {Dahlkemper}},
  \bibinfo {author} {\bibfnamefont {K.}~\bibnamefont {Jeli{\v{c}}i{\'c}}},
  \bibinfo {author} {\bibfnamefont {M.-A.}\ \bibnamefont {Geyer}}, \bibinfo
  {author} {\bibfnamefont {S.}~\bibnamefont {K{\"u}chemann}},\ and\ \bibinfo
  {author} {\bibfnamefont {A.}~\bibnamefont {Susac}},\ }\bibfield  {title}
  {\bibinfo {title} {Studying physics during the {COVID}-19 pandemic: Student
  assessments of learning achievement, perceived effectiveness of online
  recitations, and online laboratories},\ }\href
  {https://doi.org/10.1103/PhysRevPhysEducRes.17.010117} {\bibfield  {journal}
  {\bibinfo  {journal} {Physical Review Physics Education Research}\ }\textbf
  {\bibinfo {volume} {17}},\ \bibinfo {pages} {010117} (\bibinfo {year}
  {2021})}\BibitemShut {NoStop}%
\bibitem [{\citenamefont {Lahme}\ \emph {et~al.}(2022)\citenamefont {Lahme},
  \citenamefont {Klein}, \citenamefont {Lehtinen}, \citenamefont {Pirinen},
  \citenamefont {Su{\v{s}}ac},\ and\ \citenamefont {Tomrlin}}]{Lahme.2022}%
  \BibitemOpen
  \bibfield  {author} {\bibinfo {author} {\bibfnamefont {S.~Z.}\ \bibnamefont
  {Lahme}}, \bibinfo {author} {\bibfnamefont {P.}~\bibnamefont {Klein}},
  \bibinfo {author} {\bibfnamefont {A.}~\bibnamefont {Lehtinen}}, \bibinfo
  {author} {\bibfnamefont {P.}~\bibnamefont {Pirinen}}, \bibinfo {author}
  {\bibfnamefont {A.}~\bibnamefont {Su{\v{s}}ac}},\ and\ \bibinfo {author}
  {\bibfnamefont {B.}~\bibnamefont {Tomrlin}},\ }\bibfield  {title} {\bibinfo
  {title} {Digi{P}hys{L}ab: {D}igital physics laboratory work for distance
  learning},\ }\href
  {https://ojs.dpg-physik.de/index.php/phydid-b/article/view/1250/1503}
  {\bibfield  {journal} {\bibinfo  {journal} {PhyDid B - Beitr{\"a}ge zur
  Fr{\"u}hjahrstagung - virtuell 2022}\ ,\ \bibinfo {pages} {383}} (\bibinfo
  {year} {2022})}\BibitemShut {NoStop}%
\bibitem [{\citenamefont {Haller}(1999)}]{Haller.1999}%
  \BibitemOpen
  \bibfield  {author} {\bibinfo {author} {\bibfnamefont {K.}~\bibnamefont
  {Haller}},\ }\href@noop {} {\emph {\bibinfo {title} {{\"U}ber den
  Zusammenhang von Handlungen und Zielen: Eine empirische Untersuchung zu
  Lernprozessen im physikalischen Praktikum}}},\ \bibinfo {series} {Studien zum
  Physiklernen}, Vol.~\bibinfo {volume} {5}\ (\bibinfo  {publisher} {Logos},\
  \bibinfo {address} {Berlin},\ \bibinfo {year} {1999})\BibitemShut {NoStop}%
\bibitem [{\citenamefont {Girwidz}\ \emph {et~al.}(2019)\citenamefont
  {Girwidz}, \citenamefont {Thoms}, \citenamefont {Pol}, \citenamefont
  {L{\'o}pez}, \citenamefont {Michelini}, \citenamefont {Stefanel},
  \citenamefont {Greczy{\l}o}, \citenamefont {M{\"u}ller}, \citenamefont
  {Gregorcic},\ and\ \citenamefont {H{\"o}m{\"o}strei}}]{Girwidz.2019}%
  \BibitemOpen
  \bibfield  {author} {\bibinfo {author} {\bibfnamefont {R.}~\bibnamefont
  {Girwidz}}, \bibinfo {author} {\bibfnamefont {L.-J.}\ \bibnamefont {Thoms}},
  \bibinfo {author} {\bibfnamefont {H.}~\bibnamefont {Pol}}, \bibinfo {author}
  {\bibfnamefont {V.}~\bibnamefont {L{\'o}pez}}, \bibinfo {author}
  {\bibfnamefont {M.}~\bibnamefont {Michelini}}, \bibinfo {author}
  {\bibfnamefont {A.}~\bibnamefont {Stefanel}}, \bibinfo {author}
  {\bibfnamefont {T.}~\bibnamefont {Greczy{\l}o}}, \bibinfo {author}
  {\bibfnamefont {A.}~\bibnamefont {M{\"u}ller}}, \bibinfo {author}
  {\bibfnamefont {B.}~\bibnamefont {Gregorcic}},\ and\ \bibinfo {author}
  {\bibfnamefont {M.}~\bibnamefont {H{\"o}m{\"o}strei}},\ }\bibfield  {title}
  {\bibinfo {title} {Physics teaching and learning with multimedia
  applications: a review of teacher-oriented literature in 34 local language
  journals from 2006 to 2015},\ }\href
  {https://doi.org/10.1080/09500693.2019.1597313} {\bibfield  {journal}
  {\bibinfo  {journal} {International Journal of Science Education}\ }\textbf
  {\bibinfo {volume} {41}},\ \bibinfo {pages} {1181} (\bibinfo {year}
  {2019})}\BibitemShut {NoStop}%
\bibitem [{\citenamefont {Franke}\ and\ \citenamefont
  {Wegner}(2022)}]{Franke.2022b}%
  \BibitemOpen
  \bibfield  {author} {\bibinfo {author} {\bibfnamefont {J.}~\bibnamefont
  {Franke}}\ and\ \bibinfo {author} {\bibfnamefont {G.}~\bibnamefont
  {Wegner}},\ }\bibfield  {title} {\bibinfo {title} {Saxonian joint project
  d2c2 ``implementing participatory and discipline-specific approaches to
  digitalization at the university: Competencies connected'': Didactic insights
  into (partially) digitalized workshops and laboratories},\ }\href
  {https://journals.qucosa.de/ll/article/view/58/113} {\bibfield  {journal}
  {\bibinfo  {journal} {Lessons Learned}\ }\textbf {\bibinfo {volume} {2}}
  (\bibinfo {year} {2022})}\BibitemShut {NoStop}%
\bibitem [{\citenamefont {Becker}\ \emph
  {et~al.}(2020{\natexlab{a}})\citenamefont {Becker}, \citenamefont {Klein},
  \citenamefont {G{\"o}{\ss}ling},\ and\ \citenamefont {Kuhn}}]{Becker.2020c}%
  \BibitemOpen
  \bibfield  {author} {\bibinfo {author} {\bibfnamefont {S.}~\bibnamefont
  {Becker}}, \bibinfo {author} {\bibfnamefont {P.}~\bibnamefont {Klein}},
  \bibinfo {author} {\bibfnamefont {A.}~\bibnamefont {G{\"o}{\ss}ling}},\ and\
  \bibinfo {author} {\bibfnamefont {J.}~\bibnamefont {Kuhn}},\ }\bibfield
  {title} {\bibinfo {title} {Using mobile devices to enhance inquiry-based
  learning processes},\ }\href
  {https://doi.org/10.1016/j.learninstruc.2020.101350} {\bibfield  {journal}
  {\bibinfo  {journal} {Learning and Instruction}\ }\textbf {\bibinfo {volume}
  {69}},\ \bibinfo {pages} {101350} (\bibinfo {year}
  {2020}{\natexlab{a}})}\BibitemShut {NoStop}%
\bibitem [{\citenamefont {Hochberg}\ \emph {et~al.}(2018)\citenamefont
  {Hochberg}, \citenamefont {Kuhn},\ and\ \citenamefont
  {M{\"u}ller}}]{Hochberg.2018}%
  \BibitemOpen
  \bibfield  {author} {\bibinfo {author} {\bibfnamefont {K.}~\bibnamefont
  {Hochberg}}, \bibinfo {author} {\bibfnamefont {J.}~\bibnamefont {Kuhn}},\
  and\ \bibinfo {author} {\bibfnamefont {A.}~\bibnamefont {M{\"u}ller}},\
  }\bibfield  {title} {\bibinfo {title} {Using smartphones as experimental
  tools: Effects on interest, curiosity, and learning in physics education},\
  }\href {https://doi.org/10.1007/s10956-018-9731-7} {\bibfield  {journal}
  {\bibinfo  {journal} {Journal of Science Education and Technology}\ }\textbf
  {\bibinfo {volume} {27}},\ \bibinfo {pages} {385} (\bibinfo {year}
  {2018})}\BibitemShut {NoStop}%
\bibitem [{\citenamefont {Pirker}(2017)}]{Pirker.2017}%
  \BibitemOpen
  \bibfield  {author} {\bibinfo {author} {\bibfnamefont {J.}~\bibnamefont
  {Pirker}},\ }\emph {\bibinfo {title} {Immersive and Engaging Forms of Virtual
  Learning: New and improved approaches towards engaging and immersive digital
  learning}},\ \href
  {https://jpirker.com/wp-content/uploads/2013/09/dissertation-pirker.compressed.pdf}
  {\bibinfo {type} {Phd thesis}},\ \bibinfo  {school} {{Technische
  Universit{\"a}t Graz}}, \bibinfo {address} {Graz} (\bibinfo {year}
  {2017})\BibitemShut {NoStop}%
\bibitem [{\citenamefont {de~Jong}\ \emph {et~al.}(2013)\citenamefont
  {de~Jong}, \citenamefont {Linn},\ and\ \citenamefont {Zacharia}}]{Jong.2013}%
  \BibitemOpen
  \bibfield  {author} {\bibinfo {author} {\bibfnamefont {T.}~\bibnamefont
  {de~Jong}}, \bibinfo {author} {\bibfnamefont {M.~C.}\ \bibnamefont {Linn}},\
  and\ \bibinfo {author} {\bibfnamefont {Z.~C.}\ \bibnamefont {Zacharia}},\
  }\bibfield  {title} {\bibinfo {title} {Physical and virtual laboratories in
  science and engineering education},\ }\href
  {https://doi.org/10.1126/science.1230579} {\bibfield  {journal} {\bibinfo
  {journal} {Science}\ }\textbf {\bibinfo {volume} {340}},\ \bibinfo {pages}
  {305} (\bibinfo {year} {2013})}\BibitemShut {NoStop}%
\bibitem [{\citenamefont {Milgram}\ \emph {et~al.}(1994)\citenamefont
  {Milgram}, \citenamefont {Takemura}, \citenamefont {Utsumi},\ and\
  \citenamefont {Kishino}}]{Milgram.1994}%
  \BibitemOpen
  \bibfield  {author} {\bibinfo {author} {\bibfnamefont {P.}~\bibnamefont
  {Milgram}}, \bibinfo {author} {\bibfnamefont {H.}~\bibnamefont {Takemura}},
  \bibinfo {author} {\bibfnamefont {A.}~\bibnamefont {Utsumi}},\ and\ \bibinfo
  {author} {\bibfnamefont {F.}~\bibnamefont {Kishino}},\ }\bibfield  {title}
  {\bibinfo {title} {Augmented reality: A class of displays on the
  reality-virtuality continuum},\ }\href {https://doi.org/10.1117/12.197321}
  {\bibfield  {journal} {\bibinfo  {journal} {Proc. SPIE, Telemanipulator and
  Telepresence Technologies}\ }\textbf {\bibinfo {volume} {2351}},\ \bibinfo
  {pages} {282} (\bibinfo {year} {1994})}\BibitemShut {NoStop}%
\bibitem [{\citenamefont {Ma}\ and\ \citenamefont {Nickerson}(2006)}]{Ma.2006}%
  \BibitemOpen
  \bibfield  {author} {\bibinfo {author} {\bibfnamefont {J.}~\bibnamefont
  {Ma}}\ and\ \bibinfo {author} {\bibfnamefont {J.~V.}\ \bibnamefont
  {Nickerson}},\ }\bibfield  {title} {\bibinfo {title} {Hands-on, simulated,
  and remote laboratories: A comparative literature review},\ }\bibfield
  {journal} {\bibinfo  {journal} {ACM Computing Surveys}\ }\textbf {\bibinfo
  {volume} {38}},\ \href {https://doi.org/10.1145/1132960.1132961}
  {10.1145/1132960.1132961} (\bibinfo {year} {2006})\BibitemShut {NoStop}%
\bibitem [{\citenamefont {Kuhn}\ and\ \citenamefont {Vogt}(2022)}]{Kuhn.2022}%
  \BibitemOpen
  \bibinfo {editor} {\bibfnamefont {J.}~\bibnamefont {Kuhn}}\ and\ \bibinfo
  {editor} {\bibfnamefont {P.}~\bibnamefont {Vogt}},\ eds.,\ \href
  {https://doi.org/10.1007/978-3-030-94044-7} {\emph {\bibinfo {title}
  {Smartphones as Mobile Minilabs in Physics: Edited Volume Featuring more than
  70 Examples from 10 Years The Physics Teacher-column iPhysicsLabs}}},\
  \bibinfo {edition} {1st}\ ed.\ (\bibinfo  {publisher} {Springer},\ \bibinfo
  {address} {Cham},\ \bibinfo {year} {2022})\BibitemShut {NoStop}%
\bibitem [{\citenamefont {Staacks}\ \emph {et~al.}(2018)\citenamefont
  {Staacks}, \citenamefont {H{\"u}tz}, \citenamefont {Heinke},\ and\
  \citenamefont {Stampfer}}]{Staacks.2018}%
  \BibitemOpen
  \bibfield  {author} {\bibinfo {author} {\bibfnamefont {S.}~\bibnamefont
  {Staacks}}, \bibinfo {author} {\bibfnamefont {S.}~\bibnamefont {H{\"u}tz}},
  \bibinfo {author} {\bibfnamefont {H.}~\bibnamefont {Heinke}},\ and\ \bibinfo
  {author} {\bibfnamefont {C.}~\bibnamefont {Stampfer}},\ }\bibfield  {title}
  {\bibinfo {title} {Advanced tools for smartphone-based experiments:
  phyphox},\ }\bibfield  {journal} {\bibinfo  {journal} {Physics Education}\
  }\textbf {\bibinfo {volume} {53}},\ \href
  {https://doi.org/10.1088/1361-6552/aac05e} {10.1088/1361-6552/aac05e}
  (\bibinfo {year} {2018})\BibitemShut {NoStop}%
\bibitem [{\citenamefont {Stampfer}\ \emph {et~al.}(2020)\citenamefont
  {Stampfer}, \citenamefont {Heinke},\ and\ \citenamefont
  {Staacks}}]{Stampfer.2020}%
  \BibitemOpen
  \bibfield  {author} {\bibinfo {author} {\bibfnamefont {C.}~\bibnamefont
  {Stampfer}}, \bibinfo {author} {\bibfnamefont {H.}~\bibnamefont {Heinke}},\
  and\ \bibinfo {author} {\bibfnamefont {S.}~\bibnamefont {Staacks}},\
  }\bibfield  {title} {\bibinfo {title} {A lab in the pocket},\ }\href
  {https://www.nature.com/articles/s41578-020-0184-2.pdf} {\bibfield  {journal}
  {\bibinfo  {journal} {Nature Reviews Materials}\ }\textbf {\bibinfo {volume}
  {5}},\ \bibinfo {pages} {169} (\bibinfo {year} {2020})}\BibitemShut {NoStop}%
\bibitem [{\citenamefont {Klein}\ \emph {et~al.}(2014)\citenamefont {Klein},
  \citenamefont {Hirth}, \citenamefont {Gr{\"o}ber}, \citenamefont {Kuhn},\
  and\ \citenamefont {M{\"u}ller}}]{Klein.2014}%
  \BibitemOpen
  \bibfield  {author} {\bibinfo {author} {\bibfnamefont {P.}~\bibnamefont
  {Klein}}, \bibinfo {author} {\bibfnamefont {M.}~\bibnamefont {Hirth}},
  \bibinfo {author} {\bibfnamefont {S.}~\bibnamefont {Gr{\"o}ber}}, \bibinfo
  {author} {\bibfnamefont {J.}~\bibnamefont {Kuhn}},\ and\ \bibinfo {author}
  {\bibfnamefont {A.}~\bibnamefont {M{\"u}ller}},\ }\bibfield  {title}
  {\bibinfo {title} {Classical experiments revisited: {S}martphones and tablet
  {PC}s as experimental tools in acoustics and optic},\ }\href
  {https://doi.org/10.1088/0031-9120/49/4/412} {\bibfield  {journal} {\bibinfo
  {journal} {Physics Education}\ }\textbf {\bibinfo {volume} {49}},\ \bibinfo
  {pages} {412} (\bibinfo {year} {2014})}\BibitemShut {NoStop}%
\bibitem [{\citenamefont {Kaps}\ and\ \citenamefont
  {Stallmach}(2022)}]{Kaps.2022}%
  \BibitemOpen
  \bibfield  {author} {\bibinfo {author} {\bibfnamefont {A.}~\bibnamefont
  {Kaps}}\ and\ \bibinfo {author} {\bibfnamefont {F.}~\bibnamefont
  {Stallmach}},\ }\bibfield  {title} {\bibinfo {title} {Development and
  didactic analysis of smartphone-based experimental exercises for the smart
  physics lab},\ }\href {https://doi.org/10.1088/1361-6552/ac68c0} {\bibfield
  {journal} {\bibinfo  {journal} {Physics Education}\ }\textbf {\bibinfo
  {volume} {57}},\ \bibinfo {pages} {045038} (\bibinfo {year}
  {2022})}\BibitemShut {NoStop}%
\bibitem [{\citenamefont {Organtini}(2021)}]{Organtini.2021}%
  \BibitemOpen
  \bibfield  {author} {\bibinfo {author} {\bibfnamefont {G.}~\bibnamefont
  {Organtini}},\ }\href {https://doi.org/10.1007/978-3-030-65140-4} {\emph
  {\bibinfo {title} {Physics experiments with Arduino and smartphones}}}\
  (\bibinfo  {publisher} {Springer},\ \bibinfo {address} {Cham},\ \bibinfo
  {year} {2021})\BibitemShut {NoStop}%
\bibitem [{\citenamefont {Monteiro}\ and\ \citenamefont
  {Mart{\'i}}(2022)}]{Monteiro.2022d}%
  \BibitemOpen
  \bibfield  {author} {\bibinfo {author} {\bibfnamefont {M.}~\bibnamefont
  {Monteiro}}\ and\ \bibinfo {author} {\bibfnamefont {A.~C.}\ \bibnamefont
  {Mart{\'i}}},\ }\bibfield  {title} {\bibinfo {title} {Resource letter mds-1:
  Mobile devices and sensors for physics teaching},\ }\href
  {https://doi.org/10.1119/5.0073317} {\bibfield  {journal} {\bibinfo
  {journal} {American Journal of Physics}\ }\textbf {\bibinfo {volume} {90}},\
  \bibinfo {pages} {328} (\bibinfo {year} {2022})}\BibitemShut {NoStop}%
\bibitem [{\citenamefont {Thornton}\ and\ \citenamefont
  {Sokoloff}(1990)}]{Thornton.1990}%
  \BibitemOpen
  \bibfield  {author} {\bibinfo {author} {\bibfnamefont {R.~K.}\ \bibnamefont
  {Thornton}}\ and\ \bibinfo {author} {\bibfnamefont {D.~R.}\ \bibnamefont
  {Sokoloff}},\ }\bibfield  {title} {\bibinfo {title} {Learning motion concepts
  using real--time microcomputer--based laboratory tools},\ }\href
  {https://doi.org/10.1119/1.16350} {\bibfield  {journal} {\bibinfo  {journal}
  {American Journal of Physics}\ }\textbf {\bibinfo {volume} {58}},\ \bibinfo
  {pages} {858} (\bibinfo {year} {1990})}\BibitemShut {NoStop}%
\bibitem [{\citenamefont {Redish}\ \emph {et~al.}(1997)\citenamefont {Redish},
  \citenamefont {Saul},\ and\ \citenamefont {Steinberg}}]{Redish.1997}%
  \BibitemOpen
  \bibfield  {author} {\bibinfo {author} {\bibfnamefont {E.~F.}\ \bibnamefont
  {Redish}}, \bibinfo {author} {\bibfnamefont {J.~M.}\ \bibnamefont {Saul}},\
  and\ \bibinfo {author} {\bibfnamefont {R.~N.}\ \bibnamefont {Steinberg}},\
  }\bibfield  {title} {\bibinfo {title} {On the effectiveness of
  active-engagement microcomputer-based laboratories},\ }\href
  {https://doi.org/10.1119/1.18498} {\bibfield  {journal} {\bibinfo  {journal}
  {American Journal of Physics}\ }\textbf {\bibinfo {volume} {65}},\ \bibinfo
  {pages} {45} (\bibinfo {year} {1997})}\BibitemShut {NoStop}%
\bibitem [{\citenamefont {Thees}\ \emph {et~al.}(2022)\citenamefont {Thees},
  \citenamefont {Altmeyer}, \citenamefont {Kapp}, \citenamefont {Rexigel},
  \citenamefont {Beil}, \citenamefont {Klein}, \citenamefont {Malone},
  \citenamefont {Br{\"u}nken},\ and\ \citenamefont {Kuhn}}]{Thees.2022}%
  \BibitemOpen
  \bibfield  {author} {\bibinfo {author} {\bibfnamefont {M.}~\bibnamefont
  {Thees}}, \bibinfo {author} {\bibfnamefont {K.}~\bibnamefont {Altmeyer}},
  \bibinfo {author} {\bibfnamefont {S.}~\bibnamefont {Kapp}}, \bibinfo {author}
  {\bibfnamefont {E.}~\bibnamefont {Rexigel}}, \bibinfo {author} {\bibfnamefont
  {F.}~\bibnamefont {Beil}}, \bibinfo {author} {\bibfnamefont {P.}~\bibnamefont
  {Klein}}, \bibinfo {author} {\bibfnamefont {S.}~\bibnamefont {Malone}},
  \bibinfo {author} {\bibfnamefont {R.}~\bibnamefont {Br{\"u}nken}},\ and\
  \bibinfo {author} {\bibfnamefont {J.}~\bibnamefont {Kuhn}},\ }\bibfield
  {title} {\bibinfo {title} {Augmented {R}eality for presenting real-time data
  during students' laboratory work: Comparing a head-mounted display with a
  separate display},\ }\bibfield  {journal} {\bibinfo  {journal} {Frontiers in
  Psychology}\ }\textbf {\bibinfo {volume} {13}},\ \href
  {https://doi.org/10.3389/fpsyg.2022.804742} {10.3389/fpsyg.2022.804742}
  (\bibinfo {year} {2022})\BibitemShut {NoStop}%
\bibitem [{\citenamefont {Schlummer}\ \emph {et~al.}(2021)\citenamefont
  {Schlummer}, \citenamefont {Abazi}, \citenamefont {Borkamp}, \citenamefont
  {Laustr{\"o}er}, \citenamefont {Pernice}, \citenamefont {Schuck},
  \citenamefont {Schulz-Schaeffer}, \citenamefont {Heusler},\ and\
  \citenamefont {Laumann}}]{Schlummer.2021}%
  \BibitemOpen
  \bibfield  {author} {\bibinfo {author} {\bibfnamefont {P.}~\bibnamefont
  {Schlummer}}, \bibinfo {author} {\bibfnamefont {A.}~\bibnamefont {Abazi}},
  \bibinfo {author} {\bibfnamefont {R.}~\bibnamefont {Borkamp}}, \bibinfo
  {author} {\bibfnamefont {J.}~\bibnamefont {Laustr{\"o}er}}, \bibinfo {author}
  {\bibfnamefont {W.}~\bibnamefont {Pernice}}, \bibinfo {author} {\bibfnamefont
  {C.}~\bibnamefont {Schuck}}, \bibinfo {author} {\bibfnamefont
  {R.}~\bibnamefont {Schulz-Schaeffer}}, \bibinfo {author} {\bibfnamefont
  {S.}~\bibnamefont {Heusler}},\ and\ \bibinfo {author} {\bibfnamefont
  {D.}~\bibnamefont {Laumann}},\ }\bibfield  {title} {\bibinfo {title}
  {Physikalische {M}odelle erfahrbar machen: {M}ixed {R}eality im
  {P}raktikum},\ }\href
  {http://www.phydid.de/index.php/phydid-b/article/view/1128/1217} {\bibfield
  {journal} {\bibinfo  {journal} {PhyDid B - Didaktik der Physik - Beitr{\"a}ge
  zur DPG Fr{\"u}hjahrstagung - virtuell 2021}\ ,\ \bibinfo {pages} {415}}
  (\bibinfo {year} {2021})}\BibitemShut {NoStop}%
\bibitem [{\citenamefont {Haase}\ \emph {et~al.}(2018)\citenamefont {Haase},
  \citenamefont {Pfaff}, \citenamefont {Ermle}, \citenamefont {Kirstein},\ and\
  \citenamefont {Nordmeier}}]{Haase.2018}%
  \BibitemOpen
  \bibfield  {author} {\bibinfo {author} {\bibfnamefont {S.}~\bibnamefont
  {Haase}}, \bibinfo {author} {\bibfnamefont {M.}~\bibnamefont {Pfaff}},
  \bibinfo {author} {\bibfnamefont {D.}~\bibnamefont {Ermle}}, \bibinfo
  {author} {\bibfnamefont {J.}~\bibnamefont {Kirstein}},\ and\ \bibinfo
  {author} {\bibfnamefont {V.}~\bibnamefont {Nordmeier}},\ }\bibfield  {title}
  {\bibinfo {title} {Interaktive {B}ildschirmexperimente als {S}ystemkomponente
  der webbasierten {L}ernplattform tet.folio},\ }\href
  {https://docplayer.org/134232398-Interaktive-bildschirmexperimente-als-systemkomponente-der-webbasierten-lernplattform-tet-folio.html}
  {\bibfield  {journal} {\bibinfo  {journal} {PhyDid B - Didaktik der Physik -
  Beitr{\"a}ge zur DPG-Fr{\"u}hjahrstagung - W{\"u}rzburg 2018}\ ,\ \bibinfo
  {pages} {333}} (\bibinfo {year} {2018})}\BibitemShut {NoStop}%
\bibitem [{\citenamefont {{Ait Tahar}}\ \emph {et~al.}(2019)\citenamefont {{Ait
  Tahar}}, \citenamefont {Schadschneider},\ and\ \citenamefont
  {Stollenwerk}}]{AitTahar.2019}%
  \BibitemOpen
  \bibfield  {author} {\bibinfo {author} {\bibfnamefont {M.}~\bibnamefont {{Ait
  Tahar}}}, \bibinfo {author} {\bibfnamefont {A.}~\bibnamefont
  {Schadschneider}},\ and\ \bibinfo {author} {\bibfnamefont {J.}~\bibnamefont
  {Stollenwerk}},\ }\bibfield  {title} {\bibinfo {title} {Technical realisation
  of a remote-controlled forced mechanic oscillation experiment through the
  internet},\ }\href {https://doi.org/10.1088/1361-6552/aae9b0} {\bibfield
  {journal} {\bibinfo  {journal} {Physics Education}\ }\textbf {\bibinfo
  {volume} {54}},\ \bibinfo {pages} {015012} (\bibinfo {year}
  {2019})}\BibitemShut {NoStop}%
\bibitem [{\citenamefont {Thoms}(2019)}]{Thoms.2019}%
  \BibitemOpen
  \bibfield  {author} {\bibinfo {author} {\bibfnamefont {L.-J.}\ \bibnamefont
  {Thoms}},\ }\href {https://doi.org/10.1007/978-3-658-25708-8} {\emph
  {\bibinfo {title} {Spektrometrie im Fernlabor}}}\ (\bibinfo  {publisher}
  {{Springer Fachmedien}},\ \bibinfo {address} {Wiesbaden},\ \bibinfo {year}
  {2019})\BibitemShut {NoStop}%
\bibitem [{\citenamefont {Price}\ and\ \citenamefont
  {Price-Mohr}(2019)}]{Price.2019}%
  \BibitemOpen
  \bibfield  {author} {\bibinfo {author} {\bibfnamefont {C.~B.}\ \bibnamefont
  {Price}}\ and\ \bibinfo {author} {\bibfnamefont {R.}~\bibnamefont
  {Price-Mohr}},\ }\bibfield  {title} {\bibinfo {title} {Physlab: a 3{D}
  virtual physics laboratory of simulated experiments for advanced physics
  learning},\ }\href
  {https://iopscience.iop.org/article/10.1088/1361-6552/ab0005/pdf} {\bibfield
  {journal} {\bibinfo  {journal} {Physics Education}\ }\textbf {\bibinfo
  {volume} {54}} (\bibinfo {year} {2019})}\BibitemShut {NoStop}%
\bibitem [{\citenamefont {Porter}\ \emph {et~al.}(2020)\citenamefont {Porter},
  \citenamefont {Smith}, \citenamefont {Stagar}, \citenamefont {Simmons},
  \citenamefont {Nieberding}, \citenamefont {Orban}, \citenamefont {Brown},\
  and\ \citenamefont {Ayers}}]{Porter.2020}%
  \BibitemOpen
  \bibfield  {author} {\bibinfo {author} {\bibfnamefont {C.~D.}\ \bibnamefont
  {Porter}}, \bibinfo {author} {\bibfnamefont {J.~R.~H.}\ \bibnamefont
  {Smith}}, \bibinfo {author} {\bibfnamefont {E.~M.}\ \bibnamefont {Stagar}},
  \bibinfo {author} {\bibfnamefont {A.}~\bibnamefont {Simmons}}, \bibinfo
  {author} {\bibfnamefont {M.}~\bibnamefont {Nieberding}}, \bibinfo {author}
  {\bibfnamefont {C.~M.}\ \bibnamefont {Orban}}, \bibinfo {author}
  {\bibfnamefont {J.}~\bibnamefont {Brown}},\ and\ \bibinfo {author}
  {\bibfnamefont {A.}~\bibnamefont {Ayers}},\ }\bibfield  {title} {\bibinfo
  {title} {Using virtual reality in electrostatics instruction: The impact of
  training},\ }\href {https://doi.org/10.1103/PhysRevPhysEducRes.16.020119}
  {\bibfield  {journal} {\bibinfo  {journal} {Physical Review Physics Education
  Research}\ }\textbf {\bibinfo {volume} {16}},\ \bibinfo {pages} {020119}
  (\bibinfo {year} {2020})}\BibitemShut {NoStop}%
\bibitem [{\citenamefont {Brixner}\ and\ \citenamefont {von
  Mammen}(2021)}]{Brixner.2021}%
  \BibitemOpen
  \bibfield  {author} {\bibinfo {author} {\bibfnamefont {T.}~\bibnamefont
  {Brixner}}\ and\ \bibinfo {author} {\bibfnamefont {S.}~\bibnamefont {von
  Mammen}},\ }\href {https://www.uni-wuerzburg.de/en/femtopro/femtopro/}
  {\bibinfo {title} {femtopro}} (\bibinfo {year} {2021})\BibitemShut {NoStop}%
\bibitem [{\citenamefont {Samsonau}(2018)}]{Samsonau.2018}%
  \BibitemOpen
  \bibfield  {author} {\bibinfo {author} {\bibfnamefont {S.~V.}\ \bibnamefont
  {Samsonau}},\ }\bibfield  {title} {\bibinfo {title} {Computer simulations
  combined with experiments for a calculus-based physics laboratory course},\
  }\href {https://doi.org/10.1088/1361-6552/aacef5} {\bibfield  {journal}
  {\bibinfo  {journal} {Physics Education}\ }\textbf {\bibinfo {volume} {53}},\
  \bibinfo {pages} {055013} (\bibinfo {year} {2018})}\BibitemShut {NoStop}%
\bibitem [{\citenamefont {Spencer}(2005)}]{Spencer.2005}%
  \BibitemOpen
  \bibfield  {author} {\bibinfo {author} {\bibfnamefont {R.~L.}\ \bibnamefont
  {Spencer}},\ }\bibfield  {title} {\bibinfo {title} {Teaching computational
  physics as a laboratory sequence},\ }\href
  {https://doi.org/10.1119/1.1842751} {\bibfield  {journal} {\bibinfo
  {journal} {American Journal of Physics}\ }\textbf {\bibinfo {volume} {73}},\
  \bibinfo {pages} {151} (\bibinfo {year} {2005})}\BibitemShut {NoStop}%
\bibitem [{\citenamefont {Nix}(2023)}]{NixOliver.}%
  \BibitemOpen
  \bibfield  {author} {\bibinfo {author} {\bibfnamefont {O.}~\bibnamefont
  {Nix}},\ }\href {https://phet.colorado.edu/} {\bibinfo {title} {Ph{ET}
  interactive simulations for science and math}} (\bibinfo {year}
  {2023})\BibitemShut {NoStop}%
\bibitem [{\citenamefont {{American Association of Physics
  Teachers}}(1997)}]{AmericanAssociationofPhysicsTeachers.1997}%
  \BibitemOpen
  \bibfield  {author} {\bibinfo {author} {\bibnamefont {{American Association
  of Physics Teachers}}},\ }\bibfield  {title} {\bibinfo {title} {Goals of the
  introductory physics laboratory},\ }\href {https://doi.org/10.1119/1.2344803}
  {\bibfield  {journal} {\bibinfo  {journal} {The Physics Teacher}\ }\textbf
  {\bibinfo {volume} {35}},\ \bibinfo {pages} {546} (\bibinfo {year}
  {1997})}\BibitemShut {NoStop}%
\bibitem [{\citenamefont {Thoms}\ \emph {et~al.}(2021)\citenamefont {Thoms},
  \citenamefont {Meier}, \citenamefont {Huwer}, \citenamefont {Thyssen},
  \citenamefont {von Kotzebue}, \citenamefont {Becker}, \citenamefont
  {Kremser}, \citenamefont {Finger},\ and\ \citenamefont
  {Bruckermann}}]{Thoms.2021}%
  \BibitemOpen
  \bibfield  {author} {\bibinfo {author} {\bibfnamefont {L.-J.}\ \bibnamefont
  {Thoms}}, \bibinfo {author} {\bibfnamefont {M.}~\bibnamefont {Meier}},
  \bibinfo {author} {\bibfnamefont {J.}~\bibnamefont {Huwer}}, \bibinfo
  {author} {\bibfnamefont {C.}~\bibnamefont {Thyssen}}, \bibinfo {author}
  {\bibfnamefont {L.}~\bibnamefont {von Kotzebue}}, \bibinfo {author}
  {\bibfnamefont {S.}~\bibnamefont {Becker}}, \bibinfo {author} {\bibfnamefont
  {E.}~\bibnamefont {Kremser}}, \bibinfo {author} {\bibfnamefont
  {A.}~\bibnamefont {Finger}},\ and\ \bibinfo {author} {\bibfnamefont
  {T.}~\bibnamefont {Bruckermann}},\ }\bibfield  {title} {\bibinfo {title}
  {Di{K}o{LAN}: A framework to identify and classify digital competencies for
  teaching in science education and to restructure pre-service teacher
  training},\ }in\ \href {https://www.learntechlib.org/primary/p/219329/}
  {\emph {\bibinfo {booktitle} {Society for Information Technology {\&}~Teacher
  Education International Conference}}},\ \bibinfo {editor} {edited by\
  \bibinfo {editor} {\bibfnamefont {E.}~\bibnamefont {Langran}}\ and\ \bibinfo
  {editor} {\bibfnamefont {L.}~\bibnamefont {Archambault}}}\ (\bibinfo
  {publisher} {{Association for the Advancement of Computing in Education
  (AACE)}},\ \bibinfo {year} {2021})\ pp.\ \bibinfo {pages}
  {1652--1657}\BibitemShut {NoStop}%
\bibitem [{\citenamefont {Becker}\ \emph
  {et~al.}(2020{\natexlab{b}})\citenamefont {Becker}, \citenamefont
  {Me{\ss}inger-Koppelt},\ and\ \citenamefont {Thyssen}}]{Becker.2020b}%
  \BibitemOpen
  \bibinfo {editor} {\bibfnamefont {S.}~\bibnamefont {Becker}}, \bibinfo
  {editor} {\bibfnamefont {J.}~\bibnamefont {Me{\ss}inger-Koppelt}},\ and\
  \bibinfo {editor} {\bibfnamefont {C.}~\bibnamefont {Thyssen}},\ eds.,\ \href
  {https://www.joachim-herz-stiftung.de/fileadmin/Redaktion/JHS_Digitale_Basiskompetenzen_web_srgb.pdf}
  {\emph {\bibinfo {title} {Digitale Basiskompetenzen: Orientierungshilfe und
  Praxisbeispiele f{\"u}r die universit{\"a}re Lehramtsausbildung in den
  Naturwissenschaften}}},\ \bibinfo {edition} {1st}\ ed.\ (\bibinfo
  {publisher} {{Joachim Herz Stiftung}},\ \bibinfo {address} {Hamburg},\
  \bibinfo {year} {2020})\BibitemShut {NoStop}%
\bibitem [{\citenamefont {Mishra}\ and\ \citenamefont
  {Koehler}(2006)}]{Mishra.2006}%
  \BibitemOpen
  \bibfield  {author} {\bibinfo {author} {\bibfnamefont {P.}~\bibnamefont
  {Mishra}}\ and\ \bibinfo {author} {\bibfnamefont {M.~J.}\ \bibnamefont
  {Koehler}},\ }\bibfield  {title} {\bibinfo {title} {Technological pedagogical
  content knowledge: A framework for teacher knowledge},\ }\href
  {https://punyamishra.com/wp-content/uploads/2008/01/mishra-koehler-tcr2006.pdf}
  {\bibfield  {journal} {\bibinfo  {journal} {Teachers College Record}\
  }\textbf {\bibinfo {volume} {108}},\ \bibinfo {pages} {1017} (\bibinfo {year}
  {2006})}\BibitemShut {NoStop}%
\bibitem [{\citenamefont {{Arbeitsgruppe Digitale Basiskompetenzen / Workgroup
  Digital Core
  Competencies}}(2020)}]{ArbeitsgruppeDigitaleBasiskompetenzenWorkgroupDigitalCoreCompetencies.2020}%
  \BibitemOpen
  \bibfield  {author} {\bibinfo {author} {\bibnamefont {{Arbeitsgruppe Digitale
  Basiskompetenzen / Workgroup Digital Core Competencies}}},\ }\href
  {https://dikolan.de/us/competencies-american-english} {\bibinfo {title}
  {Di{K}o{LAN} digital competencies for teaching in science education:
  Comptencies}} (\bibinfo {year} {2020})\BibitemShut {NoStop}%
\bibitem [{\citenamefont {Gro{\ss}e-Heilmann}\ \emph
  {et~al.}(2021)\citenamefont {Gro{\ss}e-Heilmann}, \citenamefont {Riese},
  \citenamefont {Burde}, \citenamefont {Schubatzky},\ and\ \citenamefont
  {Weiler}}]{GroeHeilmann.2021}%
  \BibitemOpen
  \bibfield  {author} {\bibinfo {author} {\bibfnamefont {R.}~\bibnamefont
  {Gro{\ss}e-Heilmann}}, \bibinfo {author} {\bibfnamefont {J.}~\bibnamefont
  {Riese}}, \bibinfo {author} {\bibfnamefont {J.-P.}\ \bibnamefont {Burde}},
  \bibinfo {author} {\bibfnamefont {T.}~\bibnamefont {Schubatzky}},\ and\
  \bibinfo {author} {\bibfnamefont {D.}~\bibnamefont {Weiler}},\ }\bibfield
  {title} {\bibinfo {title} {Erwerb und {M}essung physikdidaktischer
  {K}ompetenzen zum {E}insatz digitaler {M}edien},\ }\href
  {https://ojs.dpg-physik.de/index.php/phydid-b/article/view/1127/1216}
  {\bibfield  {journal} {\bibinfo  {journal} {PhyDid B - Didaktik der Physik -
  Beitr{\"a}ge zur DPG Fr{\"u}hjahrstagung - virtuell 2021}\ ,\ \bibinfo
  {pages} {171}} (\bibinfo {year} {2021})}\BibitemShut {NoStop}%
\bibitem [{\citenamefont {Brandhofer}\ \emph {et~al.}(2019)\citenamefont
  {Brandhofer}, \citenamefont {Miglbauer}, \citenamefont {Fikisz},
  \citenamefont {H{\"o}fler}, \citenamefont {Kayali}, \citenamefont {Steiner},
  \citenamefont {Prohaska},\ and\ \citenamefont {Riepl}}]{Brandhofer.2019}%
  \BibitemOpen
  \bibfield  {author} {\bibinfo {author} {\bibfnamefont {G.}~\bibnamefont
  {Brandhofer}}, \bibinfo {author} {\bibfnamefont {M.}~\bibnamefont
  {Miglbauer}}, \bibinfo {author} {\bibfnamefont {W.}~\bibnamefont {Fikisz}},
  \bibinfo {author} {\bibfnamefont {E.}~\bibnamefont {H{\"o}fler}}, \bibinfo
  {author} {\bibfnamefont {F.}~\bibnamefont {Kayali}}, \bibinfo {author}
  {\bibfnamefont {M.}~\bibnamefont {Steiner}}, \bibinfo {author} {\bibfnamefont
  {J.}~\bibnamefont {Prohaska}},\ and\ \bibinfo {author} {\bibfnamefont
  {A.}~\bibnamefont {Riepl}},\ }\href
  {https://www.virtuelle-ph.at/wp-content/uploads/2021/04/Grafik-und-Deskriptoren_Langfassung_adapt-2021.pdf}
  {\bibinfo {title} {Digi.komp{P}: Digitale {K}ompetenzen f{\"u}r
  {P}{\"a}dagoginnen und {P}{\"a}dagogen}} (\bibinfo {year} {2019})\BibitemShut
  {NoStop}%
\bibitem [{\citenamefont {Redecker}\ and\ \citenamefont
  {Punie}(2017)}]{Redecker.2017}%
  \BibitemOpen
  \bibfield  {author} {\bibinfo {author} {\bibfnamefont {C.}~\bibnamefont
  {Redecker}}\ and\ \bibinfo {author} {\bibfnamefont {Y.}~\bibnamefont
  {Punie}},\ }\href {https://doi.org/10.2760/178382} {\emph {\bibinfo {title}
  {European Framework for the Digital Competence of Educators: DigCompEdu}}}\
  (\bibinfo  {publisher} {{Publications Office of the European Union}},\
  \bibinfo {address} {Luxembourg},\ \bibinfo {year} {2017})\BibitemShut
  {NoStop}%
\bibitem [{\citenamefont {Carretero}\ \emph {et~al.}(2017)\citenamefont
  {Carretero}, \citenamefont {Vuorikari},\ and\ \citenamefont
  {Punie}}]{Carretero.2017}%
  \BibitemOpen
  \bibfield  {author} {\bibinfo {author} {\bibfnamefont {S.}~\bibnamefont
  {Carretero}}, \bibinfo {author} {\bibfnamefont {R.}~\bibnamefont
  {Vuorikari}},\ and\ \bibinfo {author} {\bibfnamefont {Y.}~\bibnamefont
  {Punie}},\ }\href
  {https://publications.jrc.ec.europa.eu/repository/handle/JRC106281} {\emph
  {\bibinfo {title} {DigComp 2.1: The Digital Competence Framework for
  citizens: With eight proficiency levels and examples of use}}}\ (\bibinfo
  {publisher} {{Publications Office of the European Union}},\ \bibinfo
  {address} {Luxembourg},\ \bibinfo {year} {2017})\BibitemShut {NoStop}%
\bibitem [{\citenamefont {Ghomi}\ \emph {et~al.}(2020)\citenamefont {Ghomi},
  \citenamefont {Dictus}, \citenamefont {Pinkwart},\ and\ \citenamefont
  {Tiemann}}]{Ghomi.2020}%
  \BibitemOpen
  \bibfield  {author} {\bibinfo {author} {\bibfnamefont {M.}~\bibnamefont
  {Ghomi}}, \bibinfo {author} {\bibfnamefont {C.}~\bibnamefont {Dictus}},
  \bibinfo {author} {\bibfnamefont {N.}~\bibnamefont {Pinkwart}},\ and\
  \bibinfo {author} {\bibfnamefont {R.}~\bibnamefont {Tiemann}},\ }\bibfield
  {title} {\bibinfo {title} {Dig{C}omp{E}du f{\"u}r {MINT}: Digitale
  {K}ompetenz von {MINT}-{L}ehrkr{\"a}ften},\ }\href
  {https://doi.org/10.18716/ojs/kON/2020.1.1} {\bibfield  {journal} {\bibinfo
  {journal} {K{\"o}lner Online Journal f{\"u}r Lehrer*innenbildung}\ }\textbf
  {\bibinfo {volume} {1}},\ \bibinfo {pages} {1} (\bibinfo {year}
  {2020})}\BibitemShut {NoStop}%
\bibitem [{\citenamefont {Schmechting}\ \emph {et~al.}(2020)\citenamefont
  {Schmechting}, \citenamefont {Puderbach}, \citenamefont {Schellhammer},\ and\
  \citenamefont {Gehrmann}}]{Schmechting.2020}%
  \BibitemOpen
  \bibfield  {author} {\bibinfo {author} {\bibfnamefont {N.}~\bibnamefont
  {Schmechting}}, \bibinfo {author} {\bibfnamefont {R.}~\bibnamefont
  {Puderbach}}, \bibinfo {author} {\bibfnamefont {S.}~\bibnamefont
  {Schellhammer}},\ and\ \bibinfo {author} {\bibfnamefont {A.}~\bibnamefont
  {Gehrmann}},\ }\href
  {https://tu-dresden.de/zlsb/ressourcen/dateien/tud-sylber/Lehrkraeftebefragung_Digitalisierung_Broschuere_2020.pdf?lang=de}
  {\bibinfo {title} {Einsatz von und {U}mgang mit digitalen {M}edien und
  {I}nhalten in {U}nterricht und {S}chule: Befunde einer
  {L}ehrkr{\"a}ftebefragung zu beruflichen {E}rfahrungen und {Ü}berzeugungen
  von {L}ehrerinnen und {L}ehrern in {S}achsen 2019}} (\bibinfo {year}
  {2020})\BibitemShut {NoStop}%
\bibitem [{\citenamefont {Schulz-Zander}(2001)}]{SchulzZander.2001}%
  \BibitemOpen
  \bibfield  {author} {\bibinfo {author} {\bibfnamefont {R.}~\bibnamefont
  {Schulz-Zander}},\ }\bibfield  {title} {\bibinfo {title} {Lernen mit neuen
  {M}edien in der {S}chule},\ }in\ \href {https://doi.org/10.25656/01:7922}
  {\emph {\bibinfo {booktitle} {Zukunftsfragen der {B}ildung}}},\ \bibinfo
  {series} {Zeitschrift f{\"u}r P{\"a}dagogik, P{\"a}dagogik}, Vol.~\bibinfo
  {volume} {43},\ \bibinfo {editor} {edited by\ \bibinfo {editor}
  {\bibfnamefont {J.}~\bibnamefont {Oelkers}}}\ (\bibinfo  {publisher}
  {Beltz},\ \bibinfo {address} {Weinheim},\ \bibinfo {year} {2001})\ pp.\
  \bibinfo {pages} {181--195}\BibitemShut {NoStop}%
\bibitem [{\citenamefont {Kaiser}\ and\ \citenamefont
  {Rice}(1974)}]{Kaiser.1974}%
  \BibitemOpen
  \bibfield  {author} {\bibinfo {author} {\bibfnamefont {H.~F.}\ \bibnamefont
  {Kaiser}}\ and\ \bibinfo {author} {\bibfnamefont {J.}~\bibnamefont {Rice}},\
  }\bibfield  {title} {\bibinfo {title} {Little {J}iffy, {M}ark {IV}},\ }\href
  {https://doi.org/10.1177/001316447403400115} {\bibfield  {journal} {\bibinfo
  {journal} {Educational and Psychological Measurement}\ }\textbf {\bibinfo
  {volume} {34}},\ \bibinfo {pages} {111} (\bibinfo {year} {1974})}\BibitemShut
  {NoStop}%
\bibitem [{\citenamefont {BARTLETT}(1950)}]{BARTLETT.1950}%
  \BibitemOpen
  \bibfield  {author} {\bibinfo {author} {\bibfnamefont {M.~S.}\ \bibnamefont
  {BARTLETT}},\ }\bibfield  {title} {\bibinfo {title} {Tests of significance in
  factor analysis},\ }\href
  {https://doi.org/10.1111/j.2044-8317.1950.tb00285.x} {\bibfield  {journal}
  {\bibinfo  {journal} {British Journal of Statistical Psychology}\ }\textbf
  {\bibinfo {volume} {3}},\ \bibinfo {pages} {77} (\bibinfo {year}
  {1950})}\BibitemShut {NoStop}%
\bibitem [{\citenamefont {Guttman}(1954)}]{Guttman.1954}%
  \BibitemOpen
  \bibfield  {author} {\bibinfo {author} {\bibfnamefont {L.}~\bibnamefont
  {Guttman}},\ }\bibfield  {title} {\bibinfo {title} {A new approach to factor
  analysis: The radex},\ }in\ \href@noop {} {\emph {\bibinfo {booktitle}
  {Mathematical Thinking in the Social Sciences}}},\ \bibinfo {editor} {edited
  by\ \bibinfo {editor} {\bibfnamefont {P.~F.}\ \bibnamefont {Lazarsfeld}}}\
  (\bibinfo  {publisher} {{Free Press}},\ \bibinfo {address} {New York},\
  \bibinfo {year} {1954})\ pp.\ \bibinfo {pages} {258--348}\BibitemShut
  {NoStop}%
\bibitem [{\citenamefont {Kaiser}(1960)}]{Kaiser.1960}%
  \BibitemOpen
  \bibfield  {author} {\bibinfo {author} {\bibfnamefont {H.~F.}\ \bibnamefont
  {Kaiser}},\ }\bibfield  {title} {\bibinfo {title} {The application of
  electronic computers to factor analysis},\ }\href
  {https://doi.org/10.1177/001316446002000116} {\bibfield  {journal} {\bibinfo
  {journal} {Educational and Psychological Measurement}\ }\textbf {\bibinfo
  {volume} {20}},\ \bibinfo {pages} {141} (\bibinfo {year} {1960})}\BibitemShut
  {NoStop}%
\bibitem [{\citenamefont {Chen}\ \emph {et~al.}(2012)\citenamefont {Chen},
  \citenamefont {Lo}, \citenamefont {Lin}, \citenamefont {Liang}, \citenamefont
  {Chang}, \citenamefont {Hwang}, \citenamefont {Chiou}, \citenamefont {Wu},
  \citenamefont {Lee}, \citenamefont {Wu}, \citenamefont {Wang},\ and\
  \citenamefont {Tsai}}]{Chen.2012}%
  \BibitemOpen
  \bibfield  {author} {\bibinfo {author} {\bibfnamefont {S.}~\bibnamefont
  {Chen}}, \bibinfo {author} {\bibfnamefont {H.-C.}\ \bibnamefont {Lo}},
  \bibinfo {author} {\bibfnamefont {J.-W.}\ \bibnamefont {Lin}}, \bibinfo
  {author} {\bibfnamefont {J.-C.}\ \bibnamefont {Liang}}, \bibinfo {author}
  {\bibfnamefont {H.-Y.}\ \bibnamefont {Chang}}, \bibinfo {author}
  {\bibfnamefont {F.-K.}\ \bibnamefont {Hwang}}, \bibinfo {author}
  {\bibfnamefont {G.-L.}\ \bibnamefont {Chiou}}, \bibinfo {author}
  {\bibfnamefont {Y.-T.}\ \bibnamefont {Wu}}, \bibinfo {author} {\bibfnamefont
  {S.~W.-Y.}\ \bibnamefont {Lee}}, \bibinfo {author} {\bibfnamefont {H.-K.}\
  \bibnamefont {Wu}}, \bibinfo {author} {\bibfnamefont {C.-Y.}\ \bibnamefont
  {Wang}},\ and\ \bibinfo {author} {\bibfnamefont {C.-C.}\ \bibnamefont
  {Tsai}},\ }\bibfield  {title} {\bibinfo {title} {Development and implications
  of technology in reform-based physics laboratories},\ }\href
  {https://doi.org/10.1103/PhysRevSTPER.8.020113} {\bibfield  {journal}
  {\bibinfo  {journal} {Physical Review Special Topics - Physics Education
  Research}\ }\textbf {\bibinfo {volume} {8}},\ \bibinfo {pages} {020113}
  (\bibinfo {year} {2012})}\BibitemShut {NoStop}%
\bibitem [{\citenamefont {OpenAI}(2022)}]{OpenAI.2022}%
  \BibitemOpen
  \bibfield  {author} {\bibinfo {author} {\bibnamefont {OpenAI}},\ }\href
  {https://openai.com/blog/chatgpt} {\bibinfo {title} {Introducing
  {C}hat{GPT}}} (\bibinfo {year} {2022})\BibitemShut {NoStop}%
\bibitem [{\citenamefont {Okonkwo}\ and\ \citenamefont
  {Ade-Ibijola}(2021)}]{Okonkwo.2021}%
  \BibitemOpen
  \bibfield  {author} {\bibinfo {author} {\bibfnamefont {C.~W.}\ \bibnamefont
  {Okonkwo}}\ and\ \bibinfo {author} {\bibfnamefont {A.}~\bibnamefont
  {Ade-Ibijola}},\ }\bibfield  {title} {\bibinfo {title} {Chatbots applications
  in education: A systematic review},\ }\href
  {https://doi.org/10.1016/j.caeai.2021.100033} {\bibfield  {journal} {\bibinfo
   {journal} {Computers and Education: Artificial Intelligence}\ }\textbf
  {\bibinfo {volume} {2}},\ \bibinfo {pages} {100033} (\bibinfo {year}
  {2021})}\BibitemShut {NoStop}%
\end{thebibliography}%

\end{document}